\documentclass[singlecolumn,pre,preprintnumbers,superscriptaddress]{revtex4}
\usepackage{textcomp}
\usepackage{amsmath}
\usepackage{amssymb}
\usepackage{graphicx}
\usepackage{color}
\usepackage{soul}
\usepackage{footmisc}
\usepackage{gensymb}
\usepackage{array,makecell}  
\usepackage{multirow}  

\newcommand{\cmp}
{\affiliation{Condensed Matter Physics Division, 
Saha Institute of Nuclear Physics, 1/AF Bidhannagar, Kolkata 700064, India.}}
\newcommand{\barasat}
{\affiliation{Barasat Government College, Barasat, Kolkata 700124, India.}}
\newcommand{\snb}
{\affiliation{S. N. Bose National Centre for Basic Sciences, Kolkata 700106, India.}}
\newcommand{\isi}
{\affiliation{Economic Research Unit, Indian Statistical Institute, Kolkata-700108, India.}}

\begin{document}
\title{Econophysics Through Computation}
\author{Antika Sinha}
\email{antikasinha@gmail.com}
\affiliation{Asutosh College, Kolkata, 700026, India}
\author{Sudip Mukherjee}
\email{sudip.bat@gmail.com}
 \barasat 
 \author{Bikas K Chakrabarti}
 \email{bikask.chakrabarti@saha.ac.in}
 \cmp \snb \isi

\begin{abstract}
We introduce here very briefly, through some selective choices of problems and through the sample computer simulation programs 
(following the request of the editor for this invited review in the Journal of Physics Through Computation),  
the newly developed field of econophysics. Though related attempts could  be traced much earlier (see the Appendix), 
the formal researches in econophysics started in 1995. We hope, the readers (students \& researchers) can start themselves to 
enjoy the excitement, through the sample computer programs given, and eventually can undertake 
researches in the frontier problems, through the indicated survey literature provided. 
\end{abstract}
\maketitle

\section{Introduction}

\noindent The research field of econophysics has emerged recently as economics-inspired statistical physics. Though the 
attempts are not new and in fact almost a century old (see the Appendix), the institutionalization of this research   
field, where (statistical) physicists can and do regular researches in their own departments and publish their relevant results in traditional and contemporary physics 
journals, is new (see e.g.,~\cite{kishore_sudip19,stanley2000introduction_sudip,sinha2010econophysics_sudip,yakovenko2009colloquium_sudip,chakrabarti2013econophysics,sen2014sociophysics}). 
Indeed the term econophysics had been formally coined in 1995 (see the Appendix and Fig.~\ref{citation_hist}) 
and we are now in the silver jubilee celebration year! This review is intended for interested students for self-studies 
and self-learning through computational modelings of a few selected problems in econophysics. Some elementary 
computer programs or codes (in Fortran or Python) are added for ready support. We first introduce a few popular research problems 
in econophysics and continue discussion on them in the following.

\begin{figure}[h]
\begin{center}
\includegraphics[width=10.5cm]{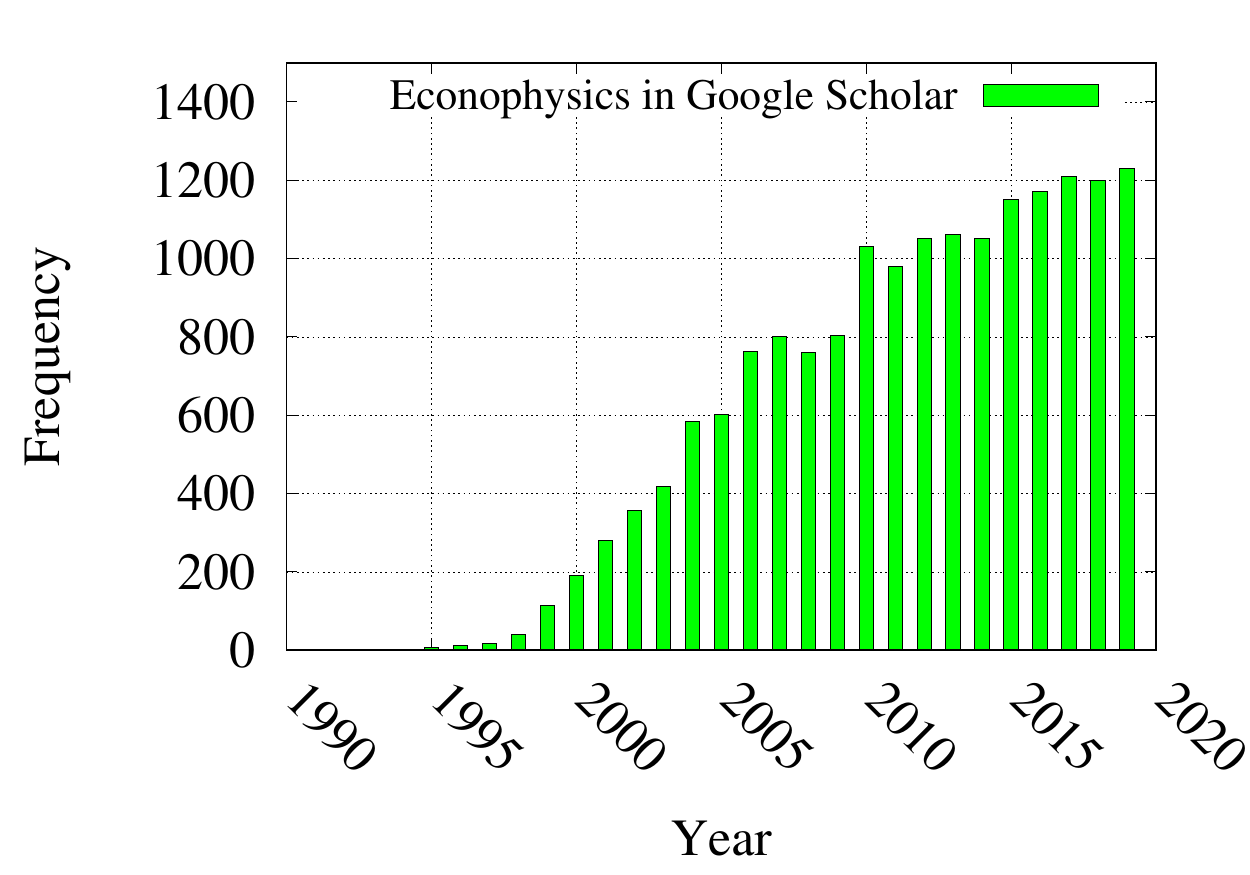} 
\end{center}
\caption{Histogram plot of numbers of entries per year containing the
term ‘econophysics’ versus the corresponding year. The
data are taken from Google Scholar (dated: 31 December, 2019). It may also 
be noted from Google Scholar that while this 24-year old econophysics has 
today typical yearly citation frequency of order $1.2 \times 10^3$, 
more than 100-year old subjects like astrophysics (Meghnad Saha published 
his ionization equation for solar chromosphere in 1920), biophysics (Karl Pearson 
coined the term in his 1892 book `Grammar of Science') and 
geophysics (Issac Newton explained planetary motion, origin of tides, etc in 
`Principia Mathematica', 1687) today have typical yearly citation 
frequencies of order $36.6 \times 10^3$, $40.2 \times 10^3$ and $45.4 \times 10^3$ 
respectively.}
\label{citation_hist}
\end{figure}

Inequality in income or acquired wealth has been ubiquitous: not only today, even in the earliest days of the human civilization. It is hard to find any society with fare amount of equality in income or wealth for everyone. It has continued to exist and sometimes have grown enormously in societies for centuries, and even today threatening our existence and wellbeing. Studies on inequality in wealth has a long history and has fascinated generations of philosophers, economists, social scientists and thinkers alike. Analysis of real data on wealth distribution is not new: with the advent of digital era, researchers from social science, even from interdisciplinary branches have been studying the recorded bulk amount of human social interaction data to explore the hidden structure of these data and also investigate the reason behind such inequality and so on. There are many ways to quantify inequality present in some social context or opportunity, e.g. income, wealth, as we know income is taken as a measure of 
economic growth of any country. The popular measures, summarized in one value, are Gini, Pietra indices etc. We will also 
discuss the recently introduced Kolkata index for measuring social inequality. All these make use of the Lorenz curve plot. On the other hand, one can also study the probability distribution and find the trend present in empirical data. Interestingly, the above mentioned indices computed from the Lorenz curve which is also plotted using the cumulative probability distribution of some context data against number of occurences. The study can also be extended to uncover the temporal pattern of social inequality from those data spread over years. 

Along with income or wealth inequality, we also put together inequality indices values measured over 
several other social context, e.g., academic citation count, city size or population, voting data, human death counts from social conflict which can be man-made like war, battle or natural disasters like earthquake, flood etc. These studies, sometimes done over couple of century-wide data (which are publicly available), have helped to uncover interesting patterns (of the presence of higher inequality but not highest inequality).

Next we study about the Kolkata Paise Restaurant (KPR) problem. KPR is a repeated many-player many-choice game played by a large number of players ($N$). Every day each player will choose a restaurant and visit there for lunch. Each restaurant prepares a single dish that costs the same at any restaurant. Player can only make single choice per day, and lunch at chosen restaurant is guaranteed if visited alone there for that day. Any day if more than one player choose the same restaurant then only one of them gets the dish and others arriving there would miss their lunch. Information about the restaurant occupations for few finite previous days (depending on the memory capacity of the players) will be publicly available. But players do not interact with others while making decision i.e. choosing a restaurant for lunch that day. In such set up, how the players should set individual choice towards socially optimal solution i.e. no food waste as well as no player staying hungry 
that day. 
A simple solution could be to hire a non playing captain (dictator) who would assign some restaurant to the players, thus all of them get their food from first day and following this setting till end. But in a democratic setup, players would like to 
make their own choice following some collectively learning strategy.
Here the objective is to achieve maximum social utilization of the scarce resources in absence of some external coordinator. This makes the KPR game interesting. Some results on statistics of KPR dynamics over variety of strategies developed, as well as an interesting phase transition phenomena have been discussed in respective section.

Another important and popular research-work deals with the economic/social networks evolved due 
to social interactions through market dynamics and game. In this context we will discuss about
 the Indian Railway network (IRN) as a complex (transport) network where each railway station is vertex and track between any two connecting stations is the edge between them. Study reveals IRN as a small world network, a popular network model where mean distance between any two nodes becomes constant as network size grows. Thus the graph like study has successfully answered several interesting questions like how many trains one passenger would require to switch to reach any destination station within country while traveling by train etc. 

We discuss the attempts made in the developments of the microscopic dynamical models (mainly based on kinetic theory), which can explore 
the underline dynamics behind the making of the real-world income or wealth distribution. We also highlight few models which 
show the possible ways of minimizing the socioeconomic inequality.

\begin{figure}[ht]
\begin{center}
\includegraphics[width=7.0cm]{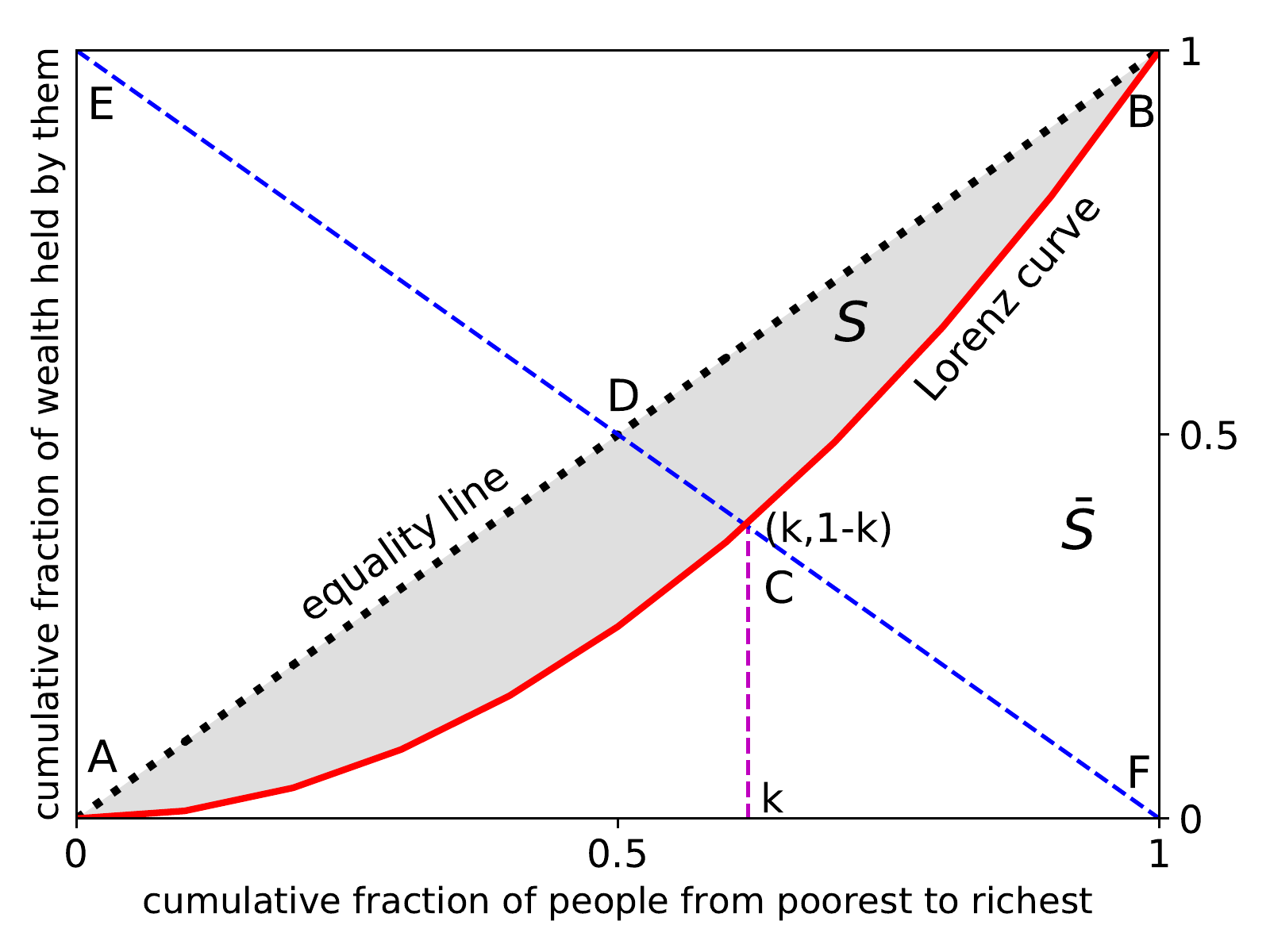}
\end{center}
\caption{Lorenz curve (in red here) plot represents the accumulated wealth against the fraction of people possessing that when arranged from poorest to richest. The diagonal from the origin represents the equality line. The Gini index ($g$)
can be measured from the (shaded) area ($S$) normalized by the area ($S+{\bar{S}} = 1/2$) of total 
area of the triangle below the equality line): $g = 2S$. The $k$ index can be measured by the 
ordinate value of the intersecting point of the Lorenz curve and the diagonal perpendicular to the 
equality line. It says that $k$ fraction of wealth is being held by $1-k$ fraction of top richest population.}
\label{LC_fig}
\end{figure}

\begin{figure}[ht]
\begin{center}
\fbox{\includegraphics[width=12.0cm] {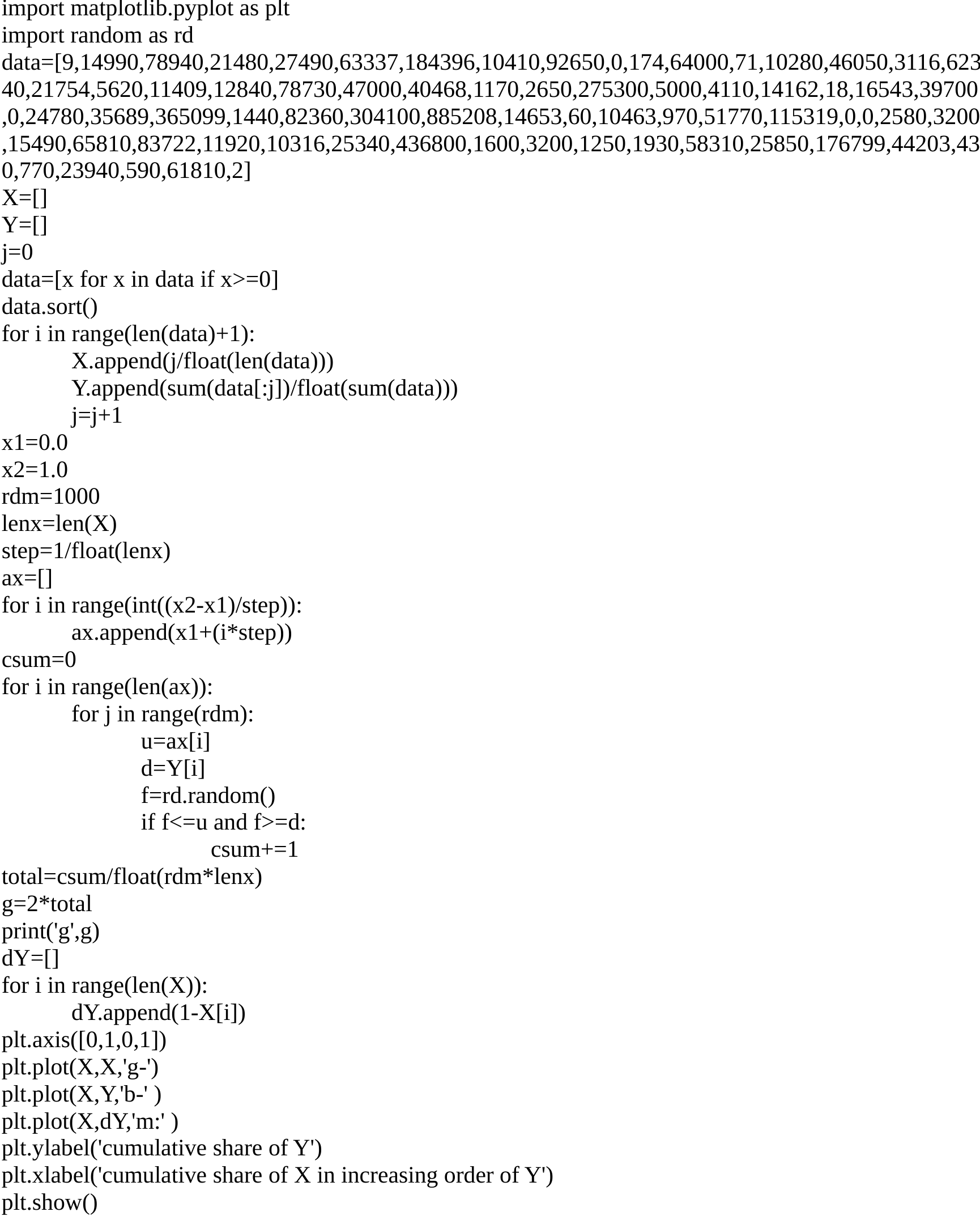}}
\end{center}
\caption{Python (version $2.7$) program to plot Lorenz curve based on hypothetical data, and also to compute gini index value. If run this program, one obtains $g$ value around $0.73$. }
\label{gini_code}
\end{figure}

\section{Measuring Social Inequality and the Kolkata Index}\label{gini-kpr}
Inequality present in socio-economic data can be quantified in several ways. Mostly used is the Gini index ($g$) introduced by Italian statistician and sociologist Corrado Gini in 1912~\cite{gini1921measurement}. There are others as well like Kolkata index ($k$), Pietra index ($p$) etc. Usually we plot histogram or frequency distribution to get the initial idea of any data in hand. And those indices mentioned above are easily measurable scalars from the cumulative probability distribution of the respective quantity. This distribution when plotted is known as Lorenz curve. Study of real data on the socio-economic context has been reported to follow fat tail, lognormal or gamma like distribution, see. Below we discuss some standard measures and techniques to get a general idea on how to measure inequality present in some quantity.

\textbf{Lorenz curve:}
Lorenz curve was proposed by Max O. Lorenz in $1905$. The curve always begins from point $(0,0)$ and ends in point $(1,1)$ as seen in Fig.~\ref{LC_fig}. 
If one plots the fraction of population or household in increasing order of $X$-axis against fraction of wealth held by them gives the cumulative probability distribution function $F$. Thus, the fraction of total wealth held by bottom x\% population is represented by $y = F(x)$. More the fractions grow close to each other, i.e. if $x$\% $\sim$ $y$\%  then the Lorenz curve becomes a straight line representing perfect equality in wealth or income distribution. Far the curve deviates from the diagonal or $45\degree$ line mean presence of greater inequality in the distribution. Intuitively, the curve never rise above the equality line ($X$ = $Y$). We also discuss about some inequality measuring indices, to be obtained using the Lorenz curve.\\
\noindent\textbf{Gini index ($g$):}
Gini index ($g$) is a standard measure of inequality, used not only by economists but also researchers across other disciplines e.g. physicist, social scientists etc. For computing the Gini index ($g$), one can fit Lorenz curve in a unit square, where the ratio of area between the Lorenz curve and equality line to the area below the equality line gives the Gini index ($g$) value. In Fig.~\ref{LC_fig}, $S$ represents the area between the Lorenz curve and equality line and $(S+{\bar{S}})$ represents the area below the equality line, then $\left(g={{S}  \over{S+{\bar{S}}}}\right)$. It ranges from $0$ to $1$ where $g = 0$ denotes perfect equality, say every individual has same income and plotting this should give the equality line (see Fig.~\ref{LC_fig}). And when $g = 1$, it represents maximum possible inequality where in a society only one person has every wealth and rest left with none. But then, Gini index ($g$) is a summary measure. Below we discuss Kolkata index ($k$), 
Pietra index ($p$) etc. to be estimated using the same Lorenz curve plot.

\textbf{Kolkata index ($k$):}
Ghosh et al. has recently proposed `Kolkata index'($k$) \cite{ghosh2014inequality}, another inequality measuring index to be obtained from the same Lorenz curve plot. The $k$ value can be estimated from the $X$-axis point where the Lorenz curve intersects the diagonal line perpendicular to the line of equality, see Fig.~\ref{LC_fig}. So, the $k$-index ranges from $0.5$ to $1$. Interestingly the complementary Lorenz function, represented as $\tilde{F}(n) = 1 - F(n)$, denotes $k$-index as a non-trivial fixed point on $x$-coordinate such that $\tilde{F}(k)$ = $k$ where $\tilde{F}(n)$ intersects the diagonal line spanning between $(0,0)$ and $(1,1)$, see Fig.~\ref{LC_fig}.

Another popular measure of inequality is the Pietra index (see, Eliazar et al.  \cite{eliazar2010measuring} for further discussion). It can be measured as the maximal vertical distance of the Lorenz curve from line of equality. So more the Lorenz curve deviates from the equality line greater is the inequality. Similar to Gini-index, the Pietra-index value also varies between $0$ to $1$, where $0$ represents complete equality and $1$ denotes extreme inequality. However we will consider Gini and Kolkata indices for our further discussion over inequality in social events.

To measure inequality, Gini index ($g$) is frequently used by the economist. Note that the $g$-index represents the  overall summary of inequality, whereas $k$-index is the indicator of the fraction of $x$ quantity held by $1-x$ fraction of population. Already mentioned that, $g$ varies between $0$ to $1$, whereas $k$ ranges from $0.5$ to $1$. One can obtain the Lorenz curve plot as well as the measure of indices from some real data using program given in Fig.~\ref{gini_code}. 

\subsection{$g$ and $k$ index values for several types of real data:}
Here we discuss results from various kinds of social events where inequality is measured (for e.g., citations, income, expenditure, vote, city size, human-death counts from social conflicts etc.) and some interesting results are observed. Chatterjee and Chakrabarti, 2016~\cite{chatterjee2017fat} studied probability distribution of many publicly available (by several universities and peace research institutes) data on human death counts occured from wars, armed-conflicts as man-made disasters as well as natural disasters like earthquake, forest fires etc. The distribution plots showed up to follow power law, in the fat-tail region, with exponent $\zeta$ around -$1.5$, see~TABLE~\ref{tab_1}. Extending their work, Sinha and Chakrabarti in~\cite{sinha2019inequality} measured the corresponding $g$ and $k$ index values by plotting Lorenz curves using many of the data discussed in~\cite{chatterjee2017fat} and the $g$ and $k$ values were found to be very high, see~TABLE~\ref{tab_2} and TABLE~\ref{tab_3}. Such high values indicates severe 
inequalities which is rarely seen in economic states of different countries (presumably because of deliberate supports to economically weaker groups, etc.). Those indices are also measured for the physical quantities (in case of natural disasters only, e.g., Richter magnitudes for earthquake, areas affected in sq. km. for floods, maximum water height in m for tsunami) causing such unequal occurences of human deaths~TABLE~\ref{tab_3.1}. Surprisingly not much inequality is observed in this case in comparison to those social effects caused by them. To establish the `similarity classes' of social inequality, these results were compared against those inequality measures found from man-made competitive societies like academic institution (paper citation counts), see TABLE~\ref{tab_4}. And the study indicated the growing recent trend of economic inequality across the world, may be because of encouraging competitiveness in societies as well as due to fast disappearance of social welfare measures in recent years. 

\begin{table}[h]
\caption{Estimated value of power law exponent $\zeta$ for man-made, natural conflicts from~\cite{chatterjee2017fat}.}
\label{tab_1}
\begin{center}
\begin{tabular}{|l||c|c|}
\hline

\textbf{Type of conflicts/disasters} & \textbf{Time period} & \textbf{$\zeta$} \\ 
\hline

conflict  & 1946-2008 & $1.54\pm0.06$  \\ \hline
war  & 1816-2007 & $1.63\pm0.03$  \\ \hline
battle  & 1989-2014 & $1.64 \pm 0.07$  \\ \hline \hline 

earthquakes  & 1900-2013 & $1.51\pm 0.05$  \\ \hline
storms  & 1900-2013 & $1.65\pm 0.03$  \\ \hline
wildfires  & 1900-2013 & $1.51\pm 0.01$  \\ \hline

\end{tabular}
\end{center}
\end{table}

\begin{table}[h]
    \caption{Estimated inequality (in death counts) index values for man-made conflicts from~\cite{sinha2019inequality} }
    \label{tab_2}
    \begin{center}
    \begin{tabular}{|l|c|c|c|} 
    \hline
      \textbf{Type of conflicts} & \textbf{Time period} & \textbf{$g$-index} & \textbf{$k$-index} \\
      \hline
		war & 1816-2007 & $0.83\pm$0.02 & $0.85\pm$0.02  \\
		\hline     
		battle & 1989-2017 & $0.82\pm$0.02 & $0.85\pm$0.02 \\
		\hline
      	  armed-conflict & 1946-2008 & $0.85\pm$0.02 & $0.87\pm$0.02  \\
      \hline
          terrorism & 1970-2017 & $0.80\pm$0.03 & $0.83\pm$0.02  \\
      \hline
       murder & 1967-2016 & $0.66\pm$0.02 & $0.75\pm$0.02 \\
		\hline	   
    \end{tabular}
    \end{center}
\end{table}

\begin{table}[h]
    \caption{Estimated inequality index values for social damages by natural disasters from~\cite{sinha2019inequality} }
    \label{tab_3}
    \begin{center}
    \begin{tabular}{|l|c|c|c|} 
    \hline
      \textbf{Type of disasters} &  \multicolumn{3}{c|} {\textbf{ Social Damage measures }}
      \\ \cline{2-4} & \textbf{Time period} & \textbf{$g$-index} & \textbf{$k$-index} \\
      
      \hline
		  earthquake & 1000-2018(July) & $0.94\pm$0.02  & $0.95\pm$0.02 \\
		\hline     
		 flood & 1900-2018(July) & $0.98\pm$0.02 & $0.98\pm$0.02 \\
		\hline
      	   tsunami & 1000-2018(July) & $0.93\pm$0.02 & $0.94\pm$0.02 \\
    
		\hline	   
		
    \end{tabular}
    \end{center}
\end{table}

\begin{table}[h]
    \caption{Estimated inequality index values for physical damages by natural disasters from~\cite{sinha2019inequality} }
    \label{tab_3.1}
    \begin{center}
    \begin{tabular}{|l|c|c|c|} 
    \hline
      \textbf{Type of disasters} &  \multicolumn{3}{c|} {\textbf{ Physical Damage measures }}
      \\ \cline{2-4} & \textbf{Time period} & \textbf{$g$-index} & \textbf{$k$-index} \\
      
      \hline
		  earthquake & 2013-2018(July) & 0.35 & 0.64 \\
		\hline     
		 flood & 1900-2018(July) & 0.76 & 0.79 \\
		\hline
      	   tsunami & 1000-2018(July) & 0.53 & 0.69 \\
    
		\hline	   
		
    \end{tabular}
    \end{center}
\end{table}

\begin{table}[h]
    \caption{Values of the inequality indices ($g$ and $k$) for some of the academic institutions 
 (from \cite{ghosh2014inequality}, see also \cite{chatterjee2017socio}; source-data taken from the Web of Science).\tiny{\textbf{[*Cambridge: University of Cambridge, Harvard: Harvard University, MIT: Massachusetts Institute of Technology, Oxford: University of Oxford, Stanford: Stanford University, Stockholm: Stockholm University, Tokyo: The University of Tokyo, BHU: Banaras Hindu University, Calcutta: University of Calcutta, Delhi: University of Delhi, IISC: Indian Institute of Science, Madras: University of Madras, SINP: Saha Institute of Nuclear Physics, TIFR: Tata Institute of Fundamental Research.]}}
 }
    \label{tab_4}
    \begin{center}
    \begin{tabular}{|c|c|c|c|c|c|c|c|} \hline    
    
    \textbf{Inst./Univ.*} & \textbf{Year}  & 
    \multicolumn{2}{ c|} {\textbf{Index values for}} & \textbf{Inst./Univ.*} & \textbf{Year} &  \multicolumn{2}{ c|} {\textbf{Index values for}} 
\\ \cline{3-4} \cline{7-8}   & &    \makecell{Gini\\($g$)} & \makecell{Kolkata\\($k$)}     &&& \makecell{Gini\\($g$)} & \makecell{Kolkata\\($k$)} \\
      
    \hline
      \multirow{4}{*}{Cambridge} &1980 &  0.74 & 0.78 & \multirow{4}{*}{BHU} &1980  & 0.68 & 0.76 \\ &1990 &  0.74 & 0.78 & &1990 &  0.71 & 0.77 \\ &2000 &  0.71 & 0.77 & &  2000 &  0.64 & 0.74\\ &2010  & 0.70 & 0.76  &  &2010 &  0.63 & 0.73 \\  
	  \hline 

      \multirow{4}{*}{Harvard} &1980 &  0.73 & 0.78 & \multirow{4}{*}{Calcutta} &1980 &  0.74 & 0.78 \\ &1990 &  0.73 & 0.78 & &1990 &  0.64 & 0.74\\ &2000 &  0.71 & 0.77 & &2000 &  0.68 & 0.74\\ &2010  & 0.69 & 0.76 & &2010 &  0.61 & 0.73   \\
	  \hline 
	   
      \multirow{4}{*}{MIT} &1980 &  0.76 & 0.79 &  \multirow{4}{*}{Delhi} &1980 &  0.67 & 0.75\\ &1990 &  0.73 & 0.78 & &1990 &  0.68 & 0.76\\ &2000 &  0.74 & 0.78 & &2000 &  0.68 & 0.76\\ &2010 &  0.69 & 0.76  & &2010 &  0.66 & 0.74\\
	  \hline 
	  
      \multirow{4}{*}{Oxford} &1980  & 0.70 & 0.77 & \multirow{4}{*}{IISC} &1980 &  0.73 & 0.78\\ &1990  & 0.73 & 0.78 & &1990 &  0.70 & 0.76 \\ &2000 &  0.72 & 0.77 & &2000 &  0.67 & 0.75\\ &2010 &  0.71 & 0.76  & &2010 &  0.62 & 0.73\\
	  \hline 
	  
      \multirow{4}{*}{Stanford} &1980 & 0.74 & 0.78  & \multirow{4}{*}{Madras} &1980 &  0.69 & 0.76 \\ &1990  & 0.70 & 0.76 & &1990  & 0.68 & 0.76 \\ &2000  & 0.73 & 0.80 & &2000  & 0.64 & 0.73\\ &2010  & 0.70 & 0.76 & &2010  & 0.78 & 0.79  \\
	  \hline 

\multirow{4}{*}{Stockholm} &1980 & 0.70 & 0.76  & \multirow{4}{*}{SINP} &1980 &  0.72 & 0.74 \\ &1990  & 0.66 & 0.75 & &1990  & 0.66 & 0.73\\ &2000  & 0.69 & 0.76 & &2000  & 0.77 & 0.79\\ &2010  & 0.70 & 0.76 & &2010  & 0.71 & 0.76  \\
	  \hline 

      \multirow{4}{*}{Tokyo} &1980  & 0.69 & 0.76 &  \multirow{4}{*}{TIFR} &1980  & 0.70 & 0.76\\ &1990  & 0.68 & 0.76 & &1990 &  0.73 & 0.77 \\ &2000  &  0.70 & 0.76 & &2000  & 0.74 & 0.77\\ &2010  & 0.70 & 0.76 & &2010 & 0.78 & 0.79 \\

	  \hline 
    \end{tabular}
    \end{center}
\end{table}
 
\begin{figure}
\includegraphics[width=7.5cm] {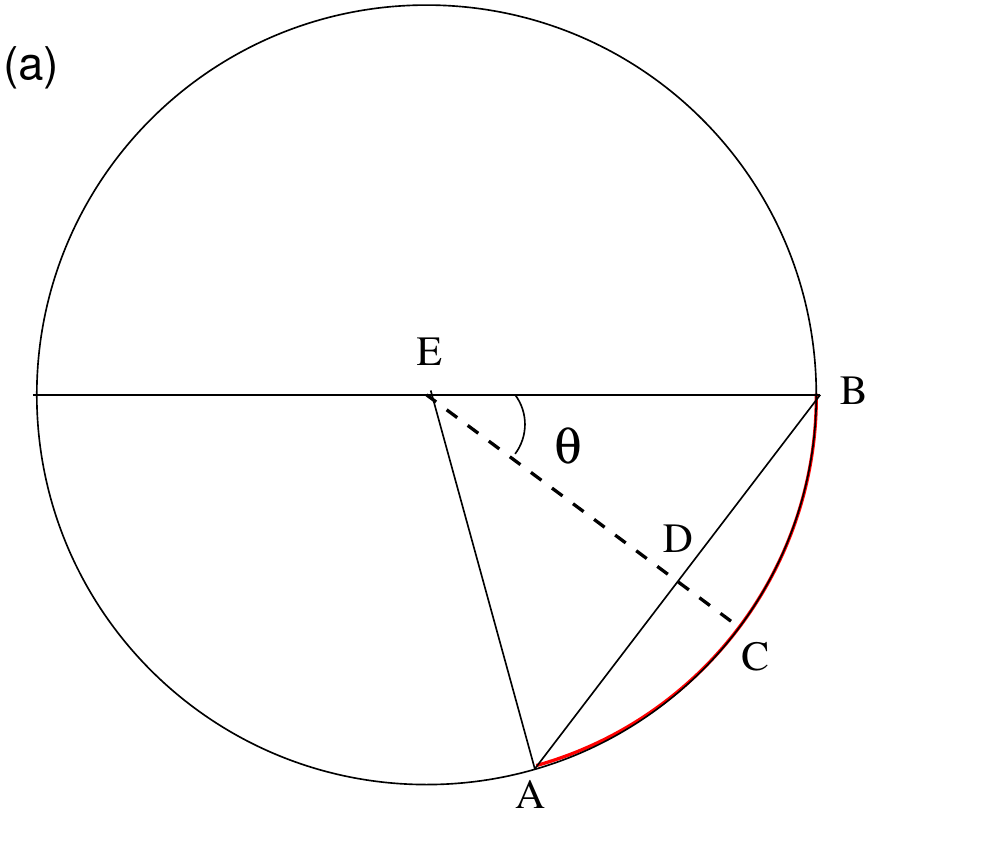}
\includegraphics[width=7.5cm]{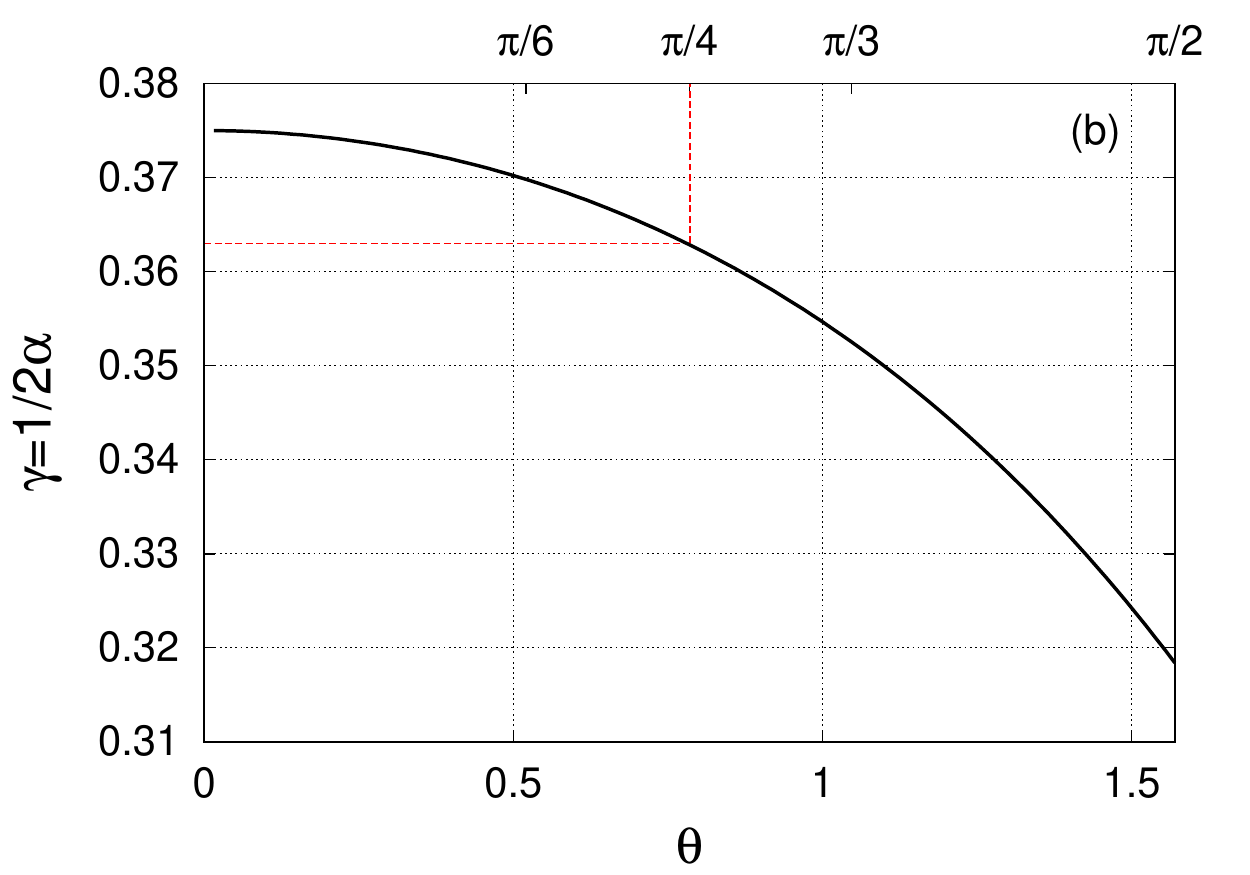}
\caption{(a) Approximating Lorenz curve as an arc of a unit circle (here ACB) and equality line as the chord AB. (b) Plot of approximated slope ($\gamma$ = $\frac{1}{2\alpha}$) of $k$-$g$ linear relation for different values of $\theta$.  }
\label{gk-a-b}
\end{figure} 

\subsection{Relationship between Gini ($g$) and Kolkata ($k$) index}
In Fig~\ref{LC_fig}, the Lorenz curve is represented by the red color line obtained using a probability distribution function $F(x)$. Assume, $X$ denotes the cumulative fraction of $x$ taking lowest to highest order and $Y$ as the cumulative fraction of $y$. The Lorenz curve cuts the anti-diagonal $Y = 1 - X$ at point C ($k,1-k$). This gives the $k$-index defined as: $k$ fraction of wealth is possessed by $1-k$ fraction of people. Here, the shaded area $\mathcal{S}$ is  enclosed by the Lorenz curve $(ACB)$ and the equality line $(ADB)$, and the The Gini index $g$ is represented by:
\begin{equation}\label{egk_1}
g = \frac{\text{ area of shaded region }}{\text{ area of the triangle $ABE$} } = 2\mathcal{S} 
\end{equation}
Here we would study some approximate ways to measure $\mathcal{S}$.
In Fig.~\ref{LC_fig}, we can assume the Lorenz curve to be given by two broken straight lines $AC$ and $CB$. Here, $AB = \sqrt{2}$, $DF = \frac{\sqrt{2}}{2}$, $CF$ = $(1-k)\sqrt{2}$, thus $CD = (DF-CF) = \frac{1}{\sqrt{2}}(2k-1)$. And the area of triangle $CAB$ is 
\begin{equation}\label{egk_2}
\mathcal{A}_1 = \frac{1}{2} \cdot AB \cdot CD = \frac{1}{2}(2k-1)
\end{equation}
Note that this is an exact result where equality in the relation holds for $g$ = $k$ at $g$ = $k$ = 1.
Thinking other way, consider that the Lorenz curve is represented by the arc $ACB$ of a circle with radius $r (=AE=BE)$, see Fig.~\ref{gk-a-b}(a). 
Find that 
$DE$ is perpendicular to base $AB$ with $\angle{BED} = \theta$. And the area of arc $ACB$ is the difference between the sector $BEAC$ and the triangle $ABE$. The area of sector $BEAC$ is $\theta r^2$. And, the area of triangle $ABE$ is given by $\frac{1}{2} \cdot$ $DE$ $\cdot AB = \frac{1}{2} \cdot r cos\theta \cdot 2r sin\theta$ = $r^{2}cos\theta sin\theta$. Thus the required area $ACBD$ representing the Lorenz curve is given by:
\begin{equation}\label{egk_3}
\mathcal{A}^{'} = r^2(\theta - sin\theta cos\theta)
\end{equation}

\begin{figure}[h]
\includegraphics[width=14.0cm] {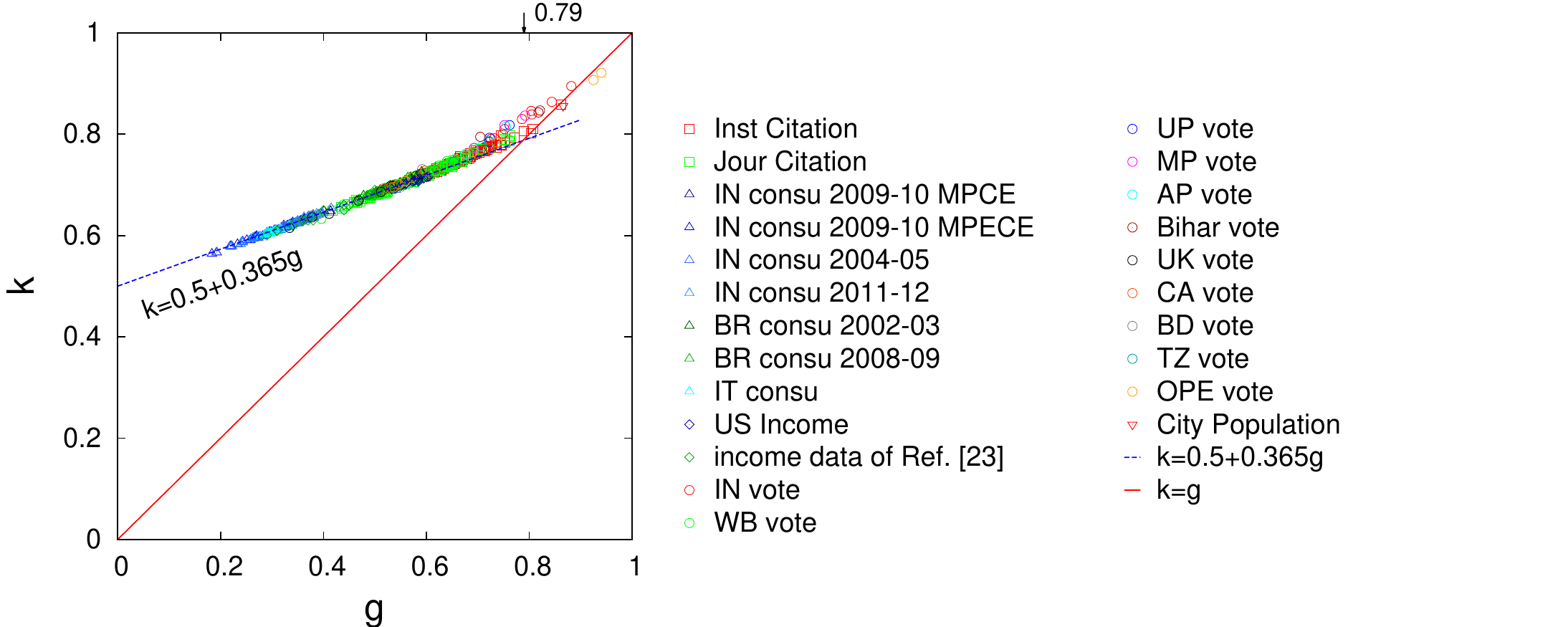}
\caption{Plot of the relationship between Kolkata index($k$) and Gini index ($g$) found from various types of data, e.g., citation, income, city population, vote share, etc. The data fits Eq.~\ref{egk_5} quite well. Above figure is taken from ref.~\cite{chatterjee2017socio}.}
\label{gk-vote}
\end{figure}

Now if we write $\mathcal{A}$ = $\mathcal{A}^{'}$ = $\frac{\alpha}{2}\cdot AB \cdot CD$, then
\begin{equation}\label{egk_4}
\alpha = \frac{\theta - sin\theta cos\theta}{sin\theta(1-cos\theta)}
\end{equation}
where $\alpha$ is a fraction incorporated to get the approximate result of $A^{'}$ as $\alpha \mathcal{A}_1$. Thus one gets $g \simeq$ 2$\mathcal{A}^{'}$ = 2$\alpha \mathcal{A}_{1}$ = $\alpha(2k-1)$ using Eq.~\ref{egk_2}, and this gives:
\begin{equation}\label{egk_5}
k = \frac{1}{2} + \frac{1}{2\alpha}g
\end{equation}
This is the general result of $g$ and $k$ relationship with slope $\gamma = \frac{1}{2\alpha}$. Fig.~4(b) also shows variation of $\frac{1}{2\alpha}$ against $\theta$. From Fig.~\ref{gk-vote}, one can find the observed approximate value of $\gamma$ = 0.365 which corresponds to $\theta$ = $\pi / 4$ 
(see Fig.~\ref{gk-a-b}(b)). This means that the Lorenz curve can be approximated as a quadrant arc of an unit circle centered at $E$ such that $2\theta = \pi/2$ (compare Fig.~\ref{gk-a-b}(a) with Fig.~\ref{LC_fig}). The readers are hereby encouraged to calculate the approximated $k$ value using Eq.~\ref{egk_5} that should come as $0.707$.

\begin{figure}[h]
\begin{center}
\includegraphics[width=8.0cm] {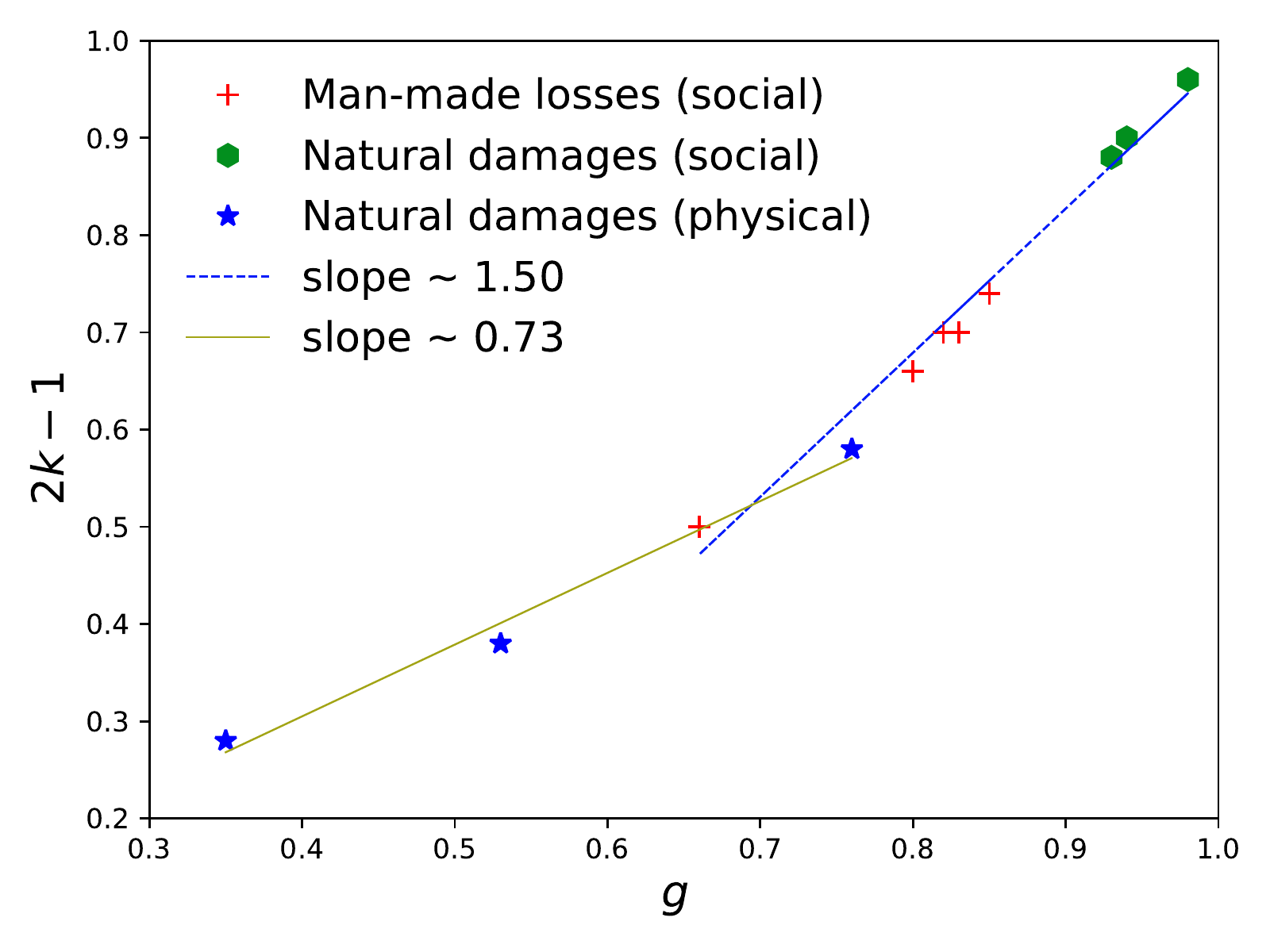}
\end{center}
\caption{Plot of $2k-1$ against $g$ measured for human death count data from social conflict, also see~TABLE \ref{tab_2} \&~TABLE \ref{tab_3}. Both the estimates range from $0$ to $1$. For both of the slopes, $2k-1$ measures are observed to be less than respective $g$ values (see Ref.~\cite{sinha2019inequality}). }
\label{gk-as}
\end{figure}

Interestingly, the results discussed in above para matches well with extensively studied real data taken from several publicly available sources e.g., citation data (from Web of Science), consumption expenditure data (for India, Brazil, Italy), income (for USA)  voting data (for Italy, Sweden, India, UK, Bangladesh, Canada etc.), city size etc. The dotted straight line represents $k = 0.5 + 0.365g$. For more details see \cite{chatterjee2017socio,inoue2015measuring}. As the $k$-index ranges from $0.5$ to $1$, a normalized estimate 2$k$-1 is also considered which represents the maximum vertical distance between the  equality line and Lorenz curve. This is how done as alike $g$-index, the $2k-1$ estimate also ranges between $[0,1]$. In \cite{sinha2019inequality}, both of $g$ and $2k-1$ are calculated for human death count data occurred from social conflict like battle, war, natural calamities etc. Two slope is observed when $2k-1$ is plotted against $g$: about $0.73$ (for lower range of $g$) and $1.5$ (for 
higher range of $g$). Though for both the cases, $g$ values are observed to be higher than corresponding $2k-1$ values; see Figs.~\ref{gk-vote} 
and~\ref{gk-as} for details.

\section{Econophysics of income and wealth distribution}\label{econo_income-wealth}

\begin{figure}[h]
\includegraphics[width=6.5cm]{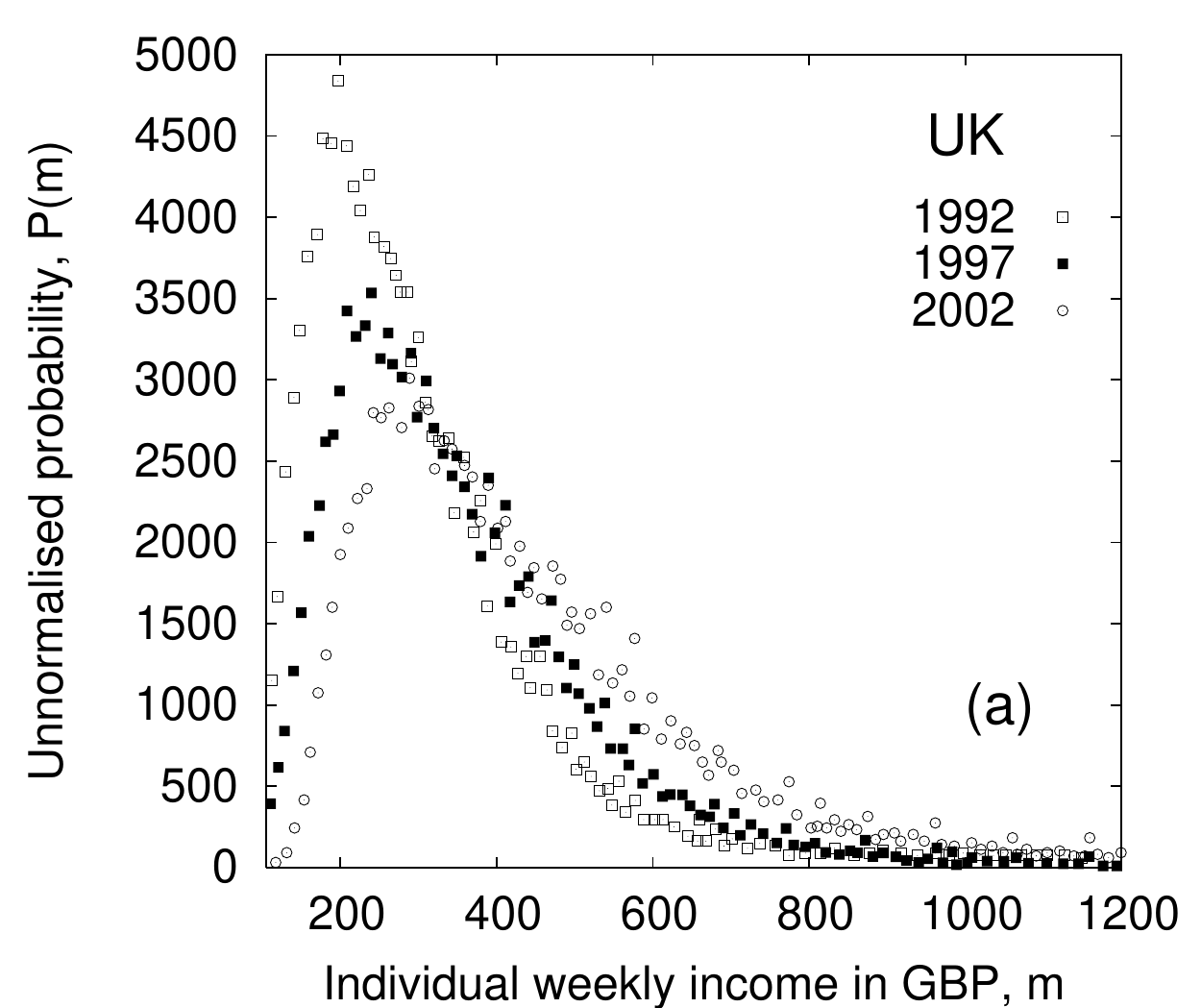}
\includegraphics[width=6.5cm]{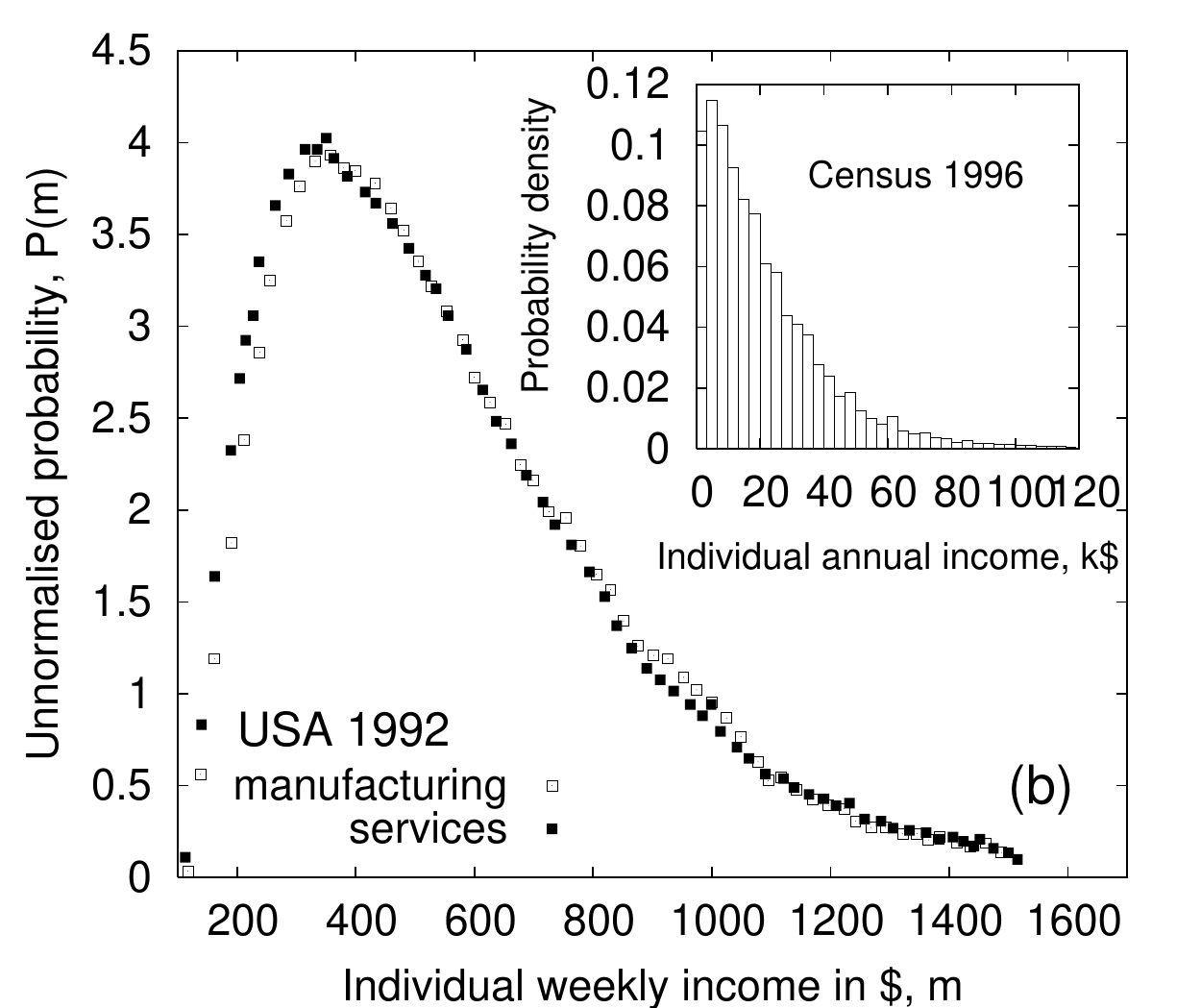}
\includegraphics[width=6.5cm]{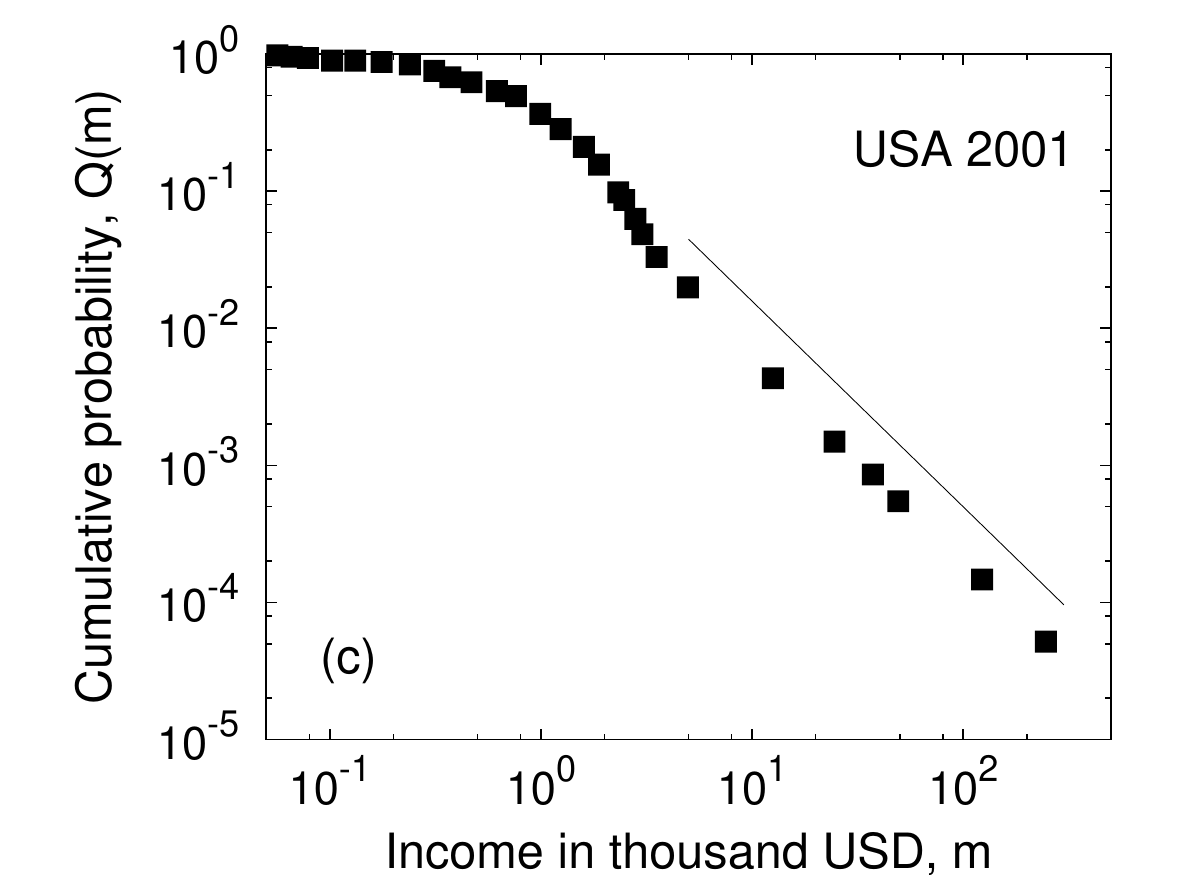}
\includegraphics[width=6.5cm]{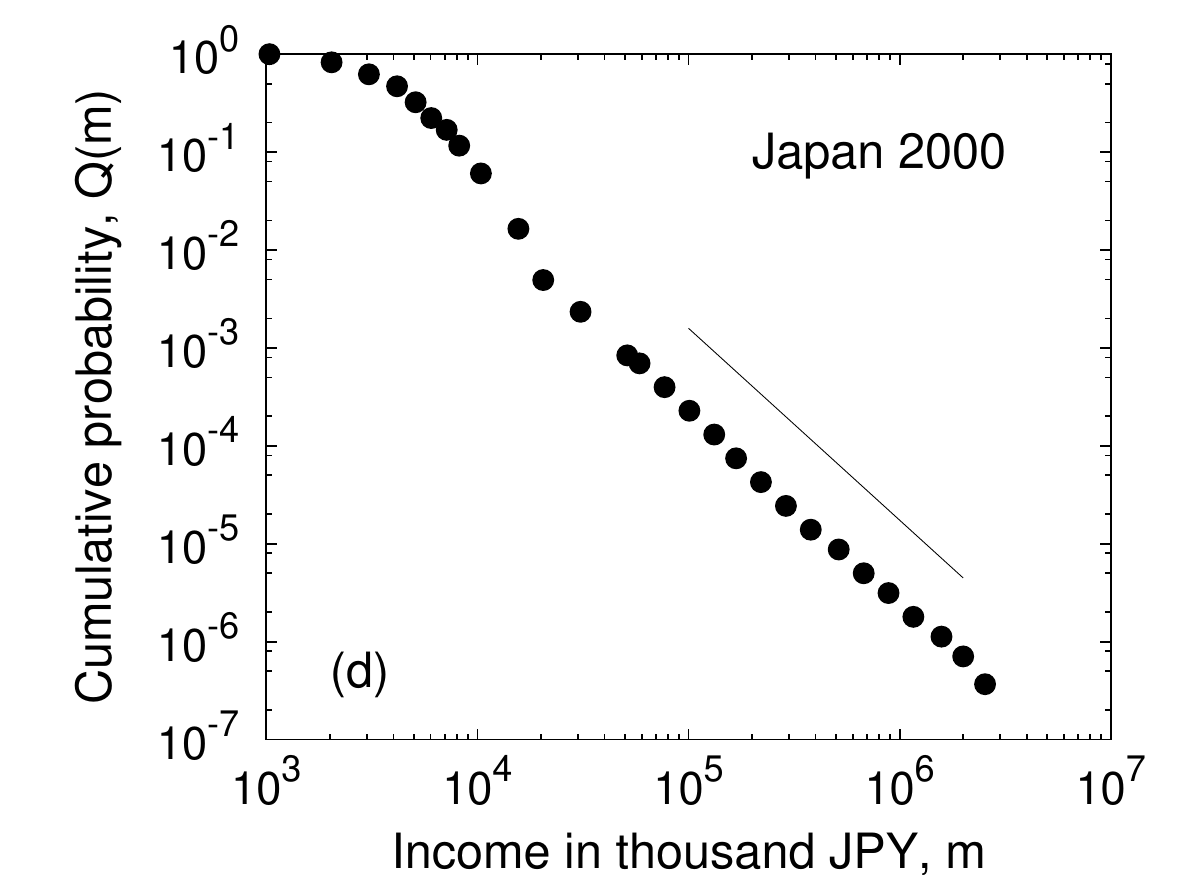}
\caption{(a) Distribution $P(m)$ of individual weekly income in UK for
1992, 1997 and 2002. 
(b) Distribution $P(m)$ of individual weekly income for
manufacturing and service sectors in USA for 1992; 
data for US Statistical survey. 
The inset shows the probability distribution of individual annual 
income, from US census data of 1996. 
(c) Cumulative probability $Q(m) = \int_m^\infty P(m) dm$ 
of rescaled adjusted gross personal annual income in US for IRS data from 2001, with Pareto exponent ${\alpha}_p \approx 1.5$
(given by the slope of the solid line).
(d) Cumulative probability distribution of Japanese personal income
in the year 2000. 
The power law (Pareto) region approximately fits to $\nu=1.96$.
The data and analysis adapted from~\cite{chatterjee2007kinetic}.}
\label{arnab_bkc_epjb2007}
\end{figure}

As discussed already, irrespective of history, culture and 
economic policies followed, socioeconomic inequalities 
are found to be omnipresent across the globe and throughout 
the ages. Indeed some robust features of such unequal 
distributions of income or wealth are already established 
(see e.g.,~\cite{chatterjee2007kinetic,chakrabarti2013econophysics}): 
While the overall the income or wealth distribution fits 
generally a Gamma-like curve (see some typical income 
distribution in Fig.~\ref{arnab_bkc_epjb2007}; whereas economists still 
like to fit it to a log-normal curve). The tail end of the
distribution (for large income or wealth) decays following 
a robust power law, called the Pareto law (see Fig.~\ref{arnab_bkc_epjb2007}).  
Physicists today have been trying to capture such generic features of 
the income or wealth distribution, using models 
based on the kinetic theory of ideal gases, where the interactions among the `social-atoms' or agents (traders) 
due to a trade (involving money exchanges), are considered as a 
two-body scattering problem where the total 
money (like energy) before and after the trade, remains conserved.  To best of our
knowledge, the first text book (`A Treatise on Heat') on the
statistical thermodynamics, which discussed the application of the
kinetic theory (of ideal gas od `social-atoms') to the derive the 
income or wealth distribution, was written by Saha
and Srivastava~\cite{sudip_saha_book}.

\subsection{Saha and Srivastava's Kinetic Theory for ideal gases with
`atoms’ and `social agents’}\label{saha_book_diss}

\begin{figure}[h]
\begin{center}
\fbox{\includegraphics[width=11.5cm, angle=-89]{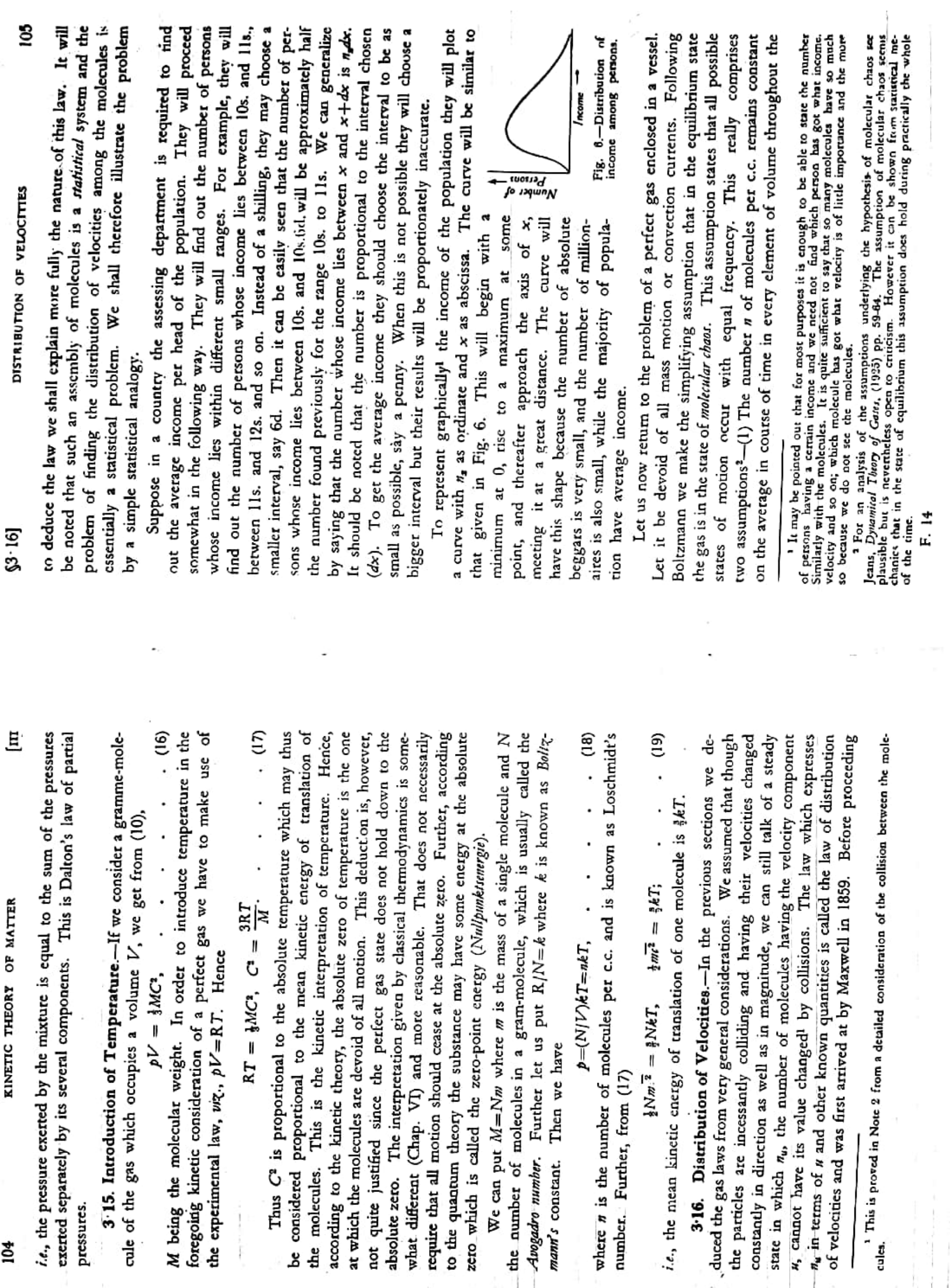}}
\end{center}
\caption{Image of the pages 104 and 105 of the book ``A Treatise on Heat'' (1931) by Saha and Srivastava~\cite{sudip_saha_book}.}
\label{saha_book}
\end{figure}

In thermodynamic systems, the number atoms or molecules is typically
of the order of Avogadro number ($\sim 10^{23}$) whereas the number of
social agents even in a global market is about $10^9$.  Still one can
imply the statistical physics principles in such economic systems. In
their famous book, Saha and Srivastava had put a discussion in the
section Maxwell-Boltzmaan velocity distribution of ideal gas, which
highlights the idea of applying kinetic theory in market to evaluate 
the income distribution of a society (see Fig.~\ref{saha_book}).  

In case of an ideal gas, at temperature $T$, the number density of particles 
$P(\epsilon)$ (atoms or molecules) having energy between $\epsilon$ to 
$\epsilon$ $+$ $d\epsilon$ is given by $P(\epsilon)d{\epsilon} = g(\epsilon)f(\epsilon)d\epsilon$. 
Here the density of states $g(\epsilon)d\epsilon$ gives the number of 
possible dynamical states between the energy $\epsilon$ to $\epsilon$ $+$ $d\epsilon$. 
The energy distribution function is denoted by $f(\epsilon)$. 
We should mention that, in case of an ideal gas, the particle can only have 
the kinetic energy $\epsilon = |\vec{p}|/2w$, where $\vec{p}$ and $w$ are the 
momentum and mass of the particle respectively. Since momentum is a vector, 
one can clearly understand that the particle can have same energy for 
different momentum vector $\vec{p}$. The density of states can be written as 
$g(\epsilon)d\epsilon = 4\pi|\vec{p}|^2d|\vec{p}|  = 2\pi(2m)^{3/2}\sqrt{\epsilon}d\epsilon$. 
As the energy is conserved during any energy exchanging process, then the energy 
distribution should satisfy $f({\epsilon}_1)f({\epsilon}_2) = f({\epsilon}_1 + {\epsilon}_2)$ 
(for any arbitrary values of energy ${\epsilon}_1$, ${\epsilon}_2$). Therefore 
one can expect the exponential form of the energy distribution function, which is 
$f(\epsilon) \sim \exp(-\epsilon/\Delta)$, where we will identify later $\Delta = K_B T$. 
This essentially gives $P(\epsilon)d{\epsilon} \sim \sqrt{\epsilon}\exp({-{\epsilon}/{\Delta}})$ 
and from such expression one can evaluate the pressure $P$ of an ideal gas 
having volume $V$ and temperature $T$. Comparing such result with the equation 
of state of ideal gas $PV = K_B T$, one can identify $\Delta = K_B T$.

In their chosen example in the section velocity distribution, Saha and
Shrivastava indicated that in a closed economic system (where no
migration of labour, no growth etc are considered) the money $m$
distribution of the agents of the system should have the similar form
of the energy distribution function of ideal gas atoms or molecules.
This is evident because in a closed economic system the money $M$
is conserved. Hence the money distribution should also satisfy the
condition $f(m_1)f(m_2) = f(m_1+m_2)$. Like the collision (the energy
exchange process) between the particles, the money exchange (due to
any kind of trading) between the agents are also random. Therefore,
one can expect the money distribution should be $f(m) \sim
\exp(-m/\sigma)$, where $\sigma$ is constant. As in the economic
system there is no quantity which is equivalent to the momentum vector
of the gas particle, in this case the density of states is constant.
Hence the number density of agents $P(m)$ having money $m$ should be
$P(m) = C\exp(-m/\sigma)$. Here $C$ is another constant. Both
$\sigma$ and $C$ can be evaluated using two conditions. One of the
condition is $N = \int_{0}^{\infty} P(m)dm$, where $N$ is the total
number of agents. There is another condition $M = \int_{0}^{\infty}
mP(m)dm$. Calculating these two integrations one would get $C =
1/\sigma$ where $\sigma = M/N$, which is actually the average money
per agent. Therefore instead of Maxwell-Bolzmann or Gamma distribution
of energy in the ideal gas, the money distribution becomes an
exponentially decaying function (like Gibbs distribution). Here most
number of agents have zero money. For the students, Saha and
Shrivastva left the task of making this exponential distribution more
like Gamma distribution.

\subsection{Data analysis of Income and Wealth distributions}

Italian scientist Pareto, from his socioeconomic study in Europe~\cite{sudip_pareto} showed 
that the income distribution contains a power law tail ($f(x) \sim x^{-(1+{\alpha}_p)}$). 
Such an observation is often called Pareto's law ($\alpha_p$ the Pareto exponent). Later another economist 
Gibrat~\cite{sudip_gibrat} showed that the part of the income distribution which 
corresponds to the higher income range actually fits well with power law type 
function whereas the rest of the distribution can be characterized by the 
log-normal function $f(x) \sim \frac{1}{x\sqrt{2\pi {\sigma}^2}}\exp{- \frac{log^2(x/x_0)}{2{\sigma}^2}}$. 
From the analysis of Japanese personal income distribution data ranging from 1887 
to 1908, Souma~\cite{souma_sudip} reported about the two type nature of income 
distribution. also indicated both Pareto index (${\alpha}_p$) and Gibrat index 
(${\beta}_g = 1/\sqrt{2{\sigma}^2}$) varies with time. 

Several physicists have been rigorously studied the financial data of many 
countries to capture the form of the income distribution. From the investigation 
of the Japanese income tax data of the year 1998, Aoyama et al.~\cite{aoyama_sudip} 
observed the power law decay of the income distribution. Dr\u{a}gulescu and 
Yakovenko analyzed the USA~\cite{druagulescu2001a_sudip} and UK~\cite{druagulescu2001b_sudip} 
income tax data. From the examinations of such data, they conjectured that the major part 
of the income distribution fits with the exponential function whereas in the higher income 
range the distribution follows power law decay. Dr\u{a}gulescu and Yakovenko explained  
the exponential nature of the income distribution in the basis of statistical mechanics~\cite{dragulescu2000_sudip}. 
They treated the closed economic system as a closed thermodynamic ideal gas system where the 
money in the economic system is equivalent to the energy in the idea gas system. They 
considered the financial transaction between any two agents is equivalent to the 
scattering between the two gas molecules or atoms in which they exchange energy. Like 
total energy of ideal gas system, the total money is also conserved in the closed economic 
system. Therefore the money distribution should follow the Gibbs distribution (exponential) 
if the density of state is constant. There are many suggestions on the shape of income 
distribution. Ferrero~\cite{ferrero_sudip} proposed that the income distribution follows 
Gamma distribution $f(x) \sim x^{n-1}\exp(-x/a)$, where $n$ and $a$ are the fitting parameters. 
Clementi~\cite{clementi_sudip} proposed $\kappa$ generalized function 
$f(x)=\frac{{\alpha}_0{\beta}_0x^{{\alpha}_0 - 1}exp_{\kappa}(-{\beta}_0x^{{\alpha}_0})}{\sqrt{1+{\kappa}^2{{\beta}_0}^2 x^{2{\alpha}_0}}}$ 
for the income distribution. Here $\kappa$ is the deformation parameter and ${\alpha}_0$, ${\beta}_0$ 
are fitting parameters. Using $\kappa$ generalized function, they found good data fit with 
USA income data. In their analysis the low income part of the distribution is exponential, 
which is retrieved by taking the limit $\kappa \to 0$. The low income part is actually 
corresponds to the low energy regime of the physical system, where one can treat such system 
non-relativistically. As a result of that the nature of the income distribution in the low income 
part is exponential. On the other hand the high income part is associated with the physical 
system in high energy scale. Therefore the physical system should be treated relativistically 
and one could not expect exponential distribution in the high income regime. The Pareto's law 
can be obtained by taking the limit $x \to \infty$ with $\kappa \ne 0$. 

Analyzing the 1996 Forbes magazine data Levy and Solomon~\cite{levy_sudip} found the existence 
of Pareto's law in the wealth distribution. The Forbes$^{[1]}$ \footnotetext[1]{Forbes magazine annually published the lists of the top 400 rich 
people of USA. The electronic data is available in www.forbes.com/lists.} magazine data from year 1988 to 2003 was 
investigated by Klass et al.~\cite{klass_sudip}. They ordered rich Americans according to their 
wealth where the wealth of the $r$-th person is $w_r$. They found a power law $w_r \sim r^{-(1/\alpha_p)}$, 
where the Pareto exponent $\alpha_p \simeq 1.43$. The exponent $\gamma$ is called Zipf exponent. The existence of power 
law in the wealth distribution of India was reported by Sinha~\cite{sinha_sudip}. 

From the analysis of Internal Revenue Service data of USA, Silva and Yakovenko~\cite{silva_sudip} 
studied the time evolution of the income distribution. They found that the form of the income 
distribution qualitatively remains similar throughout the entire period of the observation. 
Most interestingly the nature of the time evolutions of the lower and the upper parts of the 
income distribution are different. The lower income part of the distribution for all the 
years can be fitted with a single exponential curve, which actually indicates the thermal 
equilibrium in this part of the income distribution. On the other hand Silva and Yakovenko 
noticed that the power law tail in high income regime evolves significantly with time. They 
found the Pareto's exponent ${\alpha}_p$ changes from $2.8$ to $2.4$ during the year 1983-2003 
 (see also~\cite{gupta2006_sudip,sinha_sudip,saif2007_sudip,saif2009_sudip}). In this context 
 of Pareto law see, however,  Neda et al.~\cite{neda2019-sudip} for a recent review and extensive 
 analysis of income inequality data.

\subsection{Models of the Income and Wealth distributions}
There are several attempts have been made to model the observed universal income or wealth distributions 
of various countries. Physicists try to make models which can highlight the basic mechanism behind the 
formation of such universal income or wealth distribution in the society. Their another motivation is 
to explain the global economic inequality using the elementary ideas of physics.     

There are few works on the modeling of wealth distribution based on generalized Lotka-Volterra model 
(e.g.,~\cite{richmond_sudip,solomon_sudip}). Following the model, the time evolution equation of $m_{i,t}$ 
can be written as 
\begin{equation}\label{LV}
m_{i,t+1} = (1+\xi_t)m_{i,t} + \frac{a}{N}\sum m_{j,t} - c\sum m_{i,t}m_{j,t}, 
\end{equation}
where $m_{i,t}$ is the money of the $i$-th agent at time $t$. Here $N$ is the total number of agents and 
$a$, $c$ are two parameters. The variance of the distribution of random number $\xi_t$ (always positive) 
is $V$. Due to the presence of the second term in the right hand side of the equation~(\ref{LV}), the 
money of any agent should not go to zero at any instant of time. Such term may be considered as the 
effect of some kind of social security policy. The overall growth of the total money is controlled by the 
the parameter $c$. Since the total money of the system can change with time, the equation~\ref{LV} does not 
have any stationary solution. The relative money of an agent can be defined as $x_{i,t}=m_{i,t}/\langle m_t \rangle$, 
here $\langle m_t \rangle$ is the average money per agent at any instant of time $t$. The $x_{i,t}$ becomes 
independent of time if the ratio $a/V$ is constant. As a result of that, even in a non-stationary system, 
after some amount of time, one can eventually get a time invariant relative money distribution. In mean 
field approximation the distribution function $f(x)$ has the following form
 \begin{equation} \label{lotka}
  f(x)=\frac{exp[-(\nu - 1)/x]}{x^{1+\nu}}.
 \end{equation}
 Here $\nu$ (positive exponent) is the ratio of $a$ and $V$. For large value of $x$, the form of equation~(\ref{lotka}) 
 eventually becomes power-law-like. 
 
Bouchaud and M\'{e}zard~\cite{bouchaud2000_sudip} proposed a generalized model for the growth and redistribution 
of wealth. Their model (BM) able to reproduce the Pareto law. They used the physics of directed polymers in economical 
framework. In the BM model, the dynamics of the wealth $w_i$ of the $i$-th is governed by the set of stochastic 
equation, 
\begin{equation}
 \frac{dw_i(t)}{dt} = {\eta}_i(t)w_i(t) + \sum_{j \ne i}J_{ij}w_j(t) - \sum_{j \ne i}J_{ji}w_i(t)  \nonumber
\end{equation}
Here ${\eta}_i(t)$ follows a Gaussian distribution which has mean $\mu$ and variance $2{\sigma}^2$. The Gaussian 
multiplicative process simulates the investment dynamics. The $J_{ij}$ is the linear exchange rate of between $i$-th 
and $j$-th agents. Employing Fokker-Planck equation under mean field approximation, one can obtain a stationary 
solution of the distribution function $P(\overline{w})$, where $\overline{w} = \sum_{i} w_i/N$ is the mean wealth. 
Here $N$ is the total number of agents. The form of the $P(\overline{w})$ is given by
\begin{equation}
P(\overline{w}) = A\frac{\exp[(1-\nu)/\overline{w}]}{{\overline{w}}^{1+\nu}}, \nonumber
\end{equation}
where $A = (1-\nu)^{\nu}/{\Gamma(\nu)}$ and $\nu = 1 + J/{\sigma}^2$. For large value of $\overline{w}$, the 
$P(\overline{w})$ decays in a power law with exponent $\nu$. The BM model indicates about the two phases, in 
one phase only a few number of agents hold the entire amount of wealth. Such phase appears when $\nu < 1$. In 
the another phase the wealth is distributed among the finite number of agents. Under mean field approximation 
the agents in BM model exchange the same percentage of wealth they have. That means a relatively poor agent 
receives an unrealistic amount of wealth from the rich agent. In the field approximation, the wealth of the 
individual agents asymptotically converges to the mean wealth $\overline{w}$. That means after long time, all 
the agents have same amount of wealth, which is again an unrealistic situation. To introduce economic inequality, 
Scafetta et al.~\cite{scafetta2004_sudip} modulate the investment term of BM model. Garlaschelli and Loffredo~\cite{garlaschelli2008_sudip} 
accounted BM model in different types of networks.  

\begin{figure}[ht]
\begin{center}
\fbox{\includegraphics[width=8.0cm]{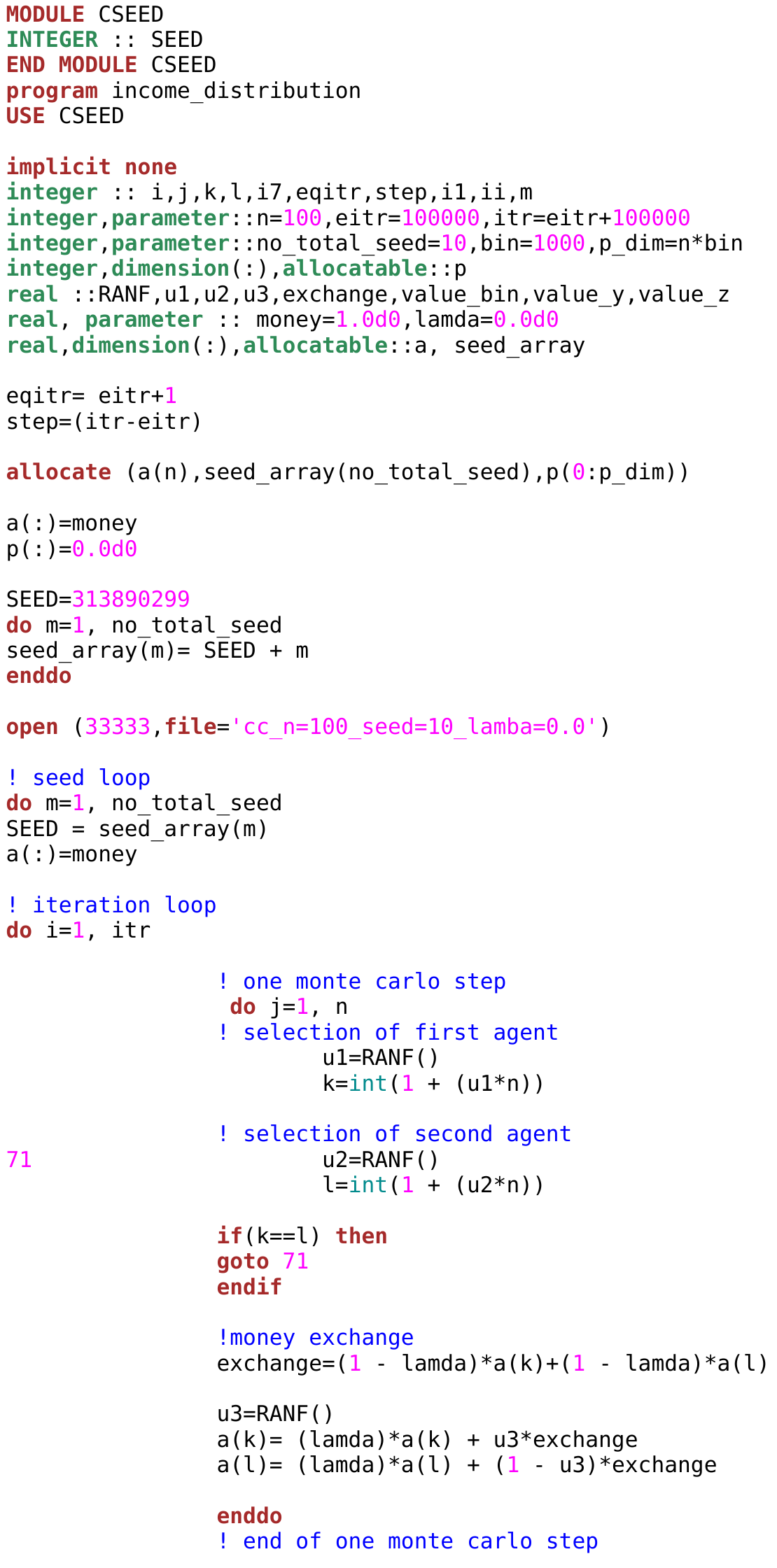} 
\vline\includegraphics[width=7.8cm]{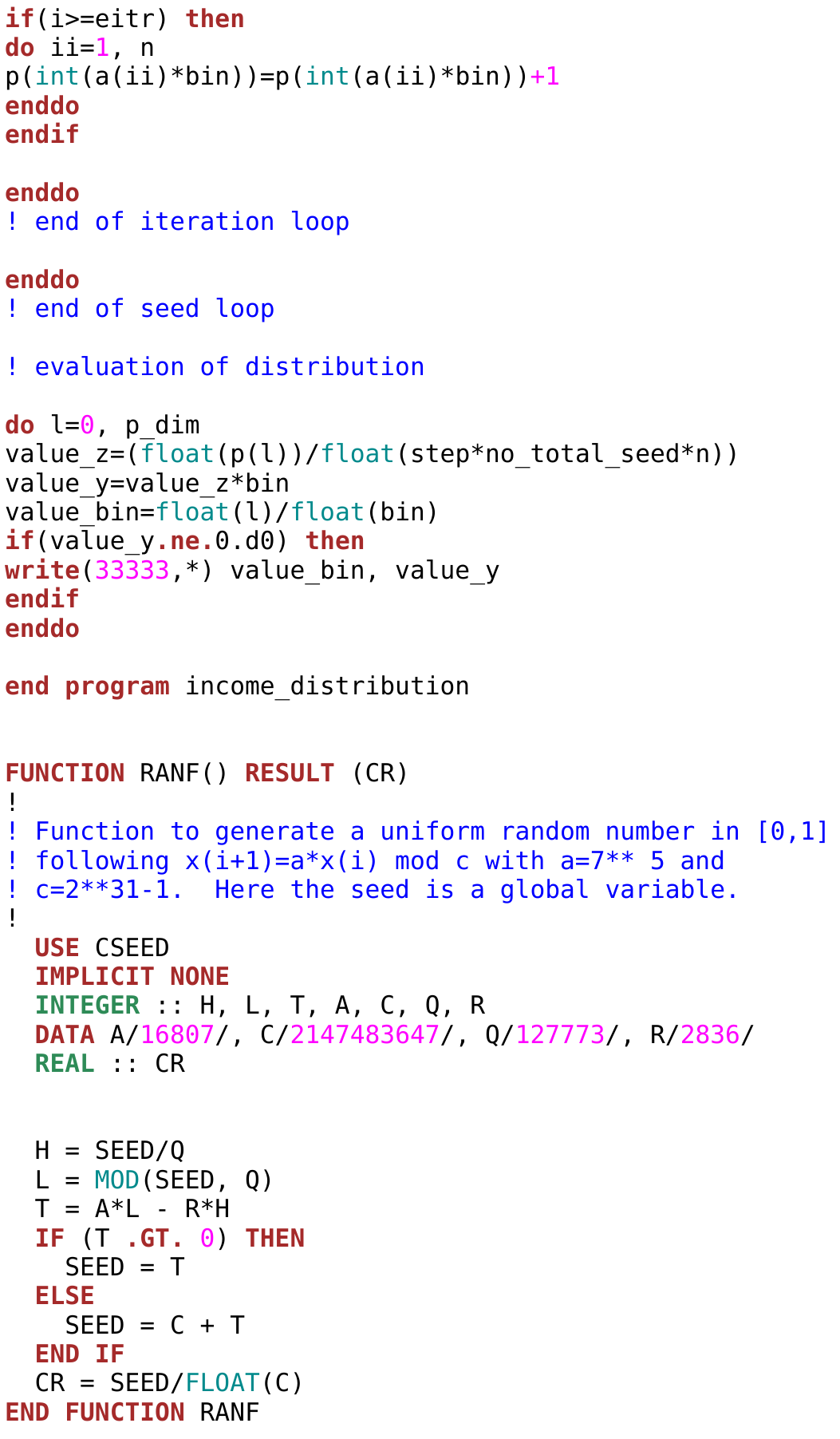}}
\end{center}
\caption{The Fortran code for simulating the dynamics of CC model.}
\label{cc_code}
\end{figure}

\begin{figure}[ht]
\begin{center}
 \includegraphics[width=10cm]{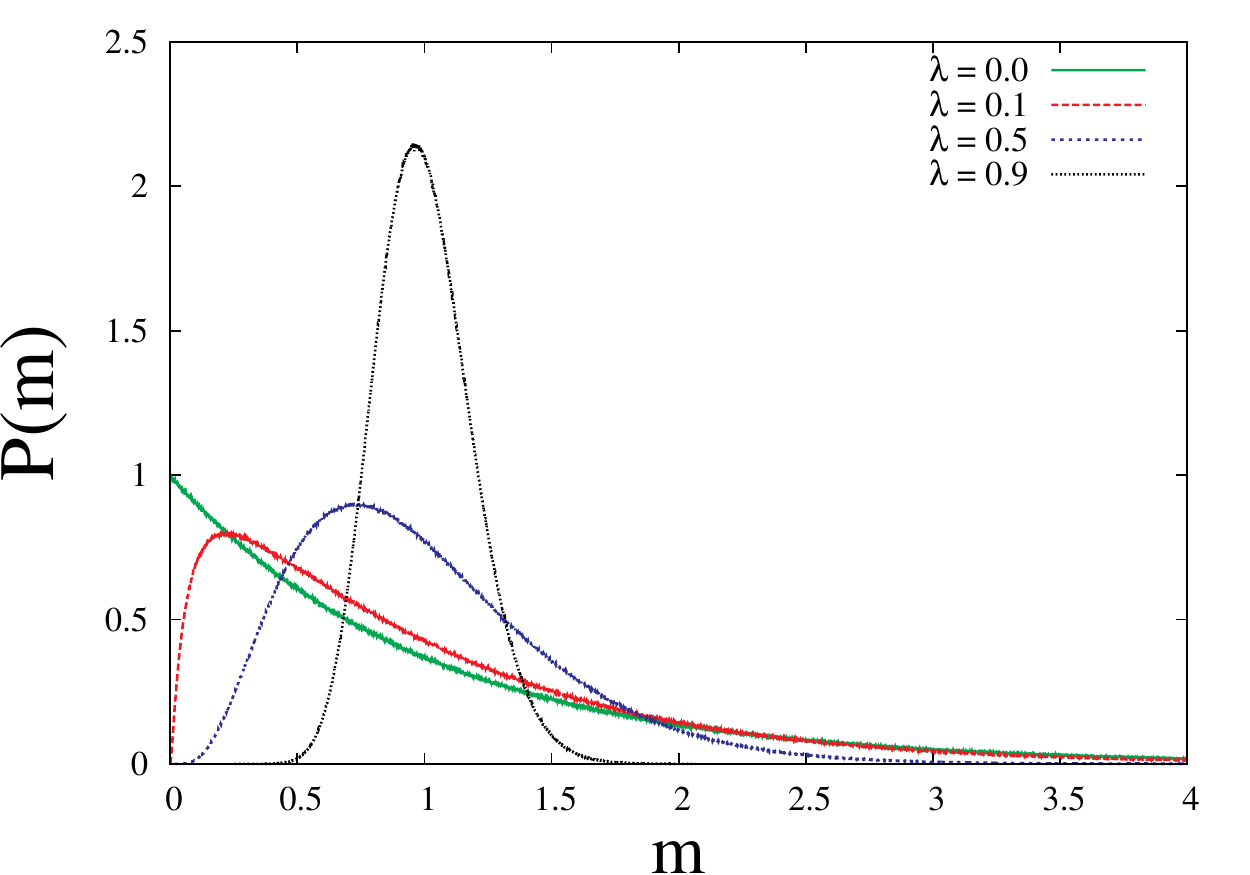}
 \end{center}
\caption{The money distribution $P(m)$ for different values of saving propensity factor $\lambda = 0, 0.1, 0.5, 0.9$. 
  Here the number of agents $N = 100$.}
  \label{cc_pdf}
 \end{figure}

Employing kinetic theory, physicists built models which can reproduce the income or wealth distribution of the 
society. Chakraborti and Chakrabarti~\cite{chakraborti_sudip} proposed a generalized model (CC model) of income 
distribution. In their proposed money exchange dynamics, during the economic transaction the participating agents 
(two people) always keep some fraction of their money and the sum of their remaining money is distributed randomly
among them. The dynamical equations of CC model are given by
 \begin{eqnarray} \label{cc_exchange}
  m_i(t+1) = \lambda m_i(t) + {\epsilon}_{ij}[(1 - \lambda)(m_i(t) + m_j(t))] \nonumber \\
  m_j(t+1) = \lambda m_j(t) + (1 - {\epsilon}_{ij})[(1 - \lambda)(m_i(t) + m_j(t))]. 
 \end{eqnarray}
Here $\lambda$ is the saving propensity and ${\epsilon}_{ij}$ is random fraction. Both $\lambda$ and ${\epsilon}_{ij}$ 
are ranging from zero to unity. One can simulate the dynamics of CC model using the Fortran code given in the Fig.~\ref{cc_code}. 
The money distribution $P(m)$ for different values of $\lambda$ are shown in the Fig.~\ref{cc_pdf}. For any non-zero 
value of $\lambda$ the most probable position of the $P(m)$ will be located at the non-zero value of the money. For zero 
value of $\lambda$ the distribution is essentially exponential whereas for non-zero value of $\lambda$ the $P(m)$ fits 
approximately to a Gamma function. Patriarca and Chakraborti~\cite{patriarca_sudip} numerically evaluated the mathematical 
form of the $P(m)$ which is given by 
\begin{eqnarray} \label{cc_exchange}
  P(m) = \frac{1}{\Gamma (n)}\Big(\frac{n}{\langle m \rangle}\Big)^n m^{n-1} \exp\Big(-\frac{mn}{\langle m \rangle}\Big), \nonumber \\
  n(\lambda) = 1+\frac{3\lambda}{1 - \lambda }.\nonumber
 \end{eqnarray}
We can see in the limit $\lambda \to 1$, the distribution function becomes sharply peaked about some non-zero value of 
money, which indicates the money is uniformly distributed among the agents. Almost simultaneous to the CC model, 
Dr\u{a}gulescu and Yakovenko~\cite{dragulescu2000_sudip} proposed another model (DY model), in which they mapped the 
two body collision process (exchange energy in the collision) into the economic system, where in each financial transaction 
(equivalent to collision) between two agents, they exchange money (equivalent to energy). The stochastic equations of money 
exchange DY model are given by, 
\begin{eqnarray} \label{money_exchange}
 m_i(t+1) = m_i(t) - \Delta m \nonumber \\
 m_j(t+1) = m_j(t) + \Delta m,
 \end{eqnarray}
where at time $t$ the $i$-th agent contains $m_i(t)$ amount of money. The financial transaction is only allowed if both
$m_i(t)$ and $m_j(t)$ are greater than zero. Here $\Delta m$ is the random fraction of the average money of the two participating 
agents. There is no provision of saving propensity in the DY model, which is the fundamental difference with the CC model. 
Using the dynamical money exchange equation of DY model, one will get a exponential money distribution, which appears as a 
special case $\lambda = 0$ in the CC model. Therefore one would get an equivalent dynamics of DY model by putting $\lambda = 0$ 
in the CC model.

\begin{figure}[ht]
\begin{center}
 \includegraphics[width=10cm]{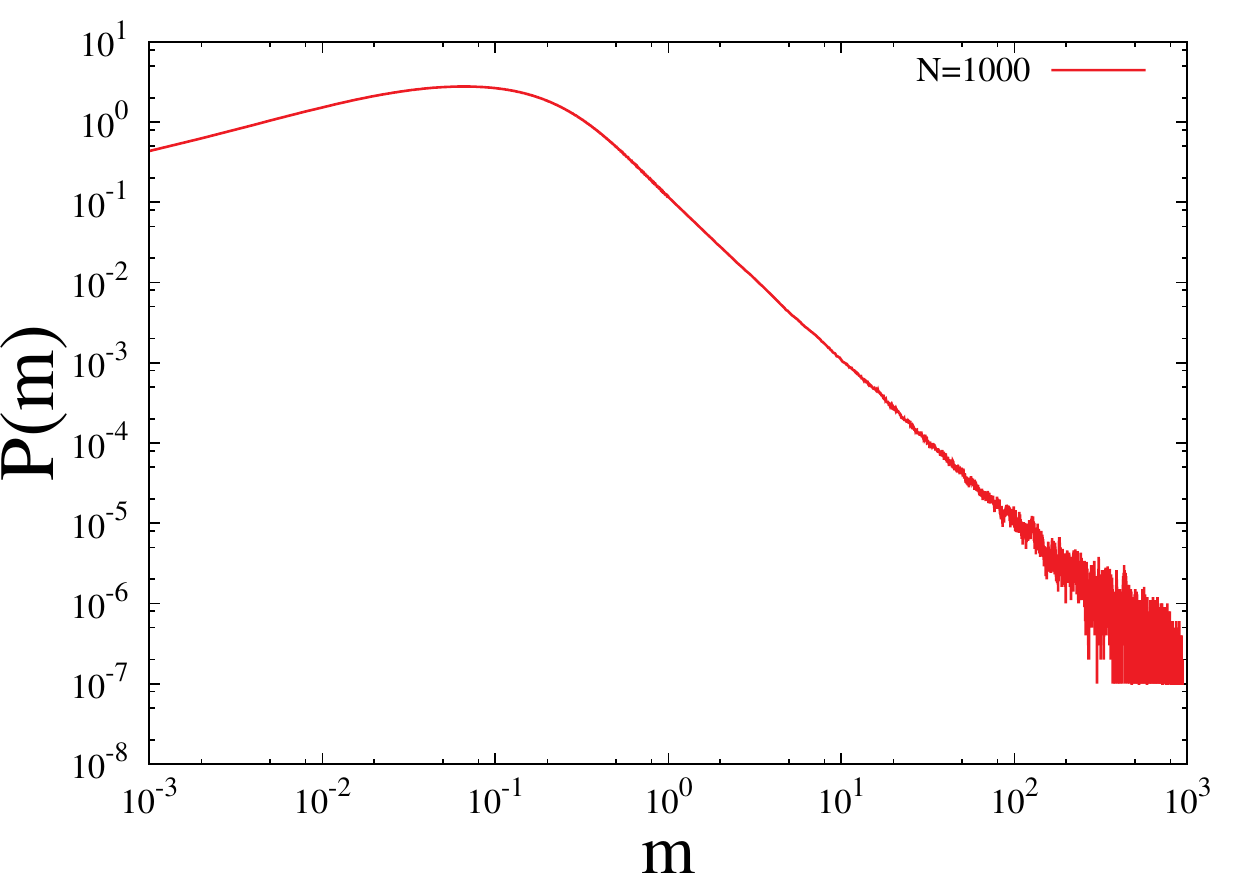}
 \end{center}
  \caption{For uniformly distributed saving propensity factor $\lambda$, the money distribution $P(m)$ contains a  
   power law tail which goes as $m^{-2}$. Here the number of agents $N = 1000$.}
  \label{ccm_pdf}
  \end{figure}

Although we observe power law tail in the income distributions of various countries, both CC and DY models fail to generate 
such power law tail by using their proposed dynamical rules. We find in the CC model, the value of $\lambda$ is same for every 
agent but in realistic situation the saving propensity should vary from agent to agent. Chatterjee et al.~\cite{chatterjee2004_sudip} 
discussed the CC model in more general way. Instead of constant value of saving propensity, they consider $\lambda$ is distributed 
among the agents. Therefore the modified dynamical equations of their model (CCM model) can be written as 
\begin{eqnarray} \label{ccm_exchange}
  m_i(t+1) = \lambda_i m_i(t) + {\epsilon}_{ij}[(1 - \lambda_i)m_i(t) + (1 - \lambda_j)m_j(t))] \nonumber \\
  m_j(t+1) = \lambda_j m_j(t) + (1 - {\epsilon}_{ij})[(1 - \lambda_i)m_i(t) + (1 - \lambda_j)m_j(t))].
 \end{eqnarray}
The values of the saving propensities of $i$-th and $j$-th agents are $\lambda_i$ and $\lambda_j$ respectively and they are 
in general different. Employing the dynamical equations~(\ref{ccm_exchange}), Chatterjee et al.~\cite{chatterjee2004_sudip}
numerically found a money distribution which contains a power tail. Such power law tail actually reveals the Pareto law. The 
existence of the power law tail is robust to the type of the distribution function $\rho(\lambda)$ of saving propensities but 
the power law exponent depends on the nature of the  $\rho(\lambda)$. In fact for the distribution function 
$\rho(\lambda) \sim |{\lambda}_0 - \lambda|^{{\alpha}_{\lambda}}$, the Pareto exponent value becomes unity for all values for $\alpha_\lambda (\ne 0)$. For uniformly distributed $\lambda$, the money distribution decays as 
$P(m) \sim m^{-2}$. The money distribution for system size $N = 1000$ is shown in Fig.~\ref{ccm_pdf}, here the relaxation time 
is of order of $10^6$. Chatterjee et al.~\cite{chatterjee2004_sudip} reported another important result regarding the 
fluctuation in money of individual agent. They showed that in case of CC model the fluctuation in the money of individual 
agent increases with the decrease in the value of $\lambda$ whereas in case of CCM model exactly opposite trend is observed. 
Patriarca et al.~\cite{patriarca2007_sudip} investigated the relaxation behavior of income or wealth distribution. They 
found the equilibrium time is proportional to the number of agents. They also noticed that for a given value of $\lambda$, 
the equilibrium time ${\tau}_{\lambda} \sim 1/(1 - {\lambda})$. Chakraborty and Manna~\cite{chakraborty2010_sudip} found 
a distribution with power law tail by using the dynamics of CC model and their power law exponent is similar with the 
result obtained in the CCM model. In contrast to the CC model,  Chakraborty and Manna~\cite{chakraborty2010_sudip} considered 
the probability of an agent in participating in a financial transaction is proportional to the positive power of his/her 
money. That means they evoked a dynamics where the richer class of people essentially get more opportunity in trading rather 
than the low-income people. Reduction of the Cobb-Douglas utilization maximization principle to the CC model of exchange dynamics form 
was shown by Chakrabarti and Chakrabarti~\cite{chakrabarti2009_bkc} (see also~\cite{huli_bkc} for a recent discussion).

Heinsalu and Patriarca~\cite{heinsalu2014kinetic_sudip} introduced another gas like model of income or wealth distribution. 
Their model is often called immediate exchange (IE) model. Their proposed dynamical equations of exchanging money between 
  $j$-th and $k$-th agents are given by 
\begin{eqnarray}\label{IE_model}
 m'_j = (1 - {\epsilon}_j) m_j  + {\epsilon}_k m_k \nonumber \\
 m'_k = (1 - {\epsilon}_k) m_k  + {\epsilon}_j m_j~~,
\end{eqnarray}
where $m$ and $m'$ are the money of any agent before and after the exchange respectively. Here ${\epsilon}_j)$ and ${\epsilon}_k)$ 
are two random numbers, uniformly distributed in $(0,1)$. Heinsalu and Patriarca~\cite{heinsalu2014kinetic_sudip} numerically 
found the equilibrium money distribution $f_{{\alpha}_{s}}(m)$ has the shape of $\Gamma$-function.
\begin{eqnarray} \label{IE}
f_{{\alpha}_{s}}(m) = \frac{{{\alpha}_{s}}^{{\alpha}_{s}}m^{{\alpha}_{s} - 1}}{\Gamma({{\alpha}_{s}})} \exp(- {\alpha}_{s} m)~.
\end{eqnarray}
Here the shape parameter ${\alpha}_{s} = 2$. Heinsalu and Patriarca~\cite{heinsalu2014kinetic_sudip} assert that for small 
values of wealth, the distribution function obtained from IE model matches better with the real data than the earlier models. 
Along with new dynamics, they introduced an acceptance criterion of trading for the agents. The acceptance probability of 
$j$-th agent for making transaction with the $k$-th agent is a function of ${\Delta}m_{jk}(={\epsilon}_k m_k - {\epsilon}_j m_j)$. 
Moreover, the equilibrium money distribution does not affected by the introduction of acceptance criterion. There are few 
analytical works on the IE model. Katriel~\cite{katriel2014immediate_sudip} performed analytical investigations on the IE 
model. He analytically showed the equilibrium money distribution of IE model converges to $\Gamma$-function in infinite 
population limit. Lanchier and Reed~\cite{lanchier2018_sudip} realized the IE model on connected graph, where agents are 
located at the vertex set of the graph and they can interact only with their neighbors. Lanchier and Reed~\cite{lanchier2018_sudip} 
analytically proved the conjectures made by Heinsalu and Patriarca~\cite{heinsalu2014kinetic_sudip}.

\begin{figure}[h]
\begin{center}
  \includegraphics[width=5.9cm]{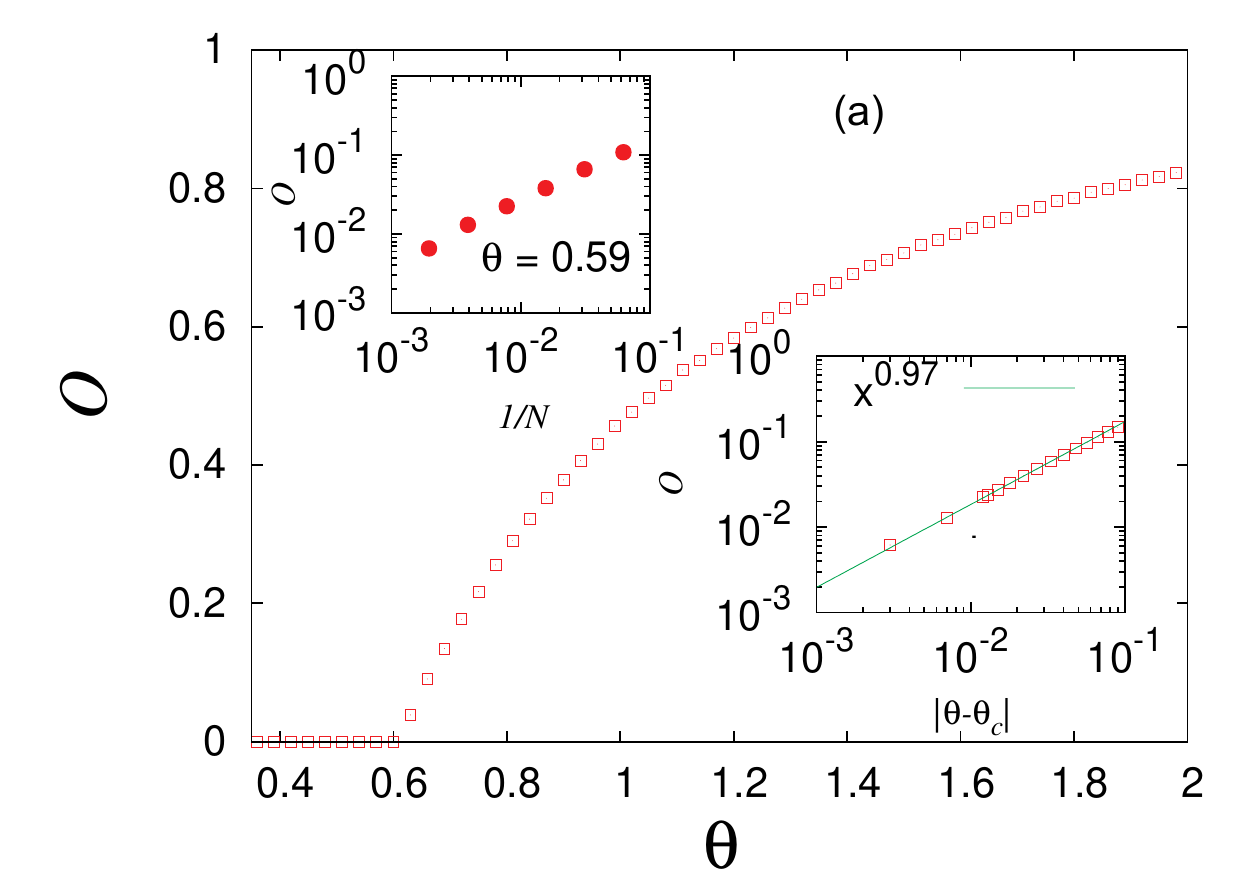}
 \includegraphics[width=5.9cm]{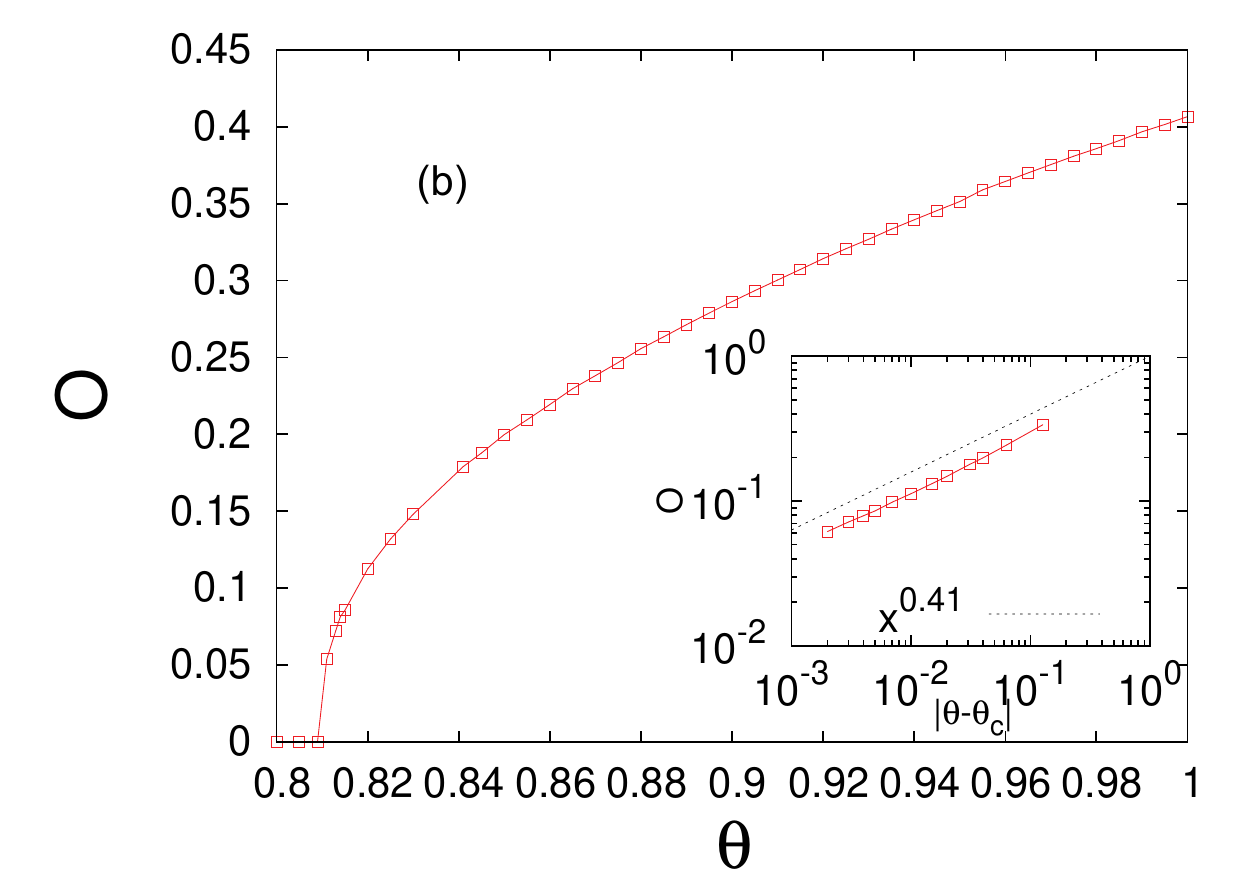}
 \includegraphics[width=5.9cm]{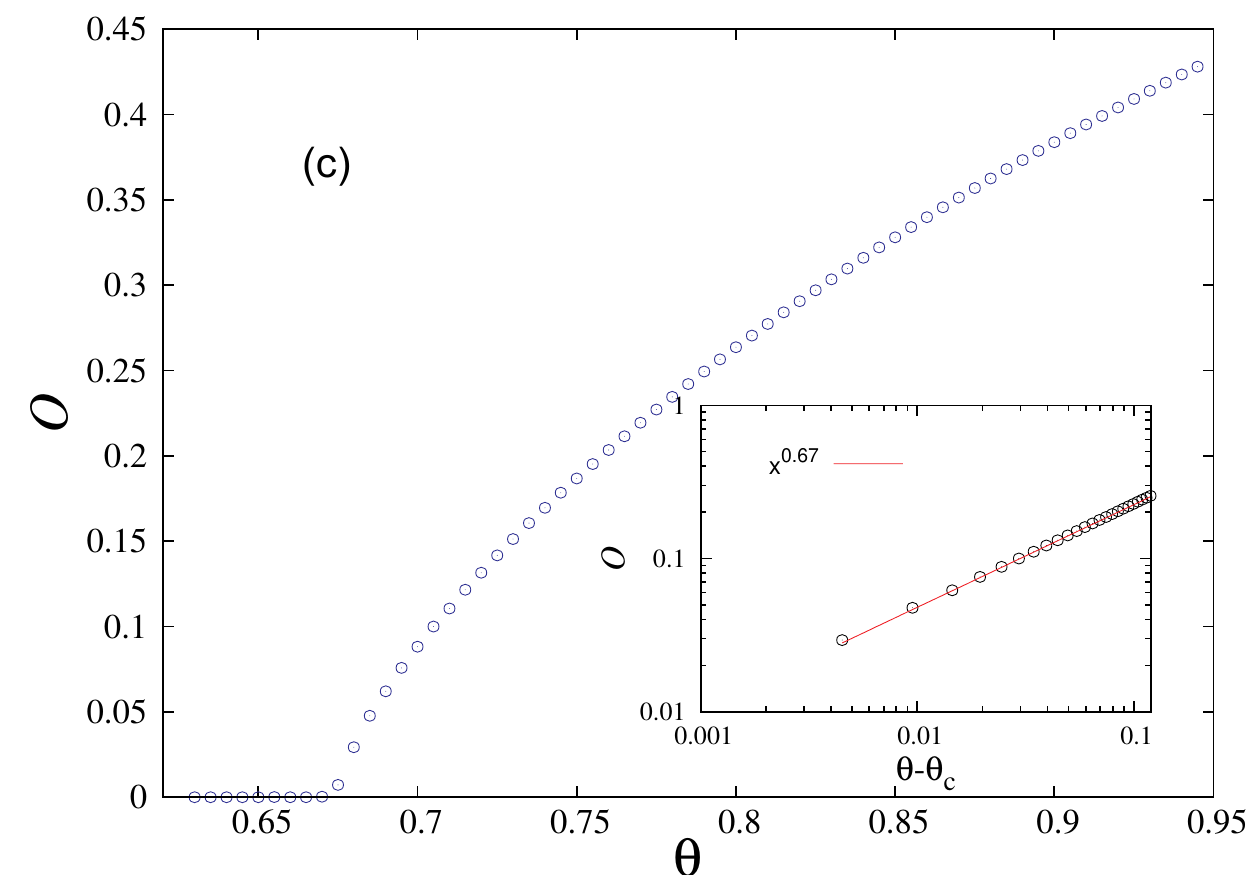}
\end{center}
\caption{(a) The numerical data of the variation of $O$ with the threshold money $\theta$ (in the mean field case), 
where $O$ is the number of agents having money below the threshold money $\theta$ in the equilibrium state. The top 
inset shows that for $\theta = 0.59$,  $O \to 0$ when $N \to \infty$. The bottom inset shows the power law fit of the 
numerical data of $O$ against $(\theta - {\theta}_c)$, which gives $\beta \simeq 0.91$, where $O \sim (\theta - {\theta}_c)^{\beta}$. 
Here the number of agents $N = 10^5$. (b) The variation of $O$ with the threshold money $\theta$ in one dimension. 
Inset shows the power law fit of the numerical data of $O$ against $(\theta - {\theta}_c)$. Such scaling fit gives $\beta \simeq 0.41$. 
Here the number of agent $N = 10^4$ (c) The variation of $O$ with the threshold money $\theta$ in two dimension. Inset 
shows the power law fit of the numerical data of $O$ against $(\theta - {\theta}_c)$. Such scaling fit gives $\beta \simeq 0.67$. 
The dimension of the lattice is $1000 \times 1000$.} 
\label{OP_fig}
\end{figure}

\subsection{A Kinetic Exchange Model with self-organized poverty-line}
Along with the modeling of the observed income or wealth data, there are few attempts made on illustrating intriguing models 
which indicate the potential way of reducing economic inequality in the society. Pianegonda et al.~\cite{pianegonda2003_sudip} 
proposed conservative exchange market (CEM) model where a group of agents is realized in one dimensional lattice. According to
their proposed dynamics, in each transaction one of the participating agent is necessarily the poorest agent of the group. Due 
to the transaction the poorest agent may gain (or lose) some amount of wealth. Such amount of wealth is equally deducted 
(or distributed) from the two nearest neighbors of the poorest agent. In the mean field version of the model (globally coupled), 
such deduction (or addition) of wealth is done from two randomly chosen agents. Considering the nearest neighbors interactions, 
they found a wealth distribution in which almost all agents are beyond a certain threshold ${\eta}_T$. Their numerical results 
indicated the value of ${\eta}_T \approx 0.4$ where in the initial configuration the wealth is a random number between $0$ to $1$, 
distributed uniformly among the agents. In the mean field case the value of ${\eta}_T \approx 0.2$. They noticed the probability 
of an agent becoming wealthier decreases with time and finally it converges to a value $\approx 0.76$. Pianegonda et al.~\cite{pianegonda2003_sudip} 
observed the fraction of rich agents is independent of the size of the market. The CEM model reveals an exponential distribution 
(beyond ${\eta}_T$) in case of nearest neighbors interactions whereas such distribution has almost linear form in the mean field 
limit. Iglesias~\cite{iglesias2010_sudip} simulated the dynamics of the CEM model through inclusion of several types of taxes. 
He considered a situation where the amount of wealth gain (or lose) by the poorest agent is equally collected from the all agents. 
Such kind of deduction of wealth can be treated as an implementation of tax (uniform for every agent) in the society for the 
development of the poor class of the people. The distribution obtained by imposing such global-uniform tax is similar with the 
distribution acquired in case of the mean field version of the CEM model but in this case the value of ${\eta}_T \approx 0.25$. 
It should be noted that in case of mean field CEM model the value of Gini coefficient $g \approx 0.1$ whereas such value becomes 
$g \approx 0.2$ for the wealth distribution extracted from the global-uniform tax version of CEM model. Iglesias~\cite{iglesias2010_sudip} 
also introduced proportional tax in the CEM model. In this case the deduction of wealth is not uniform for all the agents rather 
it is proportional to the wealth of the agents. Iglesias~\cite{iglesias2010_sudip} also simulated the effect of the proportional 
tax when it is deducted locally (i.e., from the neighbors). In the local case of the proportional tax he considered four neighbors 
and he got ${\eta}_T \approx 0.4$ whereas such value for the global case is approximately $0.32$. The value of the Gini coefficient 
is nearly $0.1$ for the wealth distribution (exponential in nature) obtained considering the local-proportional tax which is similar 
to the value of $g$ in the original local version of the CEM model. In case of global-proportional tax, the nature of the wealth 
distribution is power law and the value of $g \approx 0.16$. Ghosh et al.~\cite{ghosh2011_sudip} proposed gas-like model which shows 
an effective way of improving the financial condition of poor people. In their model they initially set an threshold $\theta$ value 
of money or wealth. Like the CEM model, in each interaction one of the agent must have the money below the $\theta$ and the other 
$j$-th agent is randomly selected (mean field case) from the rest of the agents. The dynamical equation of money exchange are given by
\begin{eqnarray}\label{Ghosh_model}
 m'_i = {\epsilon}({m_i}^{<}  + m_j) \nonumber \\
 m'_j = (1 - {\epsilon})({m_i}^{<}  + m_j)~~.
\end{eqnarray}
Here ${m_i}^{<} < \theta$, the money of the $i$-th agent before the interaction whereas $m'_i$ is the money after the interaction. 
The ${\epsilon}$ is a random number between $0$ to $1$. Each financial transaction is considered as unit time $t$. The dynamical 
exchange of money continues until the money of the all agents cross the threshold or poverty line $\theta$. After sufficiently 
long time $t > \tau$ (relaxation time) one will get a steady state distribution, in which a small perturbation cannot affect 
the distribution. Such perturbation is implemented by forcibly bringing down a agent below the level $\theta$ and to ensure the 
money conservation his/her money is given to anyone else. Addition of such perturbation cannot alter the relevant results obtained 
in the steady state. 

The equilibrium money distribution is independent of the initial states of agents which essentially 
reflects the ergodicity of the system. Ghosh et al.~\cite{ghosh2011_sudip} computed $O$, the average number of agents below the 
threshold $\theta$, which is certainly zero in steady state up to a critical value of the threshold ${\theta}_c$. That means for 
$\theta > {\theta}_c$ one will never get a distribution where $O$ is zero (see Fig.~\ref{OP_fig}). Interestingly the relaxation 
time $\tau$ divergences at $\theta = {\theta}_c$. These outcomes essentially indicate a phase transition in the system, where $O$ 
is the order parameter. In the mean field case the of ${\theta}_c \sim 0.6$ and values of the critical exponents are $\beta = 0.97 \pm 0.01$, 
$z = 0.83 \pm 0.01$ and $\delta = 0.93 \pm 0.01$. These exponents $\beta$, $z$ and $\delta$ are obtained from the power fit of 
$O \sim (\theta - {\theta}_c)^{\beta}$, $\tau \sim (\theta - {\theta}_c)^z$ and $O(t) \sim t^{-\delta}$ respectively.  Ghosh et al.~\cite{ghosh2011_sudip} studied their model in one dimension. In this case during a financial transaction one of the agent 
contains money ${m_i}^{<} < \theta$. The other agent is randomly selected from the two nearest neighbors (which can contain any 
money whatsoever) of the $i$-th agent. The variation of $O$ in the one dimension is shown in the Fig.~\ref{OP_fig}. The value of 
${\theta}_c = 0.810 \pm 0.0001$ and the obtained values of critical exponents are $\beta = 0.41 \pm 0.02$, $z = 0.810 \pm 0.0001$ 
and $\delta = 0.19 \pm 0.01$. Ghosh et al.~\cite{ghosh2011_sudip} also simulated the model in two dimension where again in a 
financial transaction one of the agent has money ${m_i}^{<} < \theta$ and the other agent is one of the four nearest neighbors of 
the $i$-th agent. The nearest neighbors can have any money whatsoever. The values of the critical point and exponents are 
${\theta}_c = 0.675 \pm 0.0005$, $\beta = 0.67 \pm 0.01$, $z = 1.2 \pm 0.01$ and $\delta = 0.43 \pm 0.02$. In should be mentioned 
that except the value of $z$, the values of $\beta$ and $\delta$ in mean field, one dimension and two dimensional cases are very 
close to the values of these exponents in the Manna model~\cite{lubeck2004_sudip,manna1991_sudip,manna1991two_sudip} 
in the respective dimensions. Ghosh et al.~\cite{ghosh2011_sudip} commented that the mismatch in the values of $z$ may arise due to 
the limitation in the system size and they conjectured that their model might belong to the Manna universality class.

\subsection{The Yard-Sale Model and effects of taxes}
Another kinetic exchange model considered had been (see Chakraborti~\cite{chak_2002_sudip}) that 
\begin{eqnarray}\label{chak_model}
 m_i(t+1) &=& m_i(t) - m_j(t) + {\epsilon}(2m_j(t)) \nonumber \\
 m'_j(t+1) &=& (1 - {\epsilon})(2m_j(t))  \nonumber 
\end{eqnarray}
when $m_i(t) \geq m_j(t)$. In many economic transactions this may be a natural feature, particularly 
for Yard-Sale model and hence the name. Here, the wealthier agent saves exactly the excess amount and 
trading takes place with double the poorer agent's money or wealth. The attractive and stable fixed 
point for the dynamic corresponds to wealth condensation in the land of one agent. This is obvious, 
as $m_j(t)$ at any $t$ becomes zero, the agent gets isolated from any further trade and this continues 
for all others until one agent grabs all! This absolute level of condensation or inequality in the 
model made it unrealistic$^{[1]}$\footnotetext[1]{The model was studied by the group members in Saha Institute of Nuclear Physics 
 during 2000-2005 and income/wealth condensation phenomenon was taken initially as a signature of the 
 absurdity of the model and the study was temporarily abandoned. Later, however, some interesting slow dynamical behavior of the model was 
 observed~~\cite{chak_2002_sudip}. It was also noted that a `mixed kinetic exchange strategy' (see e.g., Pradhan~\cite{pradhan2005random_sudip}) can  
 destabilize the condensation (with Gini coefficient $g = 1$) and can lead to some extreme but realistic distributions, having $g \lesssim 1$.}. 
 However, as discussed in the footnote, several strategies could save the model from such condensation with 
 extreme inequality ($g = 1$), and one suggested recently~\cite{bruce_2019_sudip} to impose the natural 
 tax collection by government, eventually to redistribute it among all the agents in the form of public 
 goods and services, e.g., road construction etc, is extremely successful and gives very good fit 
 to the data.

\section{Kolkata Paise Restaurant Problem}
The city Kolkata was once the capital of India and till date continues to be one of the oldest trading centre of this country. This old city has attracted large number of labors migrating from all parts of India. A century has passed since this city lost its preeminent position, but the migrant inflow has made Kolkata highly populated till date. Most of the migrants belongs to unorganized labor class, generally lacking secure wages even without fixed working hours. Sometime in Kolkata, there used to be an array of cheap restaurants at road side, namely `Paise Hotels': Paise is the smallest Indian currency. Everyday, each of those restaurant would prepare limited number of dishes, that costs very low (at rate of basic cost of cooking). And this cost matches well with the affordability of those labors. Budget is also not a constraint while choosing any restaurant. Thus the Paise hotels become much popular among the poor labors during their lunch hour. Everyday without discussing with others, they themselves 
would choose some restaurant for lunch that day. During lunch hour, they would walk down the street and visit his chosen restaurant for lunch. And, only one choice can be afforded per agent per day due to strict lunch hour.

For simplicity, we assume that only one dish would get prepared by each of those restaurants. If some day, only one labor arrives at some restaurant during lunch hour, he will be served the only dish prepared there. And he would go back to work happily. Problem arises if more than one people visit any one restaurant for lunch. Then the restaurant would choose one of them randomly and he gets the lunch. Others arrived there would miss their lunch for that day and report back to work staying hungry for rest of the working hour. Nobody would like to starve. Choosing restaurants intelligently could guarantee lunch for every labor that day. But how to choose in that way makes this problem interesting. 

So every day, these labors face a decision making problem. They do not discuss with peers while making his choice. Only information they have is the crowd distribution of every restaurants for some finite number of past days. End of the day, no one would like to continue work skipping lunch. Ideal case would be: if every labor arrives at a restaurant where nobody else visit for that day to assure his lunch. The maximum possible social utilization fraction is $1.0$ (assuming number of restaurants is same as number of labors), meaning no dish gets waste that day. 

One simple solution is: a central coordinator would ask the labors to form a queue and assign one of the restaurant to the first one in queue. Rest would follow him since then. Thus full social utilization is achieved and that too from first day. This solution is even valid if the restaurants are ranked (may be because of quality of service or taste of dish prepared there) which is commonly agreed upon by all labors though the cost of dish remain the same as previous.

But presence of a dictator is not always practical. Say in a democratic society, each individual will have his own choice and would hardly like to compromise by listening to some dictator. Rather one would prefer to choose a restaurant on his own. So how any labor would choose some restaurant without knowing the choice of others for that day, may be e.g. choosing randomly per day or evolving some strategy to improve social utilization fraction over time. The objective is to find some strategy following which maximum social utilization can be achieved. For that one needs to study the steady state dynamics of the strategy undertaken. Memory is limited and the labors only have last few days crowd distribution at every restaurant. And mutual interaction among some subset of labors i.e. grouping or some kind of fixing is not allowed.

Here we discuss Kolkata Paise Restaurant (KPR) problem as a repeated many-player many-choice game problem as introduced by Chakrabarti in \cite{chakrabarti2007kolkata}: there are $\nu N$ agents choosing among $N$ restaurants (however we will consider $\nu$ = 1) for lunch every day. Agents do not interact with others while making his decision any day. Information regarding last day's restaurant fill up statistics is available publicly. With this, agents choose and visit the chosen restaurant during lunch hour. Dish is guaranteed only if a restaurant is visited by some agent alone that day. Any day if some restaurant is visited by more than one agents then only one of them gets the food and others return remaining hungry for that day. But, lunch hour is strict. Visiting another restaurant would delay to return back to work and hence only one choice can be made per day per agent. End of the day, the social utilization fraction (number of restaurants visited by at least one agent by total number of restaurant) 
will be calculated. Maximum possible social utilization fraction is unity. This is when every agent is able get lunch at some restaurant, and no dish gets wasted that day. The dictatorial solution, as discussed in previous para, though works well though we will encourage the readers to evolve some strategy following which those agents will learn to make decision their own and also social utilization fraction can be maximized as much as possible. Below we discuss few interesting strategies along with their results developed by several econophysics researchers. 

\subsection{Learning Strategies}
Here we will study the dynamics of Kolkata Paise Restaurant game problem following several strategies proposed in \cite{chakrabarti2009kolkata,ghosh2010statistics,martin2019}. They are: No Learning (NL), Limited Learning (LL), One Period Repetition (OPR), Crowd Avoiding strategy (CA), Stochastic Crowd Avoiding strategy (SCA). Among them No Learning strategy is considered to be the base strategy throughout, often compared with other mentioned strategy.
\subsubsection{\textbf{No Learning (NL)}}
\begin{figure}[h]
\begin{center}
\fbox{\includegraphics[width=6.0cm]{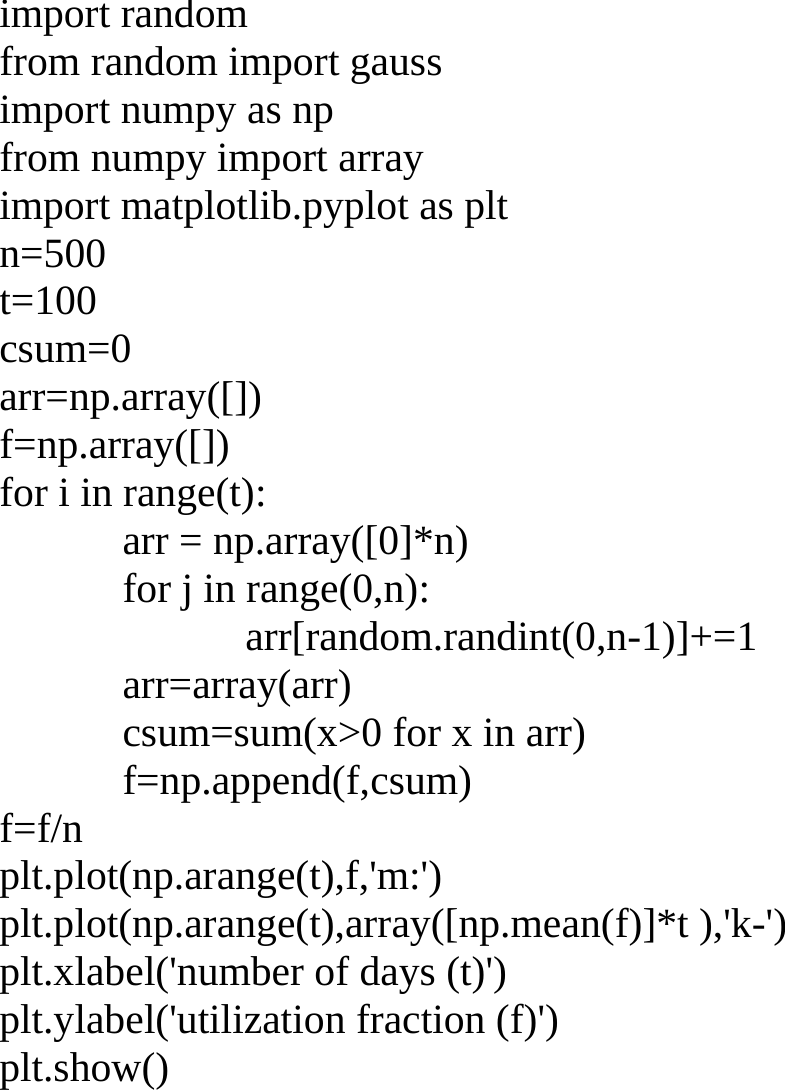}}
\end{center}
\caption{Python (version $2.7$) program for No Learning strategy. Results (see Fig.~\ref{nl_op}) can be obtained if one runs this program.
}
\label{nl_code}
\end{figure}

\begin{figure}[h]
\begin{center}
\includegraphics[width=8.0cm]{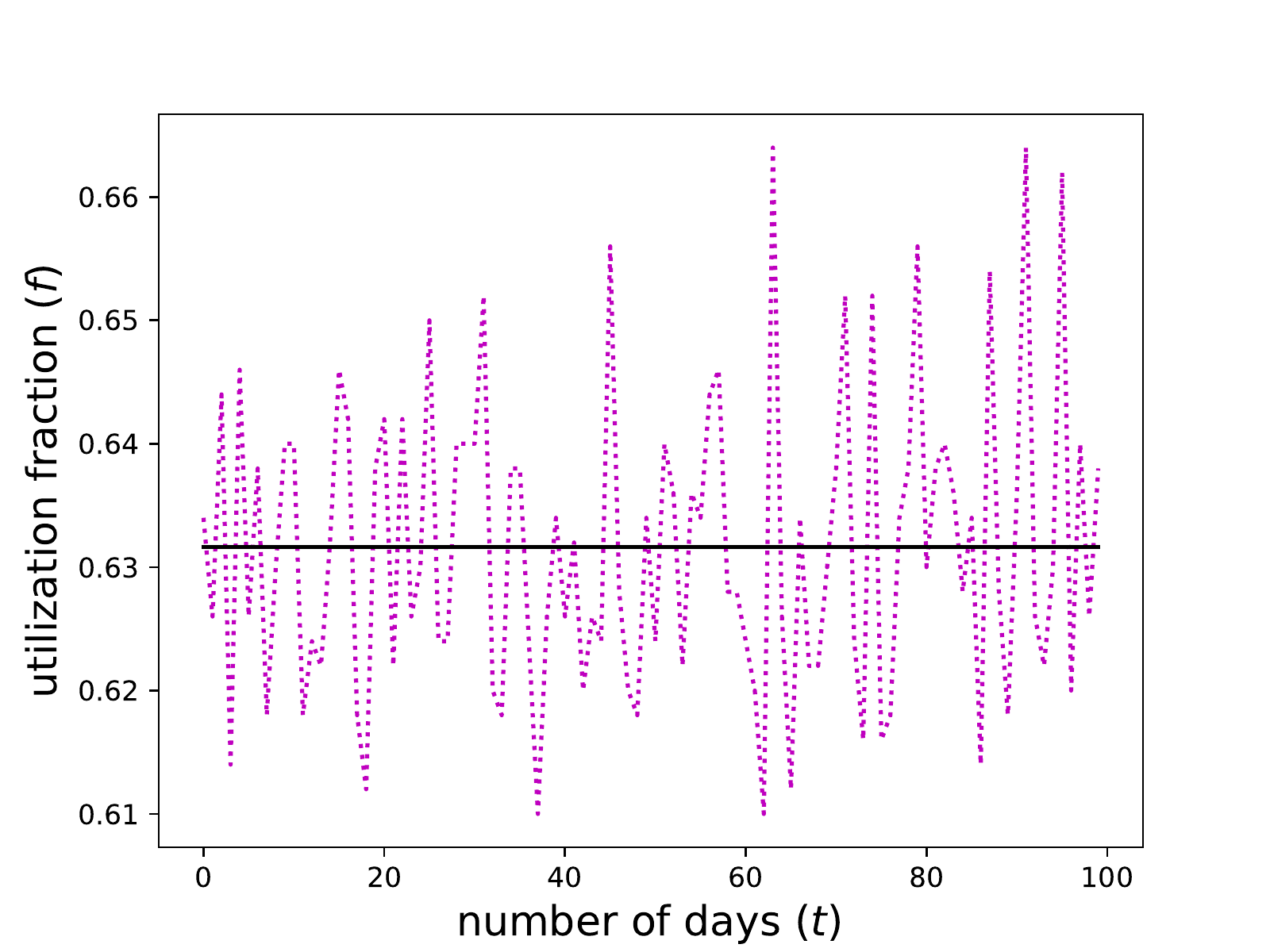}
\end{center}
\caption{Plot of social utilization fraction ($f$) in magenta color vs. days ($t$) following No Learning (NL) strategy. Average utilization fraction $\bar{f}(\simeq 0.63)$ over 100 days is shown in black color straight line.}
\label{nl_op}
\end{figure}

In this strategy, $\nu N$ agents randomly chooses among $N$ restaurants. We consider no past history i.e. memory. For simpilicity, restaurant occupying density $\nu$ is considered to be $1$ throughout study. 
The probability of choosing a restaurant by $n < N$ agents is:
\begin{equation}\label{ekpr_1}
\tilde{P}(n) = \binom{\nu N}{n} {p}^{n} {(1-p)}^{\nu N - n}
\end{equation}
The restaurants being equi-probable, the probability of choosing one restaurant among $N$ is $p = \frac{1}{N}$ and for $N\to\infty$ one gets (using Poisson Limit theorem):
\begin{equation}\label{ekpr_2}
\tilde{P}(n) = \frac{\nu^n }{n!} exp(-\nu) 
\end{equation}
So, fraction of restaurants not chosen by any agent is $\tilde{P}(n=0)=exp(-\nu)$, and this gives the average fraction of restaurants chosen by at least one agent on that day $\bar{f}$ is $1-exp(-\nu)\simeq 0.63$. This we will consider as the base strategy and will compare with remaining cases for improvement in $f$.
Results Following No Learning strategy given in Fig.~\ref{nl_op} can be obtained using program given in Fig.~\ref{nl_code}.
\subsubsection{\textbf{Limited Learning (LL)}}
On first day, agents will randomly choose some restaurant similar to No Learning strategy with $f=0.63$. Next day onwards, they will make individual choice depending upon their last day lunch availability: below we discuss the LL(1) inspired strategy proposed in \cite{chakrabarti2009kolkata}.
If an agent gets lunch from some restaurant on $t$-th day, then he opts for the best restaurant on day $t+1$.
If he did not get lunch on any $t$-th day, then next day $t+1$ he randomly choose one among the other $(N-1)$ restaurant with equal probability.
Say $x_t$ fraction of agents or rather $f_tN$ number of agents ($f_t$ is utilization fraction at $t$-th day) on getting their lunch on some day $t$ will visit the best restaurant (restaurant 1), and only one of them will get lunch there and others will not get their lunch for that day. Remaining ($N-f_tN$) agents will try from the remaining $(N-1)$ restaurant following no learning case. And the recursion relation will be:
\begin{equation}\label{ekpr_3}
f_{t+1} = \frac{1}{x_t} + (1-exp(-\nu_t)); \nu_t = 1 - f_t
\end{equation} 
The first term of the summand will contribute 0 as $N\to\infty$, and one gets the steady state utilization fraction $f \simeq 0.43$. This also matches well with numerical simulation result as reported in~\cite{chakrabarti2009kolkata}.
\subsubsection{\textbf{One Period Repetition (OPR)}}
On first day say $t-1$, agents will randomly choose a restaurant following NL strategy.   
If an agent gets his lunch on day $t-1$ from some restaurant, he will revisit there on day $t$. This is one period repeat.
If some agent got his lunch from same restaurant for two consecutive days $t-1$ and $t$, then he will compete for the best restaurant (ranking of restaurant is agreed upon by all agents) on day $t+1$. 
For any day $t$ if an agent fails to get lunch, next day $t+1$ he will randomly choose one among restaurants which remained vacant yesterday.

Here the fraction $f_t$ (representing social utilization at day $t$) is made of two parts: $x_{t-1}$ fraction of agents who will continue their lunch at last day chosen restaurant, and rest of the fraction $x_t$ who have chosen today:
\begin{equation}\label{ekpr_4}
f_t = x_{t-1} + [1-x_{t-1}](1-exp(-1))
\end{equation}
The fraction $x_t$ of agents who have chosen today is given by NL strategy where $N(1-x_{t-1})$ left out agents finds one out of $N(1-x_{t-1})$ yesterday's vacant restaurants, so ${\nu}_t = 1$.
\\On next day, the fraction $f_{t+1}$ will be:
\begin{equation}\label{ekpr_5}
f_{t+1} = \frac{1}{x_{t-1}} + x_t + (1-x_{t-1}-x_t) (1-exp(-1))
\end{equation}
fraction $\frac{1}{x_{t-1}}$ would be very small and gets ignored, and replacing $x_t$ with $(1-x_{t-1})[1-exp(-1)]$ one gets, 
\begin{equation}\label{ekpr_6}
f_{t+1} = [1-x_{t-1}](1-exp(-1)) + [1-x_{t-1}-(1-x_{t-1})(1-exp(-1))](1-exp(-1))
\end{equation}
At stable state convergence, $f_{t-1}$ = $f_t$ = $f_{t+1}$ = $f$ and $x_{t-1}$ = $x_t$ = $x_{t+1}$, so dropping the subscript $t$ and equating $f_t$ = $f_{t+1}$ one calculates $x$ to be 0.19 and $f$ = 0.71. So this strategy is an improvement over NL case, though an agent after getting lunch at some restaurant $k$ revisits there the very next day with probability 1. The fluctuations in the social utilization fraction is found to be Gaussian in the simulation results reported by \cite{chakrabarti2009kolkata}.  
\subsubsection{\textbf{Crowd Avoiding strategy (CA)}}
As the name suggests, agents following this strategy will randomly choose some restaurant on day $t$ where nobody had visited last day i.e. day($t-1$). We provide a sample program in Fig.~\ref{ca_op} to calculate the steady state utilization fraction $f$ following Crowd Avoiding strategy. Computer simulation results the distribution of social utilization fraction $f$ to be Gaussian with peak around 0.46. It can be understood following way: as the fraction of restaurants filled last day will strictly get avoided by agents today, so the number of available restaurants today is $N(1-f)$ which will be chosen randomly by all agents. With $\nu = \frac{1}{1-f}$, the recursion relation becomes:
\begin{equation}\label{ekpr_7}
f = (1-f)[1-exp(-\frac{1}{1-f})]
\end{equation}
\begin{figure}[h]
\begin{center}
\fbox{\includegraphics[width=6.9cm]{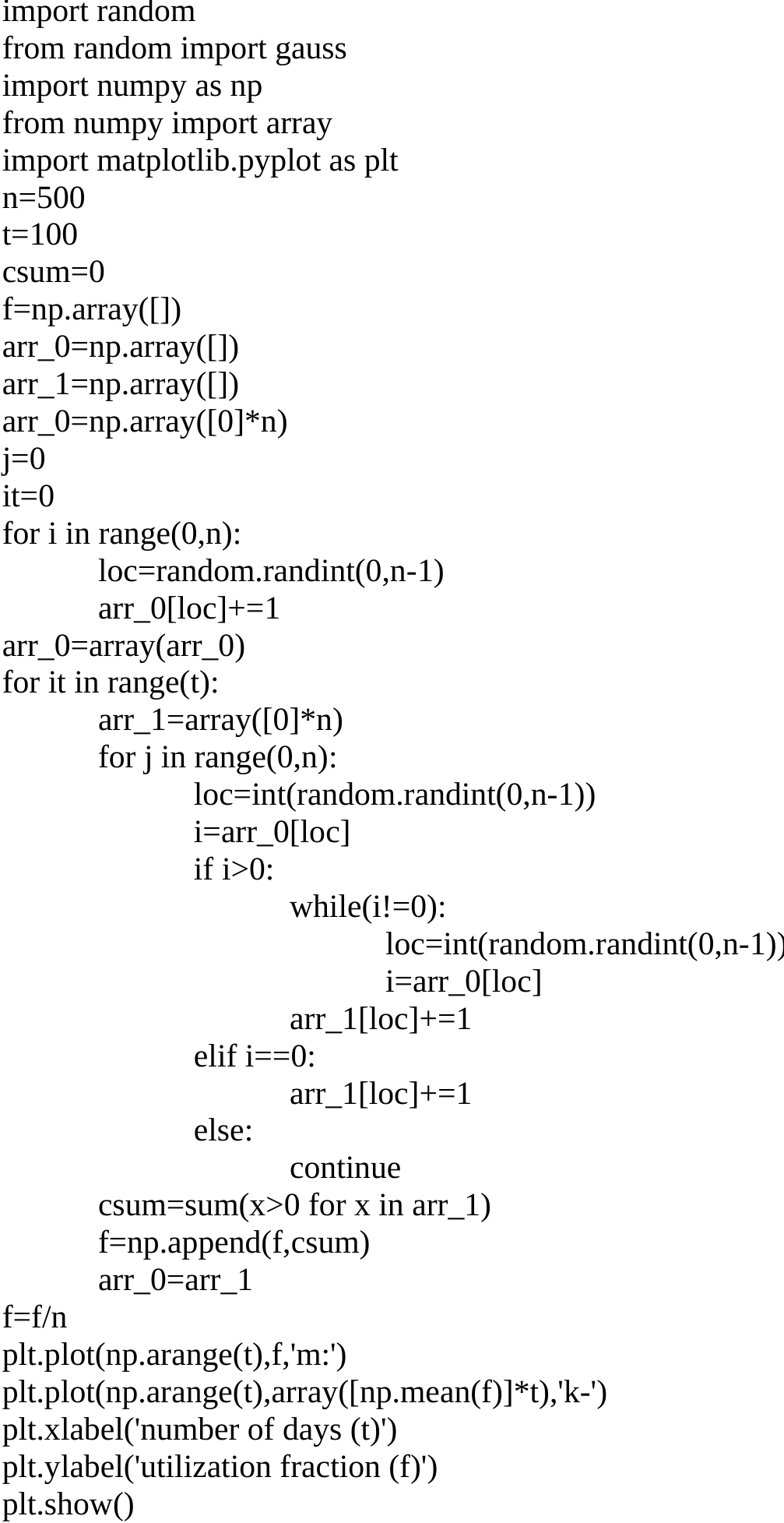}}
\end{center}
\caption{Python (version $2.7$) program for Crowd Avoiding strategy. Results (see Fig.~\ref{ca_op})  can be plotted if one runs this program.}
\label{ca_code}
\end{figure}
Solving above equation, we get $f$ = 0.46, which fits well along the simulation result (Fig.~\ref{ca_op}).
\begin{figure}[h]
\begin{center}
\includegraphics[width=8.5cm]{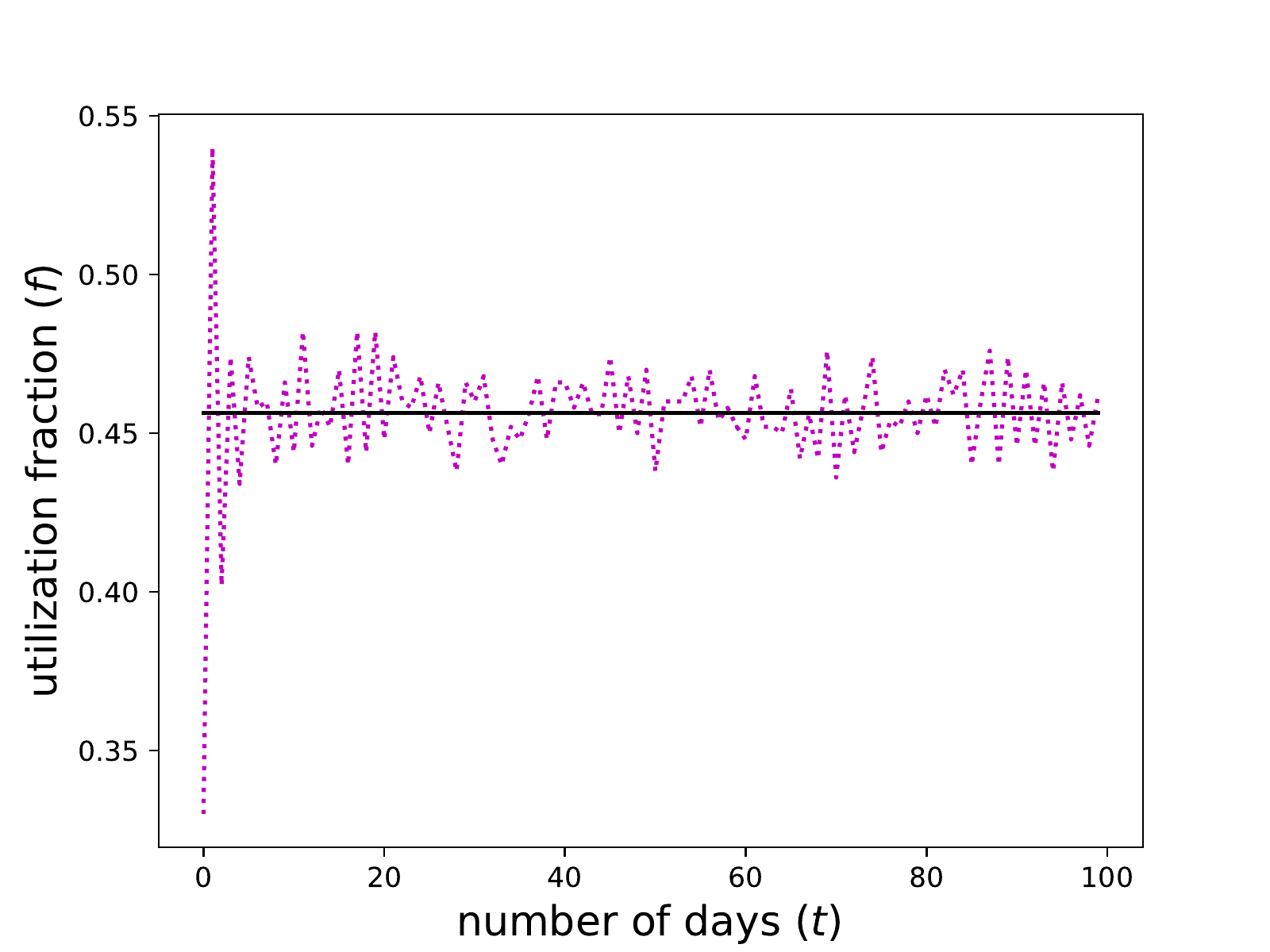}
\end{center}
\caption{Plot of social utilization fraction ($f$) in magenta color vs. days ($t$) following Crowd Avoiding strategy.  Average utilization fraction $\bar{f}(\simeq 0.46)$ over 100 days is shown in black color straight line.}
\label{ca_op}
\end{figure}

\subsubsection{\textbf{Stochastic Crowd Avoiding strategy (SCA)}}
We consider the strategy be following: if an agent arrives at restaurant $j$ for getting his lunch on day ($t$-1), then next day (i.e. day $t$) probability of visiting back to the same restaurant for that agent to have his lunch will depend on how crowded was the last day's restaurant. So, probability of visiting restaurant $j$ on day $t$ is $p_j(t)$=$\frac{1}{n_j(t-1)}$. And the probability of visiting any other restaurant $j^{'}(\neq j)$ goes as $p_{j^{'}}(t)$=$\frac{(1-p_j(t))}{(N-1)}$. Both the numerical and analytical results of average utilization fraction $\bar{f}$ following this strategy is found to be $0.8$. The distribution is Gaussian with peak around $f\simeq 0.8$~\cite{chakrabarti2009kolkata}.

\begin{figure}[h]
\begin{center}
\includegraphics[width=8.5cm]{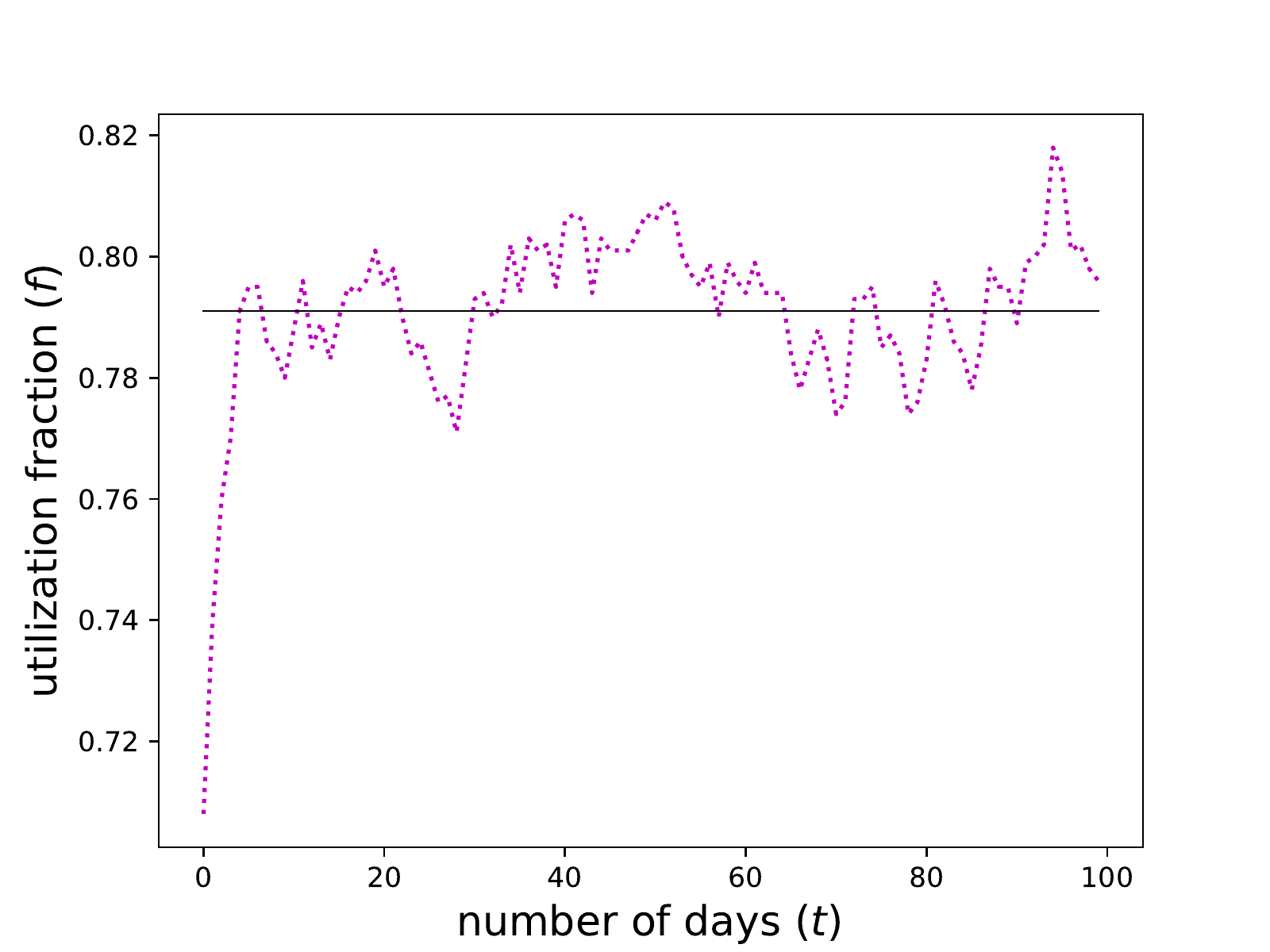}
\end{center}
\caption{Plot of social utilization fraction ($f$) in magenta color vs. days ($t$) following Stochastic Crowd Avoiding strategy. Average utilization fraction $\bar{f}(\simeq 0.79)$ over 100 days is shown in black color straight line.
}
\label{sca_op}
\end{figure}

The approximate estimation for the steady state behavior of $f$ following the above strategy can be made as: Suppose $a_i$ denotes the fraction of restaurants having exactly $i$ number of agents ($i$ = $0,1,...N$) arrived on day $t$. And also assume that $a_i$ = 0 for $i \geq$ 3 for large $t$ as the dynamics gets stabilized in steady state. Thus, $a_0+a_1+a_2$ = 1, $a_1+2a_2$ = 1 for large $t$ gives $a_0$ = $a_2$. Following the strategy, $a_2$ fraction of agents will try to leave their last day's visited (each with a probability of 1/2) restaurant on day $t+1$ and of course, no such activity will take place in some restaurant where nobody (fraction $a_0$) or only one agent (fraction $a_1$) visited last day. Fraction of $a_2$ agents with successful leave attempt will now get equally distributed into $N-1$ remaining restaurants. Out of this $a_2$, fraction visiting any vacant restaurant ($a_0$) of last day is $a_0a_2$, at present fraction of vacant restaurant is $a_0$ = $a_0 - a_0a_2$. In this process, the 
vacancy which will also pop up in those restaurants where exactly two agents visited last day is $(a_2/4) - a_2(a_2/4)$. At steady state one can write:
\begin{equation}\label{ekpr_8}
a_0 = a_0 - a_0a_2 + \frac{a_2}{4} - a_2\frac{a_2}{4}
\end{equation}
Using $a_0 = a_2$, we get $a_0=a_2=0.2$, and also $a_1=0.6$. Thus, social utilization fraction $f$ becomes $a_1 + a_2$ = 0.8. This is an approximate result considering nil contribution from $a_i$ with $i \geq 3$ over large $t$, as seen numerically. Simulation results also confirm the approximated result of steady state utilization fraction, see Fig.~\ref{sca_op}.

\subsection{Phase Transition in KPR game}
\begin{figure}[h]
\begin{center}
\includegraphics[width=8.5cm]{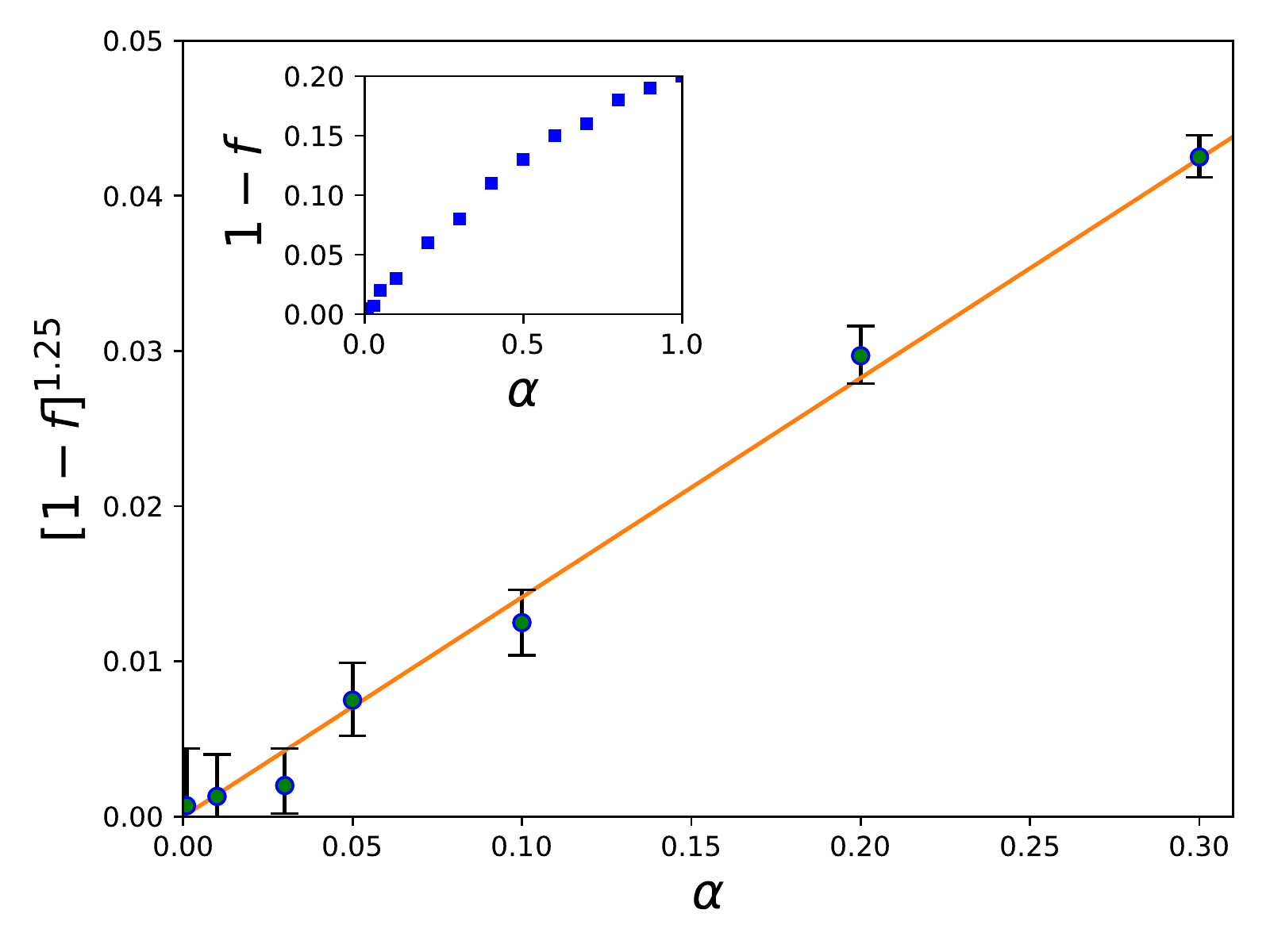}
\end{center}
\caption{Figure 18 Plot of $[1-f]$ against $\alpha$ fitted to $(\alpha - \alpha_c)^{\beta}$  where $\beta \simeq 0.8$ ($f$ denoting the utilization fraction and $\alpha$ denoting the weight factor power as defined earlier). The critical point $\alpha_c$ is fount to be at $0_+$~\cite{sinha2019phase}. 
}
\label{pt_kpr}
\end{figure}

Recently, a novel phase transition phenomena is reported in Kolkata Paise restaurant game problem~\cite{sinha2019phase} if $n_i(t-1)$ number of agents revisit their last day visited restaurant $i$ with weight $[n_i(t-1)]^\alpha$ and $1/(N-1)$ for any other restaurant, then the steady state ($t$ independent) utilization fraction becomes $f(t)$ = $[1-\sum_{i=1}^{N} [\delta(n_i(t))/N]]$. Here, the critical point $\alpha_c$ is found near point $0_+$ where $(1-f)$ is found to vary as $(\alpha - \alpha_c)^{\beta}$ where $\beta \simeq 0.8$. See Fig.~\ref{pt_kpr}, where ($1-f$) is plotted against $\alpha$, fitting is done using maximum likelihood estimation. Inset plot shows the direct functional relationship between ($1-f$) and $\alpha$.

\subsection{Application of KPR: Vehicle for Hire problem in mobility market}
KPR game model has been applied in competitive resource allocation systems where scarce resources need to be allocated effectively for repeated times. Areas like dynamic matching in mobility markets~\cite{martin2019}, mean field equilibrium study for resource competitive platform reported in~\cite{yang2018mean} are such examples. In mobility market, agents generally have their individual preference/ranking towards resources. Vehicle for hire markets is one of such platform where drivers individually decide some pick up location to increase their own utilization and accordingly offer individual transportation to customers by car. Martin \& Karaenke~\cite{martin2017vehicle} has extended the KPR game model to generalize the Vehicle for Hire Problem (VFHP) and applied playing strategies discussed in previous section assuming drivers as agents and customers as resources in hire market. Below we discuss some of the results where drivers will have their individual ranking over customers unlike original KPR game 
model having commonly agreed resource ranking. 

\subsubsection{\textbf{No Learning (NL)}}
$\nu N$ drivers,independent of their individual preference ranking and any last trip history, choosing randomly among $N$ customers for their next ride. For simplicity $\nu$ is considered as 1. The probability of choosing any particular customer by $n < N$ drivers is:
\begin{equation}\label{eakpr_1}
\tilde{P}(n) = \binom{\nu N}{n} {p}^{n} {(1-p)}^{\nu N - n}
\end{equation}
Assuming each of the customers to be equi-probable with probability $p = \frac{1}{N}$ and for $N\to\infty$ one gets (using Poisson Limit theorem):
\begin{equation}\label{eakpr_2}
\tilde{P}(n) = \frac{\nu^n }{n!} exp(-\nu) 
\end{equation}
Thus fraction of customers not chosen by any driver is $\tilde{P}(n=0)=exp(-\nu)$, and one obtains the average fraction of customers chosen by at least one driver for ride $\bar{f}$ is $1-exp(-\nu)\simeq 0.63$. This result is exactly similar to NL or base strategy of KPR and again we will compare this result with other strategies.

\subsubsection{\textbf{Limited Learning (LL)}}
Following LL strategy of KPR, drivers choose a customer randomly for ride at time $t$ and go to their most preferred customer at time $t+1$ if they got a tour at time $t$. otherwise they choose randomly again. 
One obtains the utilization fraction by following formula:
\begin{equation}\label{eakpr_3}
f_{t+1} = f_t \cdot (1-exp(-1)) + (1-exp(-\nu_t)); \nu_t = 1 - f_t
\end{equation} 
The left summand of above equation represents all those drivers who are successful in choosing their most preferred customer at time $t+1$ after choosing randomly at time $t$ or earlier. The right summand models those who choose randomly or successful in choosing their top priority customer at time $t$ and return there. On solving the above equation, one gets the steady state utilization $f = 0.702$. It is quite higher than the same strategy applied in original KPR game model. Drivers having self preference over top customer choice, may have different customer as his best choice. Thus the effective fraction of top customers becomes much higher than having one commonly agreed top rank resource. This is making LL strategy in Vehicle for hire market superior to original KPR LL strategy utilization. 

\subsubsection{\textbf{One Period Repetition (OPR)}}
At time $t-1$, drivers will randomly choose customers for trip. For the next ride he will repeat trip with the same customer of time $t-1$, if successful in previous ride choosing any customer. For drivers with two consecutive successful trip with same customer, would go for individual best ranked customer at time $t+1$. Any driver follows NL strategy for next trip if fails in getting a successful ride at any time slot.
So, overall utilization fraction at time $t$, say $f_t$, comprises of two parts: $x_{t-1}$ fraction of drivers who continues riding with last time chosen customer and rest of the fraction who choose any customer at current time slot. So $f_t$ becomes:
\begin{equation}\label{eakpr_4}
f_t = x_{t-1} + [1-x_{t-1}](1-exp(-1))
\end{equation}
The utilization fraction for $x_t$ fraction of drivers who have chosen in current trip will be given by NL strategy where $N(1-x_{t-1})$ left out drivers finds one out of the $N(1-x_{t-1})$ customers waiting for a ride.
\\On next trip, utilization fraction $f_{t+1}$ will be:
\begin{equation}\label{eakpr_4}
f_{t+1} = x_{t-1} + x_t + (1-x_{t-1}-x_t) (1-exp(-1))
\end{equation}
substituting $x_t$ with $(1-x_{t-1})[1-exp(-1)]$ one gets, 
\begin{equation}\label{eakpr_5}
f_{t+1} = x_{t-1} + [1-x_{t-1}](1-exp(-1)) + [1-x_{t-1}-(1-x_{t-1})(1-exp(-1))](1-exp(-1))
\end{equation}
At stable state convergence, $f_{t-1}$ = $f_t$ = $f_{t+1}$ = $f$ and $x_{t-1}$ = $x_t$ = $x_{t+1}$, so dropping the subscript $t$ and equating $f_t$ = $f_{t+1}$ one calculates $x$ to be $0.28$ and $f = 0.73$.

\subsection{\textbf{Application of KPR: Resource allocation in wireless IoT system} }
Recently, Park et al. has proposed and analyzed a KPR inspired learning framework \cite{park2017kolkata} for resource allocation in the IoT environment. Here, the IoT devices has been modeled as non-cooperative agents choosing their preferred resource block with limited past information made available by their neighbors. However, socially optimal solution is reported for denser as well as lesser dense IoT environment, discussed below. 
\\

Internet of Things (IoT) technology is behind many smart city application. In an IoT environment, huge number of devices gets deployed for task and given limited source of energy (battery) for each device makes it challenging to ensure efficient resource allocation per time slot. The transmissions are often random and infrequent in nature. Now, up link of such an IoT system in time division multi-access way is considered with one base station (BS) to serve $N$ IoT devices transmitting short packets whenever they want to. For each time slot $t$, the channel will be divided into $b$ resource blocks (RB), each of which to be used by exactly one transmitting device: multiple devices choosing same RB end up transmission failure. At time slot $t$, the probability of $N_t = n_t$ devices transmitting with probability $p$ is:  
\begin{equation}\label{eakpr_6}
\tilde{P}(N_t = n_t) = {N\choose n_t} p^{n_t} (1-p)^{N-n_t}
\end{equation}


Generally in an IoT system, $N_t$ is much larger than $b$. In order to obtain maximum socially optimal solution, the trivial solution could be a centralized solution where the BS allocates RBs to transmitting devices. But for several reasons this solution is impractical. The simple solution is to let the transmitting device randomly choose a RB with equal probability. Let $S_n$ be a random variable representing number of successful transmission with support function supp($S_n$) as [0,$N_t$] for $N_t \leq b$ and [0,$b$-1] when $N_t > b$. Taken $N_t = n_t$, the probability of having minimum $s$ successful transmission is:
\begin{equation}\label{eakpr_7}
Pr(S_n \geq s| N_t = n_t) = \prod_{i = 0}^{s-1} \frac{b-i}{b} [{\frac{b-s}{b}}]^{(n_t - s)}\\
\end{equation}
 
From equation  and , one gets probability of having $s$ successful transmission as:
\begin{equation}\label{eakpr_8}
Pr(S_n = s) = \sum_{i=s}^{N} Pr(S_n = s|N_t = i)Pr(N_t = i)
\end{equation}

Following \cite{park2017kolkata}, Park et al. has modeled the one to one association between RBs and IoT devices as a KPR game where each player $i \in \rho$ has a set of actions $\mathbb{A}_i$ of selecting set of $b$ RBs. Also the RBs have different channel gain $\tilde{g}_i$ and a device would prefer to transmit using a channel block with higher gain. Let the utility function for an IoT device $i$ choosing an action $a_i,_t$ of selecting RB $j$ with channel gain $\tilde{g}_i,_j$ at slot $t$ is:
\begin{equation}\label{eakpr_9}
u(a_i,_t) = \begin{cases}
\tilde{g}_i,a_i,_t & \text{if $a_i,_t \neq a_k,_t \forall k \in \rho, k \neq i,$.}\\
0 & \text{otherwise.}
\end{cases}
\end{equation}

Unlike original KPR game model, multiple transmitting devices choosing single RB results into zero utility. Within certain communication range ($r_c$), during time slot $t$ each transmitting device learns from their neighbor's RB usage; on ($t-1$) slot, if neighbor's transmission is successful then choose some better ranked RB else transmit with less preferable RB than neighbor's last preference, otherwise transmit randomly. Nash equilibrium is reported for certain values of $r_c$ for even denser IoT network with service rate of $27.1$\%, which is about threefold higher than respective baseline. 

\section{Social Networks}
Society evolves in to many complex network structures in the hyperspace of inter-agent interactions (viewed as links) among the agents (viewed as nodes). Examples may be transport networks, bank networks, etc. 
Geometrically defined, for example, if one puts $N$ random dots (nodes) on an unit area the minimum 
distance of separation between any two dots or nodes will decays as $1/\sqrt{N}$. The same problem 
in an unit three dimensional volume will give the inter-node separation decaying as $1/{N}^{1/3}$, and 
decays as $1/N^{1/d}$ for a $d$ dimensional system. 
For complex social networks such inter-node distance (for a fixed embedding volume) will decay as 
$1/{(\log N)}$, as in an effective infinite-dimensional space. This indicates the  minimum contact 
number required for the spread of disease or information in a society. In other words, the number of 
transits for transport between any two nodes can be much lower than naively expected for two or three dimensional world!

\subsection{Indian Railway network analysis}     
Here we discuss one of the highly cited research work from Kolkata: the structural properties of the $160$ years' old Indian Railway network (IRN) as complex network. Indian Railway is the largest medium of transport in the country. Question is, how many trains one passenger would need to change while reaching any destination within country while traveling by train. Passengers would not like to change many trains to avoid the hassle and latency. How would if a train runs between every stations (junction, regular as well as remote) to make journey hassle-free i.e. switching trains during journey become less; considering $579$ trains running between $587$ stations. But it will end up incurring too much latency. In~\cite{sen2003small}, for the first time Indian Railway network has been studied by Sen et al. as a graph where each railway-station is considered a node and the physical track joining any two stations as the edge so that there exist minimum one train running between them using the track. For 
simplicity, edges are considered to be one unit distant, ignoring actual geographical distance between the two stations it connect as an unweighted graph. Rigorous investigation of the structural properties of IRN as a complex network are discussed below.

\begin{figure}[h]
\begin{center}
\includegraphics[width=6.5cm]{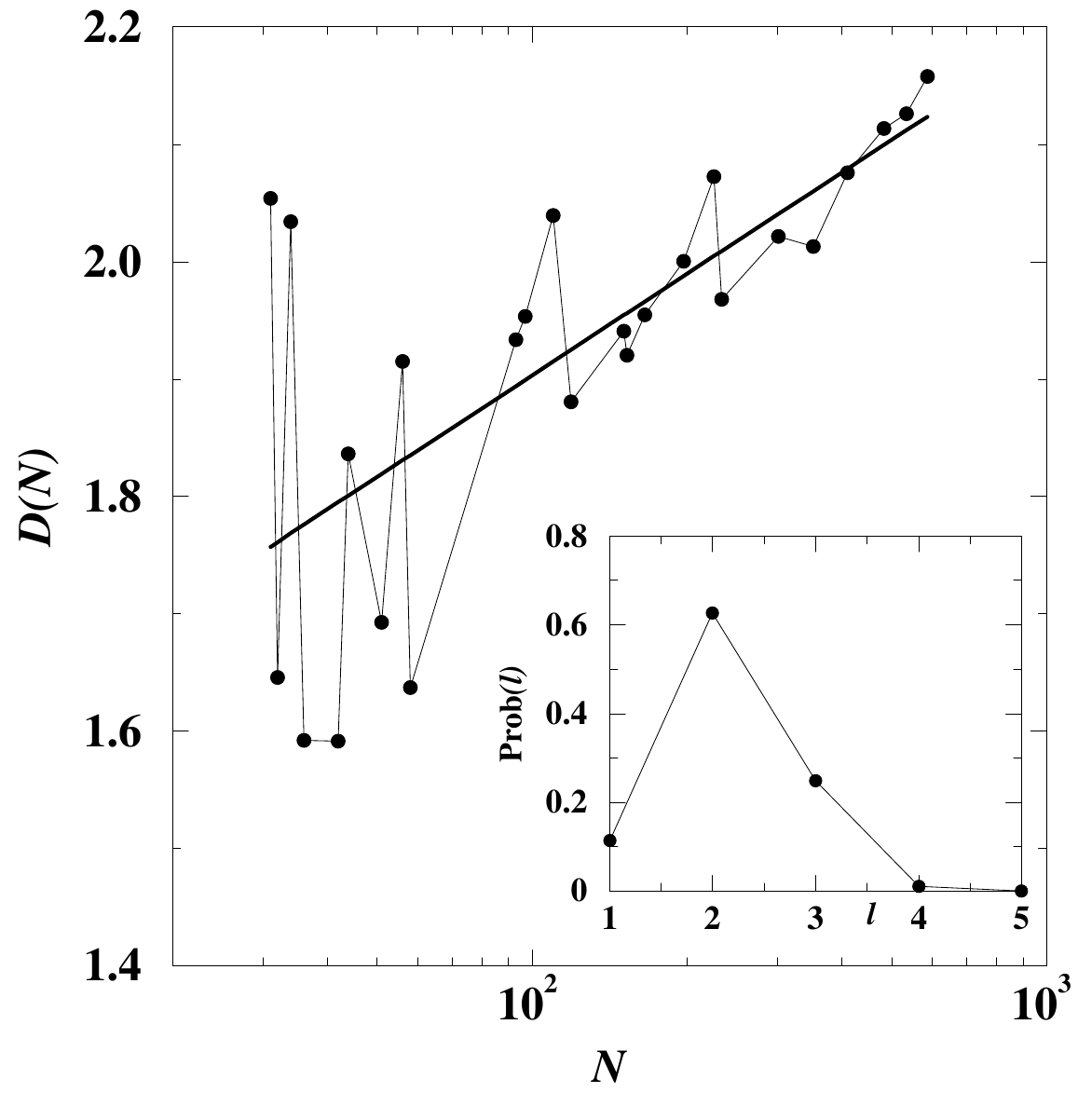}
\end{center}
\caption{Plot of the mean distance $D(N)$ considering $25$ different subsets of IRN consisting different number of nodes $(N)$. The fitted function over the whole range of nodes is $D(N) = A+B\log(N)$ where $A \approx 1.33$ and $B \approx 0.13$. The probability distribution Prob$(\ell)$ of the shortest path lengths $\ell$ on IRN is shown in the inset. The mean distance $D(N)$ of the IRN network is $\approx 2.16$, varying maximum up to $5$ links. 
}
\label{mean_distance}
\end{figure}

\begin{figure}[h]
\begin{center}
\includegraphics[width=6.5cm]{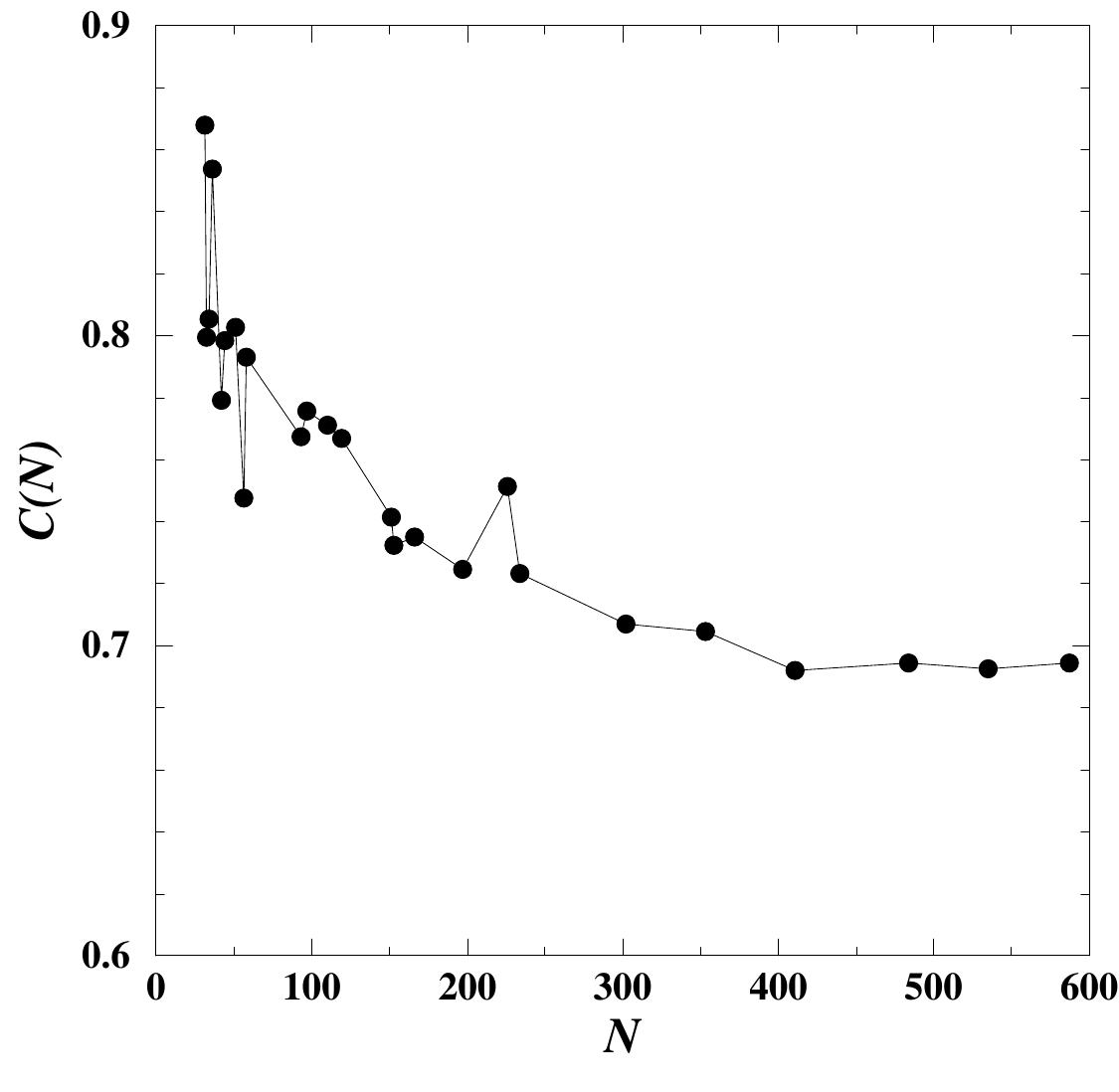}
\end{center}
\caption{Plot of the computed clustering coefficients $C(N)$ over $25$ different subsets of IRN with different number of nodes $N$. Initially very high followed by fluctuations, $C(N)$ has finally seen to stabilize at $0.69$ as $N\to\infty$, over the whole IRN. 
}
\label{cc}
\end{figure}

\begin{figure}[h]
\begin{center}
\includegraphics[width=6.5cm]{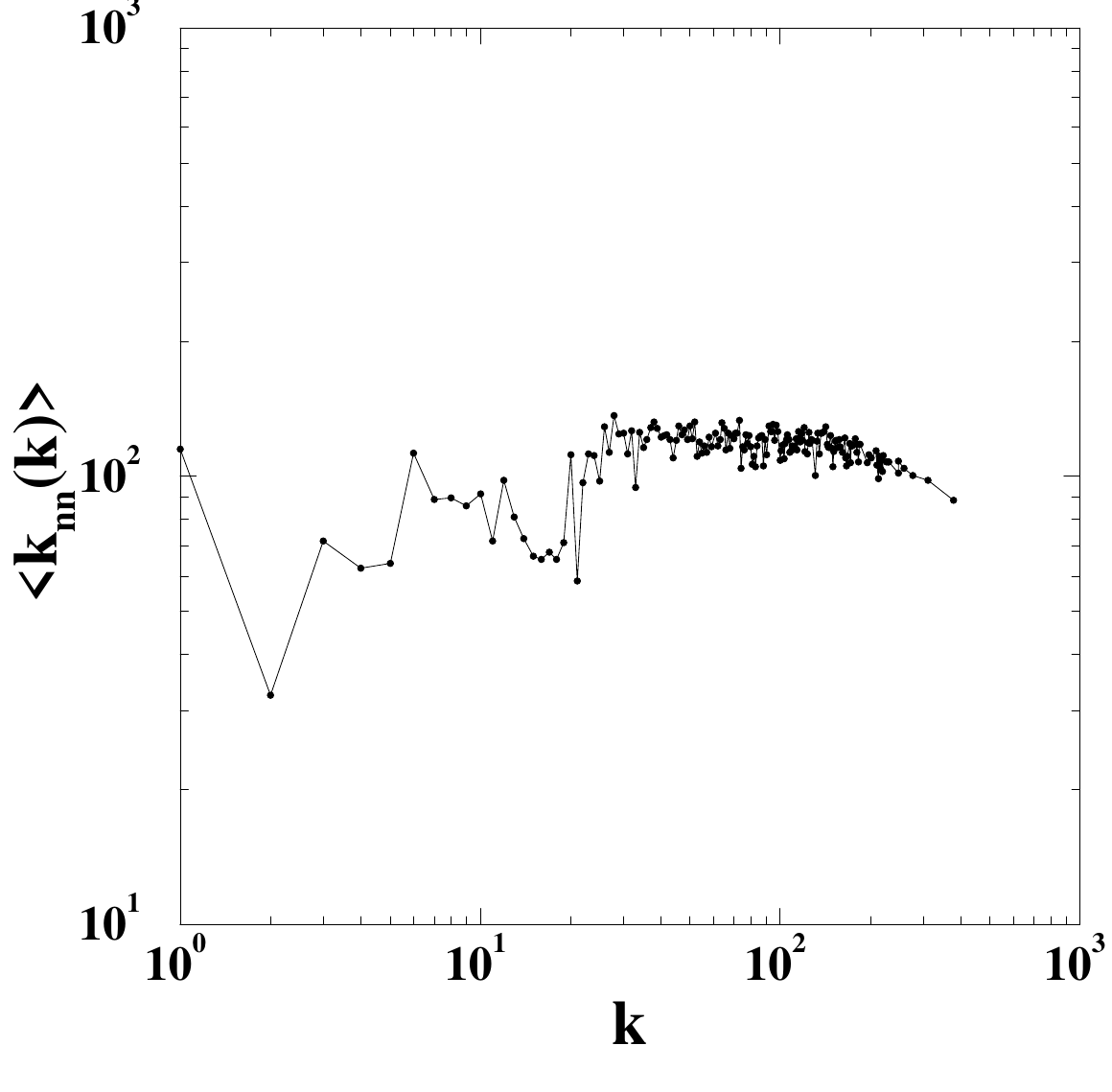}
\end{center}
\caption{Plot of the fluctuations of average degree $<k_{nn}(k)>$ of neighbors over the degree of nodes $k$ over the whole range of IRN. 
}
\label{corr_deg}
\end{figure}

With a motive of being fast and economic, railways run several trains covering short as well as long route. IRN is quite a large network consisting more than $8000$ stations over which almost $10000$ trains run over the country. For the purpose of coarse-grained study of IRN, \cite{sen2003small} has considered only $587$ stations ($N$) with $579$ tracks ($L$) represented as a grant rectangular matrix $G(N,L)$ such that $G[i,j] = 1$ if train $j$ has a stop at station $i$. In $1998$, Watts and Strogatz proposed a model of network in~\cite{watts1998collective} with properties like: diameter of the network changes very slowly as the size of the network grows over time and the network will possess dense connectivity among the neighbors of a node, termed as clustering coefficient $C(N)$. They named this network type as Small World Network (SWN) arguing it's diameter growth is similar to random network i.e. $ln(N)/N \to 0$ as $N\to\infty$ with large value of clustering coefficients $C(N) \sim 1$. Sen et al. has 
measured both of the metric over $25$ different subsets of IRN and concluded it to behave similar to a small world 
network (see Figs.~\ref{mean_distance},~\ref{cc}). Practically this implied that over the years IRN has grown up to be economic, fast i.e. very few trains need to be changed to reach any arbitrary station over the whole network. 
As per Graph theory, how well connected a node is primarily agreed upon by the number of neighbor it has i.e. degree of the node. Hence it is important to study the degree distribution $P(k)$ of IRN. The cumulative degree distribution $F(k)$ = $P(k)dk$ of IRN when plotted on a semi-logarithmic scale, has been found to fit an exponentially decaying distribution $F(k) \sim exp(-\beta k)$ with $\beta= 0.0085$. But $F(k)$ does not tell much about how well connected are the neighbors of a high degree station i.e. the correlation (whether positive or negative) of the average degree $<k_{nn}(k)>$ of neighbors to that  of the respective node. Fig.~\ref{corr_deg} shows the plot of $<k_{nn}(k)>$ against respective vertex's degree to find the general assortative behavior of the network. Using rigorous measure discussed in \cite{sen2003small}, the value came for IRN is very small ($ \sim -0.033$). 

\section{Summary and Discussion}
Unequal distribution of income, wealth and other social
features are not only common, they have been the persistent
feature throughout the world and through the history of
human civilization. Apart from the thinkers and philosophers,
the economists and other social scientists have studied
extensively about it. Recently, physicists are trying to
measure and explore the cause of such inequalities.

We first discuss (in Section~\ref{gini-kpr}) about measures
of social inequalities, including the Kolkata index $k$, giving
the fraction of wealth $k$ possessed by $(1-k)$ fraction of
the reach population: As such it generalizes the Pareto's
`80/20 law'. We have discussed in this section about the Gini
and Kolkata index values measured in various social contexts,
e.g., of the income and wealth, deaths in social conflicts and
natural disasters, citations of papers across the institutions
and journals etc.        

Next, in section~\ref{econo_income-wealth} we investigate the
nature of income and wealth distribution in various societies, 
and find a dominant feature in that typically more the ninety
percent of the population in any society has a distribution
which fits a Gamma distribution, while the upper tail part
(for the super-rich fraction of the population) fits a robust
power law or Pareto law. We show, the kinetic theory of ideal
gas, where the trading agents are like `social atoms' of the
gas as in two person trading of scattering process with conserved
money (like the conserved energy for the gas atoms) and saving
a fraction of respective money in each trading indeed gives a Gamma-like
distribution which crosses over to Pareto-like power law when
the saving fractions are inhomogeneous.

Econophysics is interdisciplinary by nature, contributed by physicists, economists, 
statisticians, social scientists, computer scientists etc. Here, an effort has been 
put to give a glimpse of recent studies from Econophysics by computation. 
Inequality measuring techniques starting from simple histogram to standard gini index as well 
as Kolkata index and others (see e.g.,~\cite{ghosh2014inequality,chatterjee2017socio,inoue2015measuring,banerjee2019kolkata}). One of the major goal of econophysics 
has been to search for a successful theory or model which can capture the behavior of the real economic data
of income or wealth distribution. In this review we briefly discuss some of the models inspired by the kinetic theory of ideal 
gases which are able to give some insights into  the mechanisms of the income distributions. 
For application of kinetic exchange models to social opinion formation, see~\cite{lallouache_bkc,biswas2012_bkc,mukherjee2016_bkc}. 
There are a few  extensive reviews and books~\cite{yakovenko2009colloquium_sudip,chakrabarti2013econophysics,pareschi2013interacting_sudip} 
highlighting these developments of the econophysics (of income and wealth distribution) as well as in sociophysics.  
Beside this field of study there are many applications of physical 
laws in financial and stock markets (see e.g.,~\cite{stanley2000introduction_sudip,sinha2010econophysics_sudip,bhadola2017_sudip,bertschinger2018reality}).
Social resource allocation models (see for e.g.,~\cite{chakrabarti2017econophysics},\cite{dhar2011_bkc}) along with several 
intelligent collective learning strategies are discussed along with their programs. We also discussed the connectivity 
structures of social networks (see e.g.,~\cite{sen2014sociophysics,chakrabarti2007econophysics}). In particular 
we discussed here for the Indian Railway network, how the minimum number of train connections (links)  one need to 
hop for going from one destination to another within India (or, for that matter, any other country), grows 
with the total number of service stations (nodes) in India (network of the respective country). Such shortest number of links in any network 
gives the idea for the time required to spread rumor or contact diseases, etc. in a networked society.

In order to give a birds'-eye-view of the developments of econophysics, we give the chronological entries in `Timeline of 
econophysics' (Appendix) for the major developments in the initial phase. For further studies and search of research 
problems in these and related fields, 
see Refs~\cite{stanley2000introduction_sudip,sinha2010econophysics_sudip,yakovenko2009colloquium_sudip,chakrabarti2013econophysics,sen2014sociophysics,chakrabarti2007econophysics,
chakrabarti2017econophysics,slanina2013essentials_sudip,pareschi2013interacting_sudip,aoyama2017macro_sudip,jovanovic2017econophysics_sudip}.

\section{Acknowledgement}
We are grateful to Muktish Acharyya for the invitation to write this review. BKC is grateful to J. C. Bose fellowship (DST, Govt. India) 
for financial support.


\appendix
\section*{Appendix}

~~~~~~~~~~~~~~~~~~~~~~~~~~~~~~~~~~~~~~{\bf{Econophysics Timeline}}
\vskip 1cm

\noindent {\bf{I. Developments up to 2000:}}
\medskip

\noindent 1931: M. N. Saha \& B. N. Srivastava in their
textbook ``A Treatise on Heat" [Indian Press,
Allahabad (1931), pp. 104-105], in the section
on Maxweall-Boltzmann velocity distribution in
ideal gases, suggested  the use of statistical
physics to explain the (assumed) Gamma-like
income/wealth distribution in any country (see
Fig.~\ref{saha_book} in subsection~\ref{saha_book_diss}, where two pages from the
section are reproduced). M. N. Saha published
several articles on scientific study of social
problems in his founded and edited journal
Science \& Culture [see, e.g., B. K. Chakrabarti,
``Econophysics as Conceived by Meghnad Saha",
Science \& Culture, Vol. 84 (2018), pp. 365-369]
and was the founder of Saha Institute of
Nuclear Physics, Kolkata (named after his death).
\medskip

\noindent 1942: E. Majorana wrote a detailed Note emphasizing on
``The Value of Statistical Laws in Physics \& in Social
Science'', published in 1942 after he went missing in
1938  [translated in to English in 2005, from Italian,
by R. N. Mantegna, published in Quantitative Finance,
Vol. 5, p. 135-140 (2005); see also R. N. Mantegna, in
Proceedings of the  XXXIII Congress of the Italian
Society for the  History of Physics and Astronomy
(SISFA 2013);  https://arxiv.org/pdf/1409.0789 and 
T. Gianfranco, ``From Galileo to Modern Economics: The 
Italian Origins of Econophysics'', Palgrave Macmillan, (2018)].
\medskip

\noindent 1960: B. Mandelbrot published ``The Pareto-Levy Law
\& the Distribution of Income", International Economic
Review, Vol. 1 (1960), pp. 79-106, arguing for
statistical mechanical modelling  of markets.
\medskip

\noindent 1963: B. Mandelbrot published ``The Variation of Certain
Speculative Prices", Journal of Business, Vol. 36 (1963),
pp. 394-419, indicating  the failure of the Random walk
statistics in  capturing the Stock price fluctuations.
\medskip

\noindent 1971: L. P. Kadanoff published the paper ``From
Simulation Model to Public Policy: An Examination of
Forrester's ``Urban Dynamics'", Transactions of
the Society for Computer Simulation, Vol. 16 (1971),
pp. 261-268, arguing for statistical physical analysis of 
social dynamics.
\medskip

\noindent 1974: E. W. Montroll \& W. W. Badger published the
book titled ``Introduction to Quantitative Aspects of
Social Phenomena", Gordon and Breach, Washington (1974).
Later, E. W. Montroll also published ``On the Dynamics \&
Evolution of Some Sociotechnical Systems", Bulletin of
the American Mathematical Society, Vol. 16 (1987), pp.
1-46, again emphasizing the importance of statistical
physics in modelling of social dynamics.
\medskip

\noindent 1995: B. K. Chakrabarti \& S. Marjit published ``Self-
Organization \& Complexity in Simple Model Systems:
Game of Life \& Economics", Indian Journal of Physics B,
Vol. 69 (1995), pp. 681-698, discussing in particular
the classical (and quantum-like) statistical mechanics
of market models and the consequent nature of income or
wealth distributions [this paper being the first Indian
one with both physicist and economist coauthors].
H. E. Stanley coined the term `Econophysics' in the
Statphys-Kolkata II Conference 1995 [Series initiated
by the Saha Institute of Nuclear Physics \& S. N. Bose
National Centre for Basic Sciences; X-th Conference held in
2019] and published in its Proceedings [H. E. Stanley et
al., ``Anomalous Fluctuations  in the Dynamics of Complex
Systems: From DNA \& Physiology to Econophysics", Physica
A: Statistical Mechanics and its Applications, Vol. 224
(1996), pp. 302-321].
\medskip

\noindent 1999: R. Mantegna \& H. E. Stanley published ``Scaling
Behaviour in the Dynamics of an Economic Index", Nature,
Vol. 376 (1999), pp. 46-49. Also, R. Mantegna \& H. E. Stanley 
``An Introduction to Econophysics", Cambridge
University Press, Cambridge (1999). 
\medskip

\noindent 2000:  A. Chakraborti
and B. K. Chakrabarti published ``Statistical
Mechanics of Money: How Saving Propensity Affects
its Distribution" European Physical Journal B,
Vol. 17 (2000), pp. 167-170; A. A. Dragulescu \& V.
M. Yakovenko published ``Statistical Mechanics of
Money", European Physical Journal B, Vol. 17 (2000),
pp. 723-729; and J.-P. Bouchaud and M. Mézard
published ``Wealth Condensation in a Simple Model of
Economy",  Physica A: Statistical Mechanics and its
Applications, Vol. 282 (2000), pp. 536-545. All these
three papers employed  ideal gas-like model in which
each agent is represented by a gas molecule in a
trading (modeled as elastic collision) with money
(modeled as kinetic energy) to investigate the
emerging features of income \& wealth distributions.
\vskip 0.7cm

\noindent{\bf{II. Developments beyond 2000:}}
\medskip

\noindent As discussed in the Introduction (see Fig.~\ref{citation_hist}) the
number of published  papers, reviews, books,
conference proceedings, reporting on several important
advances, started growing with an unprecedented rate
and it has become impossible for us to choose among
them. Instead, we give below list of a few selected
reviews, books, thesis and popular science writings,
journal special issues and conference proceedings.
\vskip 0.7cm

\noindent a) Reviews:
\medskip

\noindent 2007: A. Chatterjee \& B. K. Chakrabarti (2007), ``Colloquium: Kinetic
      Exchange Models for Income \& Wealth Distributions",
      European Physical Journal B, Vol. 60, pp. 135-149.
\medskip

\noindent 2009: V. M. Yakovenko \&  J. Barkley Rosser Jr. (2009), ``Colloquium:
      Statistical Mechanics of Money, Wealth, and Income",
      Reviews of Modern Physics, Vol. 81, pp. 1703-172.
\medskip      
      
\noindent 2011: A. Chakraborti, I. Muni Toke, M. Patriarca \& F. Abergel (2011),
      ``Econophysics Review I: Empirical Facts", Quantitative
      Finance, Vol. 11, pp. 991-1012; ``Econophysics Review II:
      Agent-based Models", Quantitative Finance, Vol. 11,
      pp. 1013-1041.
\vskip 0.7cm

\noindent b) Books:
\medskip

\noindent 2005: D. Challet, M. Marsili \& Y.-C. Zhang (2005), Minority Games, Oxford University Press, Oxford. 
\medskip      
      
\noindent 2010: S. Sinha, A. Chatterjee, A. Chakraborti \& B. K. Chakrabarti (2010)
         Econophysivcs: An Introduction, Wiley-VCH, Weinheim.
\medskip          

\noindent 2013: B. K. Chakrabarti, A. Chakraborti, S. R. Chakravarty \& A.
         Chatterjee (2013), Econophysics of Income \& Wealth
         Distributions, Cambridge University Press, Cambridge.
\medskip         
         
\noindent 2013: F. Slanina (2013), Essentials of Econophysics Modelling, Oxford
         University Press, Oxford.
\medskip      
      
\noindent 2013: P. Richmond, J. Mimkes \& S. Hutzler (2013), Econophysics \& Physical Economics,  
          Oxford University Press, Oxord.    
\medskip

\noindent 2014: Pareschi \& Toscani (2014), Interacting Multiagent Systems,
         Oxford Univ. Press, Oxford.
\medskip         
         
\noindent 2016: M. Shubik \& E. Smith (2016), The Guidance of an Enterprise Economy,
          MIT Press, Massachusetts.
\medskip         
         
\noindent 2017: H. Aoyama, Y. Fujiwara, Y. Ikeda, H. Iyetomi, W. Souma \& H.
         Yoshikawa (2017), Macro-Econophysics, Cambridge
         University Press, Cambridge.
\medskip

\noindent 2017: F. Jovanovic \& C. Schinckus (2017), Econophysics and Financial
         Economics, Oxford Univ. Press, New York.
\medskip    

\noindent 2017: B. K. Chakrabarti, A. Chatterjee, A. Ghosh, S. Mukherjee,  \&
         B. Tamir (2017), Econophysics of the Kolkata Restaurant
         Problem and Related Games: Classical and Quantum
         Strategies for Multi-agent, Multi-choice Repetitive
         Games, New Economic Windows, Springer, Cham.
\vskip 0.7cm

\noindent c) Thesis:
\medskip

\noindent 2018: C. Schinckus (2018), Thesis entitled ``When Physics Became
Undisciplined: An Essay on Econophysics", Department of
History and Philosophy of Science, University of Cambridge (Freely available online).\\
\vskip 0.7cm

\noindent d) Popular Reports and Wrtings:
\medskip 

\noindent 2002: B. Hayes (2002), ``Follow the Money", American Scientist,
             Vol. 90, pp. 400-405.
\medskip

\noindent 2005: J. Hogan (2005), ``Why it is hard to share the wealth?",
             New Scientist, 12 March 2005, p. 6.
\medskip

\noindent 2007: A. Chatterjee, S. Sinha \& B. K. Chakrabarti (2007), ``Economic
             inequality: Is it natural?", Current Science,
             Vol. 92 (2007) pp. 1383-1389. 
\medskip

\noindent 2007: M. Buchanan (2007), ``The Socail Atom", Bloomsbury, New York.
\medskip

\noindent 2012: S. Battersby (2012), ``Physics of our Finances",
             New Scientist, 28 July 2012, p. 41.
\medskip

\noindent 2012: S. Sinha \& B. K. Chakrabarti (2012), ``Econophysics: An
             Emerging  Discipline", Economic \& Political
             Weekly, Vol. 46 (2012), pp. 44-65.
\medskip  

\noindent 2019: K. C. Dash (2019), ``The Story of Econophysics'', Cambridge Scholars Publishing, Newcastle upon Tyne.
\medskip

\noindent 2019: B. M. Boghosian (2019), ``Is Inequality Inevitable?", Scientific
             American, Vol. 321, pp. 70-77.
\vskip 0.7cm

\noindent e) Special Issues:
\medskip

\noindent 2010: Fifteen Years of Econophysics Research, Science \& Culture,
     Vol. 76 (2010), pp. 293-488, Eds. B. K. Chakrabarti \&
     A. Chakraborti
     [Contributions by: B. K. Chakrabarti \& A. Chakraborti;
     B. K. Chakrabarti \& A. Chatterjee; B. M. Roehner; D.
     Helbing; F. Abergel; T. Preis; S. R. Bentes; P. Ormerod;
     R. Savit; K. Kaski; P. Richmond; H. Aoyama, Y. Fujiwara,
     Y. Ikeda; H. Iyetomi \& W. Souma; M. Patriarca, H.
     Heinsalu \& J. Kalda; M. Ausloos; J. Perello; T. Yu \& H.
     Li; D. Helbing \& S. Balietti; G. Toscani; L. Krapivsky
     \& S. Redner; V. M. Yakovenko; J. R. Iglesias; J. C.
     Ferrero; D. Maldarella \& L. Pareschi; S. Sinha; T.
     Kaizoji; J.-I. Inoue, N. Sazuka \& E. Scalas; S. Franz,
     M. Marsili \& P. Pin; M. Lallouache, A. Jedidi \& A.
     Chakraborti; M. Lallouache,  A. Chakrabarti \& B. K.
     Chakrabarti]
\medskip

\noindent 2016: Can Economics be a Physical Science?, European Physical
     Journal: Special Topics, Vol. 225 (2016), pp. 3087-3344,
     Eds.: S. Sinha, A. S. Chakrabarti \& M. Mitra
     [Contributions by: S. Sinha; A. S. Chakrabarti \& M. Mitra;
     J. Barkley Rosser Jr.; R.J. Buonocore, N. Musmeci, T. Aste
     \& T. Di Matteo; A. S. Chakrabarti \& R. Lahkar; B. K.
     Chakrabarti; A. Chakraborti, D Raina \& K. Sharma; D.
     Challet; A. Chatterjee; K. Chatterjee \& S. Roy; S. A.
     Cheong; D. K. Foley; M. Gallegati;  T. A. Huber \& D.
     Sornette;  J. P. Jericó \& M. Marsili; T. Kaizoji; K.-K.
     Kleineberg \& D. Helbing; B. LeBaron; T. Lux; R. N.
     Mantegna; S. Marjit; J. Mimkes; P. Ormerod; E. Scalas;
     C. Schinckus; V. M. Yakovenko; H. Yoshikawa]
\vskip 0.7cm

\noindent f) Conference Proceedings:
\medskip

\noindent So far nine Proc. Volumes of the Econophysics-Kolkata Series
of conferences (held in Kolkata \& Delhi) have been published
in the New Economic Windows Series of Springer. Details of
the Volumes and of the contributors are given below:
\medskip

\noindent 2005: Econophysics of Wealth Distributions, Eds. A. Chatterjee,
S. Yarlagadda \& B. K. Chakrabarti, Springer (2005), pp. 1-248
[Contributions by: F. Clementi \& M. Gallegati; V. M. Yakovenko
\& A. C. Silva; Y. Fujiwara; W. Souma \& M. Nirei; S. Sinha \&
R. K. Pan; T. Lux; J. Mimkes \& G. willis; J. Mimkes \& Y. Aruka;
A. Chatterjee \& B. K. Chakrabarti; M. Patriarca, A. Chakraborti,
K. Kaski \& G. Germano; K. Bhattacharya, G. Mukherjee \& S. S.
Manna; P. Richmond, P. Repetowicz \& S. Hutzler; Y. Wang \& N.
Ding; S. Yarlagadda \& A. Das; S. R. Gusman, M. F. Laguna \& J.
R. Iglasias; J. C. Ferrero; G. Willis; S. Sinha; M. Milakovic;
I. Bose \& S. Banerjee; A. Mehta, A. S. Majumdar \& J. M. Luck;
D. Bagchi; A. Sarkar \& P. Barat; U. K. Basu; D. P. Pal \& H. K.
Pal; B. K. Chakrabarti; P. M. Anglin; V. M. Yakovenko].
\medskip

\noindent 2006: Econophysics of Stock \& Other Markets, Eds. A.Chatterjee
\& B.K. Chakrabarti, Springer (20006), pp 1-253
[Contributions by: T. Kaizoji; A. Chakraborti; S. Sinha \& R. K.
Pan; V. Kulkarni \& N. Deo; Z. Eisler \& J. Kertesz; H. Li \& Y.
Gao; A. Sarkar \& P. Barat; D. Bagchi; U. K. Basu; D. Challet;
G. Raffaelli \& M. Marsili; C. Gao; Y. Aiba \& N. Hatano; R.
Stinchcombe; B. K. Chakrabarti, A. Chatterjee \& P. Bhattacharyya;
S. Sinha; S. Hayward; M. Mitra; P. Manimaram, J. C. Parikh, P.
K. Panigrahi, S. Basu, C. M. Kishtawal \& M. B. Porcha; M. S.
Santhanam; Y. Wang; A. Sarkar; B. K. chakrabarti; J. Barkley
Rosser Jr.; J. Barkley Rosser Jr.; M. Marsili; P. Richmond,
B. K. Chakrabarti, A. Chatterjee \& J. Angle]
\medskip

\noindent 2007: Econophysics of Markets \& Business Networks, Eds. A.
Chatterjee \& B. K. Chakrabarti, Springer (2007), pp. 1-266
[Contributions by: S. Sinha \& R. K. Pan; G. Bottazzi \& S.
Sapio; D. P. Ahalpara, P. K. Panigrahi \& J. C. Parikh; A.
Chakraborti, M. Patriarca \& M. S. Santhanam; K. B. K. Mayya
\& M. S. Santhanam; P. Repetowicz \& P. Richmond; N. Gupta,
R. Hauser \& N. F. Johnson; A.-H. Sato \& K. Shintani; A.
M. Chmiel, J. Sienkiewicz, K. Suchecki \& J. A. Holyst; K.
Bhattacharya, G. Mukherjee \& S. S. Manna; W. Souma; G. De
Masi \& M. Gallegati; M. Li, J. Wu, Y. Fan \& Z. Di; J. Angle;
S. Sinha \& N Srivastava; A. Sarkar; M. Mitra; B. K.
Chakrabarti; M. Gallegati, B. K. Chakrabarti; T. Kaizoji,
Y. Wang, Y. Fujiwara, S. Sinha, P. Richmond \& J. Mimkes]
\medskip

\noindent 2010: Econophysics \& Economics of Games, Social
Choices \& Quantitive Techniques, Eds. B. Basu, B. K.
Chakrabarti, S. R. Chakravarty \& K. Ganguly, Springer
(2010), pp. 1-394
[Contributions by: A. Ghosh, A. S. Chakrabarti \& B. K.
Chakrabarti; D. Mishra \& M. Mitra; E. W. Piotrowski, J.
Sladkowski \& A. Szczypinska; G. Szabo, A. Szolnoki \&
J. Vukov; T. Hogg, D. A. Fattal, K.-Y. Chen \& S. Guha;
S. Yarlagadda; J.-i. Inue \& J. Ohkubo; D. Faischi \&
M. Marsili; G. Toscani \& C. Brugna; B. During; K.
Gangopadhyay \& B. Basu; S. V. V. Vikram \& S. Sinha;
K. Panigrahi, S. Ghosh, P. Manimaram \&  D. P. Ahalpara;
A. K. Ray; J. Angle, F. Nielsen \& E. Scalas; V. A. Singh,
P. Pathak \& P. Pandey; J. Basu, B. Sarkar \& A.
Bhattacharya; A. sarkar, S. Sinha, B. K. Chakrabarti,
A. M. Tishin \& V. I Zverev;  K. Banerjee \& R. S. Dubey;
S. Subramanian, S. R. Chakravarty \& S. Ghosh; D.
Mukherjee; S. Maitra; V. K. Ramachandran, M. Swaminathan
\& A. Bakshi; A. Majumdar \& M. Chakrabarty; S. Das \& M.
Chakrabarty; S. Datta \& A. Mukherji; B. Chakraborty \&
M. R. Gupta; S. Lahiri; B. S. Chakraborty  \& A. Sarkar;
M. Mitra \& A. Sen; A. Kar; C. Ganguly \& I. Ray; A. Kar,
M. Mitra \& S. Mutuswami; S. Roy; K. G. Dastidar]
\medskip

\noindent 2011: Econophysics of Order-driven Markets, Eds. F.
Abergel, B. K. Chakrabarti, A. Chakraborti \& M. Mitra,
Springer (2011), pp. 1-308
[Contributions by: F. Pomponio \& F. Abergel; V. S.
Vijayaraghavan \& S. Sinha; R. Delassus \& S. Tyc; I.
Muni Toke; T. Preis; C.-A. Lehalle, O. Gueant, J.
Razafinimanana; F. Abergel \& A. Jedidi; V. Alfi, M.
Cristelli, L. Pietronero \& A. Zaccaria; F. Ghoulmie;
L. Foata, M. Vidhamali \& F. Abergel; K. Al Dayri, E.
Bacry \& J. F. Muzy; G. La Spada, J. Doyne Farmer, F.
Lillo; N. Huth \& F. Abergel; C. Y. Robert \& M.
Rosenbaum; J. Gatheral, A. Schied \& A. Slynko; F.
Baldovin, D. Bovina, F. Camana \& A. L. Stella; S.
R. Chakravarty \& D. Chakrabarti; K. Gangopadhayay \&
B. Basu; M. Mitra, M.-T. Wolfram; A. Chakraborti
\& B. K. chakrabarti]
\medskip

\noindent 2013: Econophysics of Systemic Risk \& Network Dynamics,
Eds. F. Abergel, B. K. Chakrabarti, A. Chakraborti \& A.
Ghosh, Springer (2013), pp. 1-298
[Contributions by: G. Demange; D. Lautier \& F. Raynaud;
Y. Fujiwara; F. Baldovin, F. Camana, M. Caraglio, A. L.
Stella \& M. Zamparo; S. Sinha, M. Thess \& S. Markose;
G. Tilak, T. Szell, R. Chicheportiche \& A. Chakraborti;
D. Challet \& D. M. de Lachapelle; K. Gangopadhayay \& B.
Basu; A. Zaccaria, M. Cristelli \& L. Pietronero; A.
Mehta; H. Chen \& J.-i. Inoue; A. Ghosh, S. Biswas, A.
Chatterjee, A. S. Chakrabarti, T. Naskar, M. Mitra \&
B. K. Chakrabart; P. Banerjee M. Mitra \& C. Mukherjee;
P. Sharif \& H. Heydari; T. Ibuki, S. Suzuki \& J.-i.
Inoue; S. Kumar \& N. Deo; K. C. Dash \& M. Dash; P.K.
Panigrahi, S. ghosh, A. Banerjee, J. Bahadur \& P.
Manimaran]
\medskip

\noindent 2014: Econophysics of Agent-Based Models, Eds. F.
Abergel, H. Aoyama,  B. K. Chakrabarti, A. Chakraborti
\& A. Ghosh, Springer (2014) pp. 1-301
[Contributions by: H. Kyan \& J.-i. Inoue; K.
Gangopadhayay \& K. Guhathakurata; S. Laruelle \& G.
Pages; M. Ausloos; M. Murota \& J.-i. Inoue; A. Ghosh,
A. S. Chakrabarti, A. K. Chandra, \& A. Chakraborti;
K. Roy Chowdhury, A. Ghosh, S. Biswas \& B. K.
Chakrabarti; M. Cristelli, A. Tacchella \& L. Pietronero;
T. Squartini \& D. Garlaschelli; Y. Aruka, Y. Kichikawa \&
H. Iyetomi; S. Sinha \& U. Kovur;  H. Aoyama; M. B.
Gordon, J.-P. Nadal, D. Phan \& V. Semeshenko; K. C. Dash;
A. Ghosh \& A. S. Chakrabarti; H. Aoyama]
\medskip

\noindent 2015: Econophysics \& Data Driven Modelling of Market
Dynamics, Eds. F. Abergel, H. Aoyama,  B. K. Chakrabarti,
A. Chakraborti \& A. Ghosh, Springer (2015), pp. 1-352
[Contributions by: M. Anane \& F. Abergel; S. El Aoud
\& F. Abergel; C. Kuyyamudi \& A. S. Chakrabarti; S.
Easwaran, M. Dixit \& S. Sinha; S. Fujimoto, A. Ishikawa
\& T. Mizuno; K. Guhathakurta; T. Hishikawa \& J.-i.
Inoue; A. Chaplot \& R. Sen; V. Saidevan \& S. Sinha;
S. Sridhar, T. Thiagarajan \& S. Sinha; A. Chakraborti,
Y. Fujiwara, A. Ghosh, J.-i. Inoue \& S. Sinha; A.
Chatterjee; K. C. Dash; A. Ghosh]
\medskip

\noindent 2019: New Prerspectives \& Challenges in  Econophysics
\& Sociophysics, Eds. F. Abergel,  B. K. Chakrabarti,
A. Chakraborti, N. Deo \& K. Sharma, Springer (2019),
pp. 1-272
[Contributions by: D. Challet; H. K. Pharasi, K. Sharma,
A. Chakraborti \& T. H. Seligman; I. Muni Toke; A. Bansal
\& D. Mukherjee; M. K. Verma; D. Sanyal; A. Sengupta \&
A. S. Chakrabarti; I. Vodenska \& A. P. Becker; K. Sharma,
A. S. Chakrabarti \&  A. Chakraborti; P. Bhadola \& N. Deo;
A. Chakrabarti \& R. Sen; S. Das; S. Marjit, A. Mukherji \&
S. Sarkar; A. Sawhney  \& P. Majumdar; A. Flache; U.
Niranjan, A. Singh \& R. K. Agarwal; S. S. Husain \& K.
Sharma; B. K. Chakrabarti; K. Sharma \& A. Chakraborti]

\end{document}